\documentclass[aps,prd,showpacs,amsfonts,notitlepage,amssymb,amsmath,nofootinbib,superscriptaddress]{revtex4-1}
\usepackage[utf8]{inputenc}
\usepackage[english]{babel}
\usepackage[titletoc,toc,title]{appendix}
\usepackage{amsmath}
\usepackage{amsfonts}
\usepackage{amssymb}
\usepackage[colorlinks=true,linkcolor=blue,citecolor=blue,urlcolor=blue]{hyperref}
\usepackage{graphicx}
\usepackage{enumerate}
\usepackage{csquotes}
\usepackage[caption=false]{subfig}
\usepackage{dsfont}
\usepackage{color}
\usepackage{bbold}
\usepackage{mathtools}
\usepackage{tensor}

\newcommand{\s}{\nobreak\hspace{.08em plus .04em}}
\newcommand{\ep}{\epsilon}

\newcommand{\om}{\omega}
\newcommand{\abs}[1]{\ensuremath{\left \lvert #1 \right\rvert}}

\newcommand{\eq}[1]{Eq.~(\ref{#1})}

\newcommand{\dd}{\ensuremath{\mathrm{d}}}
\newcommand{\rad}{\ensuremath{\mathrm{rad}}} 

\begin{document}

\title{
Nonlinearities induced by parametric resonance
\\ in effectively 1D atomic Bose condensates}
\date{\today}

\author{Scott Robertson\thanks{scott.robertson@th.u-psud.fr}}
\affiliation{Laboratoire de Physique Th\'eorique (UMR 8627), CNRS, Univ. Paris-Sud, Universit\'e Paris-Saclay, 91405 Orsay, France}
\author{Florent Michel\thanks{florent.michel@th.u-psud.fr}}  
\affiliation{Center for Particle Theory, Durham University, South Road, Durham, DHA 3LE, UK} 
\author{Renaud Parentani\thanks{renaud.parentani@th.u-psud.fr}}
\affiliation{Laboratoire de Physique Th\'eorique (UMR 8627), CNRS, Univ. Paris-Sud, Universit\'e Paris-Saclay, 91405 Orsay, France}

\begin{abstract}
We present a numerical study of the dynamical effects following a sudden change of the transverse trapping frequency in an elongated Bose-Einstein condensate, which induces periodic oscillations of the radial density.  
At early times, we observe an exponential growth of the number of resonant longitudinal phonons, in agreement with the predictions of the Bogoliubov-de Gennes treatment. We then observe an ordered sequence of phenomena induced by the nonlinearities of the system.  The first is a loss of the nonseparability of the resonant phonon pairs. 
This is followed by the saturation of the exponential growth and a strong depletion of condensed atoms. 
Notably, these effects are well-described by effective 1D dynamics, and are hardly affected by the damping of the radial oscillations. Finally, the atomic spectrum becomes broad, featureless and almost incoherent, in agreement with experimental results. The link between this sequence of events and the preheating scenario in inflationary cosmology is striking, as is the similarity of techniques used to study them. 
\end{abstract}

\maketitle

\section{Introduction}

Quantum field theory predicts the production of correlated pairs of particles due to temporal variations of a background field.  The particle production can be seeded either by an initial presence of particles (such as a thermal bath), or by vacuum fluctuations~\cite{Parker-1968,Birrell-Davies,Fulling}.  The latter mechanism generates pairs of opposite wave vectors $(k,-k)$ that are quantum mechanically entangled, at least when neglecting interactions with other degrees of freedom, see e.g.~\cite{Grishchuk-Sidorov,Albrecht-et-al,Campo-Parentani-2006,Campo-Parentani-2008-I}.  In a cosmological context, one generally considers a monotonically expanding universe and the particle production (mode amplification) mainly occurs when the wavelength of the excitations crosses the Hubble radius during the inflationary era.  The amplified modes give rise to the so-called Sakharov oscillations when re-entering the horizon in the radiation-dominated era.  Their observation in condensed matter has recently been reported in~\cite{Hung-Gurarie-Chin}, see also~\cite{Jaskula-et-al} for an earlier work along the same lines and~\cite{Carusotto-DCE,Fedichev-Fischer,Jain-et-al-DCE,
Volovik:1996qw, Volovik:2000ua, VolovikBook, Berges} for theoretical works where the analogy between cosmology and condensed matter is presented.  

However, the mode amplification process is more efficient and better controlled when the modifications of the background field are periodic in time, for this sets up a parametric resonance between the oscillating field and pairs of modes belonging to a finite resonant frequency window~\cite{Busch-Parentani-Robertson,Robertson-Michel-Parentani-1}, see Refs.~\cite{Jaskula-et-al, Carsten-Klempt2018} for experiments performed with atomic condensed clouds. 
In addition, parametric amplification closely corresponds to the exponential growth of the resonant modes of matter fields induced by oscillations of the inflaton field, a process normally referred to as ``preheating''~\cite{Kofman-Linde-Starobinsky,Micha-Tkachev-1} as it precedes the standard thermalization process (``reheating'') giving rise to a radiation-dominated universe at the end of the inflationary era, see~\cite{Figueroa2017JCAP} 
for a recent review.\footnote{While finishing this work, we became aware of~\cite{Eckel-et-al} where the dynamical processes in a supersonically expanding ring-shaped Bose-Einstein condensate are studied. It also leads to processes 
tending towards thermalization, although these appear not to be analogous to the preheating scenario as 
they are not driven by oscillations.} 

The efficiency of the exponential growth associated to the preheating mechanism implies that, at some point during the process, the nonlinearities of the system can no longer be neglected, i.e., 
the linear treatment used to derive the parametric resonance 
is no longer sufficient to describe the behavior of the system.  In fact, when working beyond this 
description, one faces 
{\it two} types of nonlinearity. 
The first concerns the interactions between 
the produced particles, which propagate 
in the homogeneous geometry 
described by the scale factor $a(t)$ and interacting 
with the mean value of the inflaton field $\varphi(t)$.  
The second concerns the backreaction 
of the produced particles on the equations of motion for $a$ and $\varphi$. 
Importantly, to obtain 
these ordinary differential equations, one has to take
either a spatial average over a large volume of the energy contribution of the produced particles, 
or an 
ensemble average over a set of statistically homogeneous configurations. 
One then finds the expected result that the amplitude of the coherent inflaton oscillations decreases in time, see e.g. Fig.~2 in~\cite{Figueroa2017JCAP}.  

In this paper, we shall adopt the same theoretical framework 
to study the nonlinear effects in an elongated atomic cloud which is put out of equilibrium by a sudden and large 
increase of the trapping frequency, thus inducing large and coherent oscillations of the radial density. In this context, the coherent oscillations of the atomic density 
in the narrow transverse directions act as the oscillations of the inflaton field in primordial cosmology, 
and longitudinal density fluctuations propagate on top of the homogeneous time-dependent condensed cloud. Moreover, 
the first kind of nonlinearity neglected in the Bogoliubov-de Gennes (BdG) approximation concerns the interactions between longitudinal phonons. These are governed by an effective one-dimensional Gross-Pitaevskii (GP) equation that we shall solve numerically using the truncated Wigner approximation (TWA)~\cite{Sinatra-Lobo-Castin,Mora-Castin,Walls-Milburn}~\footnote{After having completed this work, we were made aware of Ref.~\cite{Zache-Kasper-Berges} where an unstable two-component (one-dimensional) BEC is studied using the TWA.  Although the instability is not triggered by resonant oscillations, features very similar to ours (and those of the preheating scenario) are obtained. 
Namely, the growth of the occupation numbers shown in their Figures~11-13 behaves essentially as that of our phonon modes.  It would be interesting to further clarify the nature of the correspondence between the two systems. \label{fn:ZKB}}.   
The second kind of nonlinearity 
concerns the backreaction of these longitudinal phonons on the coherent radial oscillations, and as in cosmology, it
shall be calculated by taking the spatial average (over the length of the cloud) 
of their energy density. 
As a result, the radial oscillations are progressively damped,
just like 
those of the inflaton. 

Our observations can be summarized thus. 
At very early time, the system behaves essentially 
in accordance with 
the BdG formalism,
whose predictions 
concerning the occupation number and the nonseparability 
of produced phonon 
pairs in this particular context were previously studied in detail in~\cite{Robertson-Michel-Parentani-1}. 
We observe the first deviations from BdG to occur rather early, where they manifest as the loss of entanglement of the produced pairs. Interestingly, this loss occurs while the occupation numbers of the resonant modes are still increasing exponentially. 
It is induced by phonon-phonon interactions, and 
thus belongs to the first kind of nonlinearity 
in the above classification. As the system evolves further, the resonant modes become saturated and there is a fairly sudden transition during which the longitudinal part of the total energy is exchanged between all longitudinal modes in a broad band centered at $k=0$, resulting in a relatively featureless and incoherent distribution. 
Concurrent with this broadening is a large increase in the entropy of the phonon state. 
We also observe a 
reduction of the energy stored in the coherent radial excitations which is caused by the second kind of nonlinearity, 
but our simulations suggest that this is a separate effect, with the reduction of the radial oscillation energy occurring at a slower rate whenever the entropy is rapidly increasing. 
We can thus conclude that most of the nonlinear effects involving longitudinal phonons are essentially described by the effective 1D dynamics. 
Finally we observe that, 
near 
the end of our simulations, the entropy of the longitudinal phonons remains much lower than that of the thermal state with the same total 
energy. 
This means that we only observe the first steps towards thermalization. We make no claim about the time the system would take to thermalize, as the TWA is inapt to describe this properly, see e.g.~\cite{Mora-Castin}.

The paper is organized as follows.
In Section~\ref{sec:system} we outline the equations of motion used to model the system and the approximations made in their derivation,
explicitly obtaining 
the two kinds of nonlinear effect 
described above. 
In Section~\ref{sec:early} we focus on the behavior of the system at early time, i.e., up to the saturation of the resonant modes. 
The first deviations from BdG are observed, and the 
dissipative effects caused by phonon-phonon interactions are described.
In Section~\ref{sec:late} we turn our attention to the longer view, with particular emphasis on the broadening of the atomic spectrum, the 
rapid loss of the spatial coherence in the longitudinal direction, 
and the accompanying increase of the entropy encoded in the covariance matrix. 
We summarize our findings in Section~\ref{sec:conclusion}.

\section{System and approximations
\label{sec:system}}

This section is devoted to the description of our system, namely an elongated cylindrically symmetric atomic condensate with a longitudinal length $L \gg a_\perp$, where $a_\perp$ is the characteristic 
radius of the cylindric cloud. The system is assumed homogeneous in the longitudinal direction, 
and taken to be a torus of length $L$. 
It is put out of equilibrium by a sudden increase of the radial trapping frequency $\omega_\perp$, as in the first experiment of~\cite{Jaskula-et-al}. To describe the dynamical evolution of this system in a tractable manner, we shall rely on the hierarchy of various scales, and restrict our attention to phonon states which are statistically homogeneous in the longitudinal direction.

Under these conditions, to work beyond the mean field and BdG approximations, 
we shall proceed as in cosmology. (For previous works concerning backreaction 
effects in condensed matter systems, we refer the interested reader to~\cite{Kardar,CDeLGC,Stringari-2018}.) 
The first kind of nonlinearity concerns self-interactions of longitudinal excitations and will be described 
by an effective one-dimensional equation. The evolution of the state will be done using the truncated Wigner approximation (TWA)~\cite{Sinatra-Lobo-Castin,Mora-Castin,Walls-Milburn}. This method amounts to considering 
an ensemble of initial configurations described by the Wigner distribution function, and evolving each 
realization according to the effective one-dimensional Gross-Pitaevskii equation. Ensemble averages of field functions 
are then identified with expectation values of the corresponding symmetrized quantum operators. 
The second kind of nonlinearity, which concerns the backreaction of longitudinal excitations on the radial oscillations, will be described by an ordinary differential equation (ODE) driven by the spatial average of the energy carried by the former, as in studies of reheating~\cite{Figueroa2017JCAP}. In our settings, this ODE accounts for the conservation of the total energy of the system. 

As we shall now see, the implementation of this program relies on the use of a factorization of the three-dimensional wave function. For reasons of clarity, we shall present this factorization as an ansatz, then justify its legitimacy step by step. 

\subsection{The factorization ansatz
\label{sub:factorization}} 

We start with the standard three-dimensional Gross-Pitaevskii equation~\cite{Pitaevskii-Stringari-BEC}
\begin{equation}
i \hbar \, \partial_{t}\Psi = \left[ -\frac{\hbar^{2}}{2m} \nabla^{2} + V_{\rm ext} + g \left| \Psi \right|^{2} \right] \Psi \,,
\label{eq:GPE-3D}
\end{equation}
and its associated energy functional
\begin{equation}
E_{\rm 3D} = \int \mathrm{d}^{3}x  \, \left[ \frac{\hbar^{2}}{2m} \left| \nabla \Psi \right|^{2} + V_{\rm ext} \left| \Psi \right|^{2} + \frac{g}{2} \left| \Psi \right|^{4} \right] \,.
\label{eq:GPE-3D-energy}
\end{equation}
Here, $\Psi$ is the classical ($c$-number) atomic field, $m$ is the mass of a single atom, $V_{\rm ext}$ is the externally applied potential, and $g$ is the atom-atom coupling constant, related to the scattering length $a_{s}$ by the relation $g = 4 \pi \hbar^{2} a_{s} / m$.  Whenever $m$, $V_{\rm ext}$ and $g$ are independent of time, $E_{\rm 3D}$ is a constant of motion. It shall thus be 
constant after the sudden increase of the radial trapping frequency, which we shall use to put 
the system out of equilibrium. 

To study elongated (cigar-shaped) condensates which are cylindrically symmetric, we use a trapping potential of the form
\begin{equation}
V_{\rm ext} = \frac{1}{2} m \omega_{\perp}^{2} r^{2} \,,
\label{eq:Vext}
\end{equation}
where $r^2 = x^2 + y^2$.
Within our scheme of approximations, an initially cylindrically symmetric condensed cloud will remain 
so, 
having no dependence on the azimuthal angle.
As far as the longitudinal coordinate $z$ is concerned, we assume periodic boundary conditions so that the cloud effectively lives on a torus of fixed length $L$.

As explained above, in order to distinguish the two kinds of nonlinearity to be handled, we assume the following factorization of the three-dimensional wave function:
\begin{equation}
\Psi(r,\theta,z, t) = \frac{1}{\sqrt{2\pi}} \psi(r, t) \times 
\phi(z, t) \, .
\label{eq:factorization}
\end{equation}
For definiteness, we choose the following normalization conditions:
\begin{equation}
\int_{0}^{\infty} \mathrm{d}r \, r \, \left|\psi(r,t)\right|^{2} = 1 \,,  
\qquad \int_{0}^{L} \mathrm{d}z \, \left|\phi(z,t)\right|^{2} = N \,, 
\label{eq:normalization}
\end{equation}
$N$ being the total number of atoms.
$\left|\phi(z,t)\right|^{2}$ is thus the effective one-dimensional atom number density, and its spatial average, $n_{1} = N/L$, is a constant.

It is quite clear that there exist no exact $z$-dependent solutions of Eq.~(\ref{eq:GPE-3D}) which are factorized as in Eq.~(\ref{eq:factorization}). 
(This is just as in cosmology: one cannot assume that the geometry is homogeneous when the matter field configurations are not.) It behooves us to justify the use of the above factorization. Its validity rests on several conditions, which we now make explicit. 

\begin{figure}
	\includegraphics[width=0.45\columnwidth]{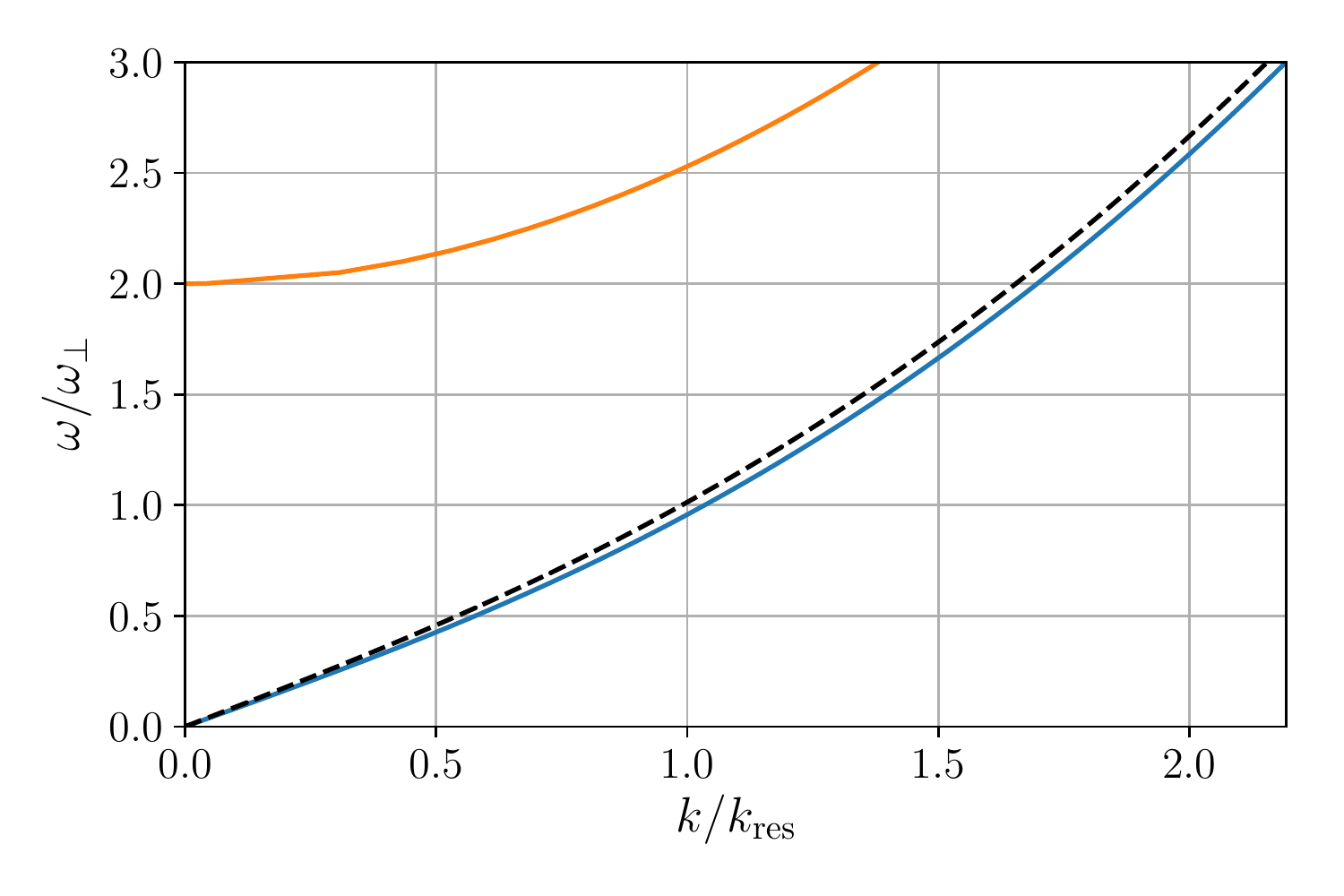} 
	\caption{Dispersion relations of the first two phonic branches with azimuthal symmetry and with $n_{1} a_{s} = 0.6$, 
		as functions of the wave number $k$
		adimensionalized by that of the resonant mode, $k_{\rm res}$, 
		see Sec.~\ref{sub:ExponentialGrowth} for its precise 
		definition. 
		The two lower curves describe, respectively, 
		the 
                numerically computed frequency without using 
		\eq{eq:factorization} (continuous blue line) and that obtained using this factorization 
		(dashed black line). Their relative difference 
		remains smaller than $8 \%$ for all $k$,
		and for $k = k_{\mathrm{res}}$ it is close to $5 \%$. 
		The upper branch (in orange) has been computed without using \eq{eq:factorization}. 
		We observe that it starts with $\omega/\omega_\perp = 2$, 
		which is a known result~\cite{Dalfovo-cited-by-Jeff}. 
	\label{fig:0}} 
\end{figure}

First of all, we shall completely neglect the longitudinal phonic excitations with nodes in the radial direction. 
The reason for this neglect is simple: these excitations 
all have a frequency which is higher than twice the trapping frequency $\omega_\perp$, see~\cite{Zaremba,Dalfovo-cited-by-Jeff} 
and Fig.~\ref{fig:0}. Moreover, $2 \omega_\perp$ is the frequency of the ``breathing'' 
(i.e. the unforced radial oscillations). 
Hence their occupation number cannot significantly increase as they cannot enter into resonance with the radial oscillations.
It should also be noted that their initial occupation is insignificant since we shall work with an initial temperature which is half of the chemical potential. In brief we work in the regime where ``the radial motion of particles is essentially frozen'' as in~\cite{Petrov}. Furthermore, when considering (at the linear level) longitudinal excitations on the lowest branch, the above factorization offers a very good description; see Fig.~\ref{fig:0}, and Fig.~13 in~\cite{Robertson-Michel-Parentani-1}. As a final comment, we should add that excitations with nodes could participate to the thermalization of the system, but we shall stop our numerical integration before their effects can become significant. 

The second condition concerns the set of phonon states we shall consider. All our initial states are 
taken to be statistically homogeneous in the longitudinal direction, and hence will remain so at all times. 
Moreover, since they are characterized by a low temperature, each realization of the ensemble 
is homogeneous to a good approximation. In our simulations, the typical value of the root-mean-square relative 
density fluctuation in the initial state is around $7\%$ for the ``benchmark'' case (presented below). 
Hence, at early times at least, and as in the BdG treatment, one can safely assume that the radial density profile does not get significantly modulated in the longitudinal direction. 

The third justification comes from an exact property of homogeneous cylindrically symmetric solutions of 
\eq{eq:GPE-3D}. Namely, for {\it any} time-dependent trapping potential of \eq{eq:Vext} quadratic in $r$, the exact evolution of the radial wave function is governed by a single scale factor we shall call $\sigma(t)$, as was shown in~\cite{Kagan-Surkov-Shlyapnikov} and exploited to study the present system in the BdG 
approximation~\cite{Robertson-Michel-Parentani-1}. Explicitly, one has
\begin{equation}
\psi(r,t) = e^{i\theta(t)} \, \frac{\sigma_0}{\sigma(t)} \, \psi(r \sigma_0/\sigma(t),0)  \ , 
\label{eq:gorasol}
\end{equation}
where $\sigma_0 \equiv \sigma(t= 0)$, 
$\psi(r,0)$ is an arbitrary exact solution of the radial GPE, 
see \eq{eq:GPE-radial}, 
and $\theta(t)$ is a phase governed by $\sigma(t)$ whose expression can be found in~\cite{Kagan-Surkov-Shlyapnikov,Robertson-Michel-Parentani-1}.
Hence the evolution of $\psi(r, t)$ is governed by the ODE obeyed by $\sigma(t)$, see below for its expression. (This equation will play the role of the Friedmann equation in cosmology). 

Finally, two scale separations ensure the stability of the results. First, owing to the fact that $L \gg a_\perp$ 
there are many longitudinal modes involved in our simulations (typically their number is on the order of $256$). 
Therefore the value of the integrated energy they carry is well-defined and hardly varies when comparing two different realizations of longitudinal modes in the TWA. Second, the damping of the radial oscillations caused by the resultant decrease of their energy is adiabatic, in the sense that the relative reduction of the amplitude of $\sigma(t)$ per oscillation is much smaller than one. 

In brief, we shall adopt the following self-consistent scheme. 
Using Eqs.~(\ref{eq:GPE-3D}), (\ref{eq:factorization}) and~(\ref{eq:gorasol}), 
we first derive an effective one-dimensional equation for $\phi(z,t)$ 
for an {\it arbitrary} function $\sigma(t)$ entering \eq{eq:gorasol}. 
This field equation governs the nonlinearities of the first kind, namely interactions between 
longitudinal excitations. Secondly, to get the modified ODE governing $\sigma(t)$ which takes into account the energy growth of these excitations, we ensemble average their energy density, which 
(because of the statistical homogeneity of the state) is equivalent to their spatial average~\footnote{\label{sigma-fluctuations} We should here point out that in our simulations, $\sigma(t)$ evolves separately for each realization, 
with the influence of inhomogeneities having been averaged over space but {\it not} over the ensemble.  
Importantly, this procedure does not lead to large fluctuations in $\sigma(t)$. In fact 
the relative variance $\left\langle \left( \sigma(t)/ \left\langle \sigma(t) \right\rangle - 1 \right)^{2} \right\rangle $ remains less than $2.5 \times 10^{-3}$ at all times for our simulations based on the benchmark parameters we discuss later in the text.}.  
Then, using the fact that $E_{\rm 3D}$ is conserved, we obtain the sought-after ODE obeyed by $\sigma(t)$. The equations for $\phi(z,t)$ and $\sigma(t)$ are thus self-consistently solved by the standard numerical recipes, much like those used in early cosmology~\cite{Figueroa2017JCAP}~\footnote{The validity of separately considering the two kinds of nonlinearity should be better in our settings than in cosmology. The reason is that the atom-atom self-interactions are repulsive while gravity is attractive. Hence there should be less clustering in our simulations than in cosmology, thereby validating our approach for longer periods of time.\label{BECvsCOSMO}}${}^{\text{,}}$\hspace*{-0.3em} 
\footnote{There is a strong
analogy between this scheme and that used in~\cite{Massar} to study black hole evaporation.
Namely, rather than working in a fixed background geometry as originally done by Hawking~\cite{Hawking75}, the outgoing flux of radiation is computed for an {\it arbitrary} slowly evaporating metric. Then the expectation value of the emitted radiation flux is computed for this geometry and used as the source term for the semi-classical Einstein equation in order to compute the mass loss of the evaporating black hole. In that case as well, the adiabaticity of mass loss is a crucial ingredient for guaranteeing the validity of the scheme.}.

\subsection{$z$-independent case} 

We start by briefly recalling the main results of~\cite{Kagan-Surkov-Shlyapnikov} which concern homogeneous cylindric condensates described in the mean field approximation. We pay 
special attention to the energy carried by such solutions.
Working in the frame in which the condensate is at rest and 
using $\phi(z,t) = \sqrt{n_{1}}$, where $n_{1}$ is the longitudinal density, 
the wave equation~(\ref{eq:GPE-3D}) (exactly) reduces to
\begin{eqnarray}
i \hbar \, \partial_{t}\psi = \hbar \omega_{\perp} a_{\perp}^{2} \left[ - \frac{1}{2r} \partial_{r} r \partial_{r} + \frac{r^{2}}{2 a_{\perp}^{4}} + 2 n_{1} a_{s} \left| \psi \right|^{2} \right] \psi \,,
\label{eq:GPE-radial}
\end{eqnarray}
where $a_{\perp} = \sqrt{\hbar/m\omega_{\perp}}$. A remarkable property of this equation is that, given any stationary solution $\psi_0(r)$, when $\omega_\perp$ is time-dependent 
the corresponding exact solution can be written 
at any time in the form of 
\eq{eq:gorasol}.
Hence the density at time $t$ is related to the initial density $\rho_0(r)$ by
\begin{equation}
\rho(t,r) = \frac{\sigma_0^{2}}{\sigma^{2}(t)} \, \rho_{0}\left( r \frac{\sigma_0}{\sigma(t)} \right) \,. 
\end{equation}
The description of the system is thus reduced to a single parameter, the scale factor $\sigma(t)$, which behaves like the position of a point particle of mass $m$ with total energy~\cite{Kagan-Surkov-Shlyapnikov,Robertson-Michel-Parentani-1} 
\begin{equation}
E_{\rm eff} = \frac{1}{2} m \dot{\sigma}^{2} + V_{\rm eff}(\sigma) = \frac{1}{2} m \dot{\sigma}^{2} + \frac{1}{2} m \omega_{\perp}^{2} \sigma^{2} + \frac{\hbar^{2}}{2 m \sigma^{2}} \,.
\label{eq:Eeff}
\end{equation}
The effective potential includes a term in $\sigma^{2}$, due to the quadratic potential $V_{\rm ext}$, and a term in $1/\sigma^{2}$, which includes repulsive forces between the atoms and the ``quantum pressure'' term that resists localization of the cloud in space~\footnote{This differs from the corresponding expression in Eq.~(5) of Ref.~\cite{Robertson-Michel-Parentani-1}, where the last term in $1/\sigma^{2}$ appears multiplied by $1+4n_{1}a_{s}$.  The resolution of this apparent paradox is that Eq.~(\ref{eq:Eeff}) allows us to renormalize $\sigma \rightarrow \sigma/l$ by dividing the last term of~(\ref{eq:Eeff}) by $l^{4}$ and multiplying the total energy in Eq.~(\ref{eq:GPE-radial-energy}) by $l^{2}$.  Here, we have chosen $\sigma$ in such a way that $E_{\rm eff}$ becomes independent of $n_{1}a_{s}$, while in Ref.~\cite{Robertson-Michel-Parentani-1} $\sigma$ had an ``absolute'' normalization as the width of the Gaussian profile we assumed for $\left|\psi\right|^{2}$.}. 
Using our conventions, in a stationary state (i.e. $\dot{\sigma} = 0$), the value of $\sigma$ is exactly $a_{\perp}$, as this is the position at which the potential $V_{\rm eff}$ is minimum. 

Using Eq.~(\ref{eq:GPE-3D-energy}), a careful analysis (see Appendix~\ref{app:cylind_solns}) shows that the energy of the oscillating cloud 
is proportional to $E_{\rm eff}$.  Since the total energy is an extensive quantity, 
it is 
also 
proportional to the total number of atoms $N$.  There remains a dimensionless factor which depends on $n_{1}a_{s}$, i.e. on the coefficient of the nonlinear term in Eq.~(\ref{eq:GPE-radial}).  We can thus write
\begin{equation}
E_{\rm rad} = A\left(n_{1}a_{s}\right) \, N \, E_{\rm eff} \,.
\label{eq:GPE-radial-energy}
\end{equation}
The calculation of 
$A\left(n_{1}a_{s}\right)$ is done at the end of Appendix~\ref{app:cylind_solns}. 

\subsection{Effective one-dimensional
equation for $\phi(z)$}

To proceed, we use the form of 
the radial function $\psi$ 
described in the previous subsection.  Now, using Eq.~(\ref{eq:factorization}), we can subtract from the full GPE~(\ref{eq:GPE-3D}) the equation of motion satisfied by $\psi(r,t)$, see Eq.~(\ref{eq:GPE-radial}). 
We then multiply the remainder by $r \, \psi^{\star}(r,t)$ and integrate over $r$, leaving the following equation for $\phi(z)$:
\begin{equation}
i \hbar \, \partial_{t}\phi = - \frac{\hbar^{2}}{2m} \, \partial_{z}^{2}\phi + g_{1}(t) \left( \left| \phi \right|^{2} - n_{1} \right) \phi \,,
\label{eq:GPE-1D}
\end{equation}
where
\begin{equation}
g_{1}(t) = \frac{2 \hbar^{2} a_{s}}{m} \int_{0}^{\infty} \mathrm{d}r \, r \, \rho^2(t,r) \,.
\label{eq:g1-defn}
\end{equation}
It is 
straightforward to show that the integral over $ r \rho^2(t,r)$ is proportional to $1/\sigma^{2}$, with a dimensionless constant of proportionality that depends on 
$n_{1}a_{s}$:
\begin{equation}
g_{1}(t) = \frac{2 \hbar^{2} a_{s}}{m} \frac{G\left(n_{1}a_{s}\right)}{\sigma^{2}(t)} \,. 
\label{eq:g1-G}
\end{equation}
As shown in Appendix~\ref{app:cylind_solns}, $G\left(n_{1}a_{s}\right)$ is related to $A\left(n_{1}a_{s}\right)$
entering in \eq{eq:GPE-radial-energy}.

We can also insert the factorization ansatz~(\ref{eq:factorization}) into the energy functional~(\ref{eq:GPE-3D-energy}) and subtract the energy due to the radial motion, leaving just that part of the energy which is due to longitudinal excitations.  The result (using the fact that, by definition, $n_{1}$ is the spatial average of $\left|\phi\right|^{2}$) is
\begin{equation}
E_{\rm long} = \int_{0}^{L} \mathrm{d}z \, \left[ \frac{\hbar^{2}}{2m} \left| \partial_{z}\phi \right|^{2} + \frac{g_{1}(t)}{2} \left( \left| \phi \right|^{2} - n_{1} \right)^{2} \right] \,.
\label{eq:GPE-1D-energy}
\end{equation}
This clearly vanishes in the $z$-independent case, where $\phi = \sqrt{n_{1}}$.  It is also straightforward to show that Eq.~(\ref{eq:GPE-1D}) follows from treating $E_{\rm long}$ of Eq.~(\ref{eq:GPE-1D-energy}) as the energy functional.  Thus, whenever $g_{1}$ is constant in time, $E_{\rm long}$ is a constant of motion.  However, in the case of interest to us, $g_{1}$ 
varies in time due to the radial oscillations, and $E_{\rm long}$ is not conserved. Then, because of resonant phonons, $E_{\rm long}$ will grow exponentially at early times, thereby correspondingly reducing $E_{\rm rad}(t)$, the energy stored in radial oscillations. 
Accounting 
for this backreaction 
(which is the second kind of nonlinearity in our classification) is the goal of the next subsection.

\subsection{Backreaction -- determining the ODE obeyed by $\sigma(t)$}

The longitudinal energy of Eq.~(\ref{eq:GPE-1D-energy}) can 
be written in a synthetic form as
\begin{equation}
E_{\rm long}(t) = \frac{\hbar^{2}}{2m} \int_{0}^{L} \mathrm{d}z \, \left| \partial_{z}\phi(z,t) \right|^{2} + \frac{1}{2} N g_{1}(t) V_{\rm long}(t) \,, 
\end{equation}
where $V_{\rm long}(t)$ is given by
\begin{equation}
V_{\rm long}(t) = \frac{1}{N} \int_{0}^{L} \mathrm{d}z \, \left( \left| \phi(z,t) \right|^{2} - n_{1} \right)^{2} \,.
\label{eq:Vz-defn}
\end{equation}
By direct inspection, one sees that $V_{\rm long}(t)$
 quantifies the departure from translation invariance along the longitudinal direction. 
Then, by using Eq.~(\ref{eq:GPE-1D}), one obtains
\begin{eqnarray}
\partial_{t} E_{\rm long} &= &\partial_{t}g_{1} \frac{N V_{\rm long}(t)}{2} \nonumber\\
&=& - a_{s} G\left(n_{1}a_{s}\right) N \frac{2 \hbar^{2} V_{\rm long}(t)}{m} \frac{\dot{\sigma}}{\sigma^{3}} \,,
\label{eq:dotElong}
\end{eqnarray}
where we have used Eq.~(\ref{eq:g1-G}) to get the second line. 

On the other hand, the time-derivative of $E_{\rm rad}$ of Eq.~(\ref{eq:GPE-radial-energy}) gives 
\begin{equation}
\partial_{t}E_{\rm rad} = A\left(n_{1}a_{s}\right) N \partial_{t}E_{\rm eff} = A\left(n_{1}a_{s}\right) N \dot{\sigma} \left( m \ddot{\sigma} + m \omega_{\perp}^{2} \sigma - \frac{\hbar^{2}}{m \sigma^{3}} \right) \,.
\end{equation}
Imposing energy conservation $\left(E_{\rm rad} + E_{\rm long}\right)= \mathrm{cst}.$ 
implies that $\sigma$ obeys $m \ddot{\sigma} = -\partial_{\sigma}V_{\rm eff}^{\rm BR}$, where the corrected effective potential is 
\begin{equation}
V_{\rm eff}^{\rm BR}\left(\sigma\right) = \frac{1}{2} m \omega_{\perp}^{2} \sigma^{2} + \frac{\hbar^{2}}{2 m \sigma^{2}} \left( 1 + 2 a_{s} \frac{G\left(n_{1}a_{s}\right)}{A\left(n_{1}a_{s}\right)} V_{\rm long}(t) \right) \,,
\label{eq:Veff_backreaction}
\end{equation}
the backreaction of longitudinal phonons being governed by the last term. For benchmark values discussed below, we get $G\left(n_{1}a_{s}\right)/A\left(n_{1}a_{s}\right)= 0.43$ . 

An important result of our simulations is that this 
backreaction plays hardly any role in the observed 
early deviations with respect to BdG predictions. 
This implies that they are essentially governed by \eq{eq:GPE-1D}. 

\subsection{Describing the initial phonon state, benchmark values
\label{sec:state_prep}}

Because of the homogeneity of the background condensate, the quantum phonon state is conveniently expressed in terms of the longitudinal momenta $\hbar k$ of the atoms, which correspond to the Fourier modes of the quantum field $\hat{\phi}$:
\begin{equation}
\hat{\phi}(z) = \frac{1}{\sqrt{L}} \sum_{k \in 2 \pi \mathbb{Z} / L} \hat{\phi}_{k} \, e^{i k z} \,.  
\end{equation}
The normalization factor $1/\sqrt{L}$ is chosen so that the $\hat{\phi}_{k}$ are standard bosonic amplitude operators: $\hat{\phi}_{k}$ and $\hat{\phi}_{k}^{\dagger}$ destroy and create, respectively, an atom of momentum $\hbar k$, and obey the commutation relation $\left[ \hat{\phi}_{k} \,, \,\hat{\phi}_{k^{\prime}}^{\dagger} \right] = \delta_{k,k^{\prime}}$. 
In the BdG formalism, the Hamiltonian is not diagonalized by the atom operators $\hat{\phi}_{k}$, but by the phonon operators $\hat{\varphi}_{k}$, these being related by the $SU(1,1)$ linear transformation
\begin{equation}
\left[ \begin{array}{c} \hat{\phi}_{k} \\ \hat{\phi}_{-k}^{\dagger} \end{array} \right] = \left[ \begin{array}{cc} u_{k} & v_{k} \\ v_{k} & u_{k} \end{array} \right] \left[ \begin{array}{c} \hat{\varphi}_{k} \\ \hat{\varphi}_{-k}^{\dagger} \end{array} \right] \, ,
\label{eq:phi_varphi}
\end{equation}
where $u_k$ and $v_k$ are normalized so that $u_k^2 - v_k^2 = 1$. 
In strict analogy to the atom operators, the phonon operators $\hat{\varphi}_{k}$ and $\hat{\varphi}_{k}^{\dagger}$ destroy and create, respectively, a phonon of momentum $\hbar k$, and obey the commutation relation $\left[ \hat{\varphi}_{k} \,,\, \hat{\varphi}_{k^{\prime}}^{\dagger} \right] = \delta_{k,k^{\prime}}$.  
When neglecting phonon-phonon interactions, the BdG formalism is exact. In addition,
when the two-mode state $\left(k,-k\right)$ 
is homogeneous and Gaussian, it is completely determined by the expectation values 
\begin{equation}
n_{\pm k}^{\rm ph} = \left\langle \hat{\varphi}_{\pm k}^{\dagger} \hat{\varphi}_{\pm k} \right\rangle \,, \qquad c_{k}^{\rm ph} = \left\langle \hat{\varphi}_{k} \hat{\varphi}_{-k} \right\rangle \,,
\end{equation}
with analogous expressions for the atomic expectation values $n_{\pm k}^{\rm at}$ and $c_{k}^{\rm at}$. 
As is well known, when the phonons are initially in a thermal state at temperature $T$, their Wigner function takes the form~\cite{LeonhardtQO} 
\begin{equation}
W\left(\varphi_{k} \,, \varphi_{k}^{\star}\right) = \frac{1}{2 \pi \left(n_{k}^{\rm ph} +1/2 \right)} \, \mathrm{exp}\left( - \frac{\left|\varphi_{k}\right|^{2}}{n^{\rm ph}_{k}+1/2} \right) \,, \qquad 2 n^{\rm ph}_{k}+1 = \mathrm{coth}\left(\frac{\hbar \omega_{k}}{2 k_{B} T}\right) \,. 
\label{eq:Wigner_thermal}
\end{equation}

When using the truncated Wigner approximation (TWA), the initial state is prepared by randomly selecting the phonon amplitudes $\varphi_{k}$ (for $k \neq 0$) distributed according to the probability distribution of Eq.~(\ref{eq:Wigner_thermal}). 
We then transform these into (initial) atomic amplitudes via the Bogoliubov transformation of Eq.~(\ref{eq:phi_varphi}). 
Notice that 
the $k=0$ component $\phi_{0}$ is 
determined by imposing a fixed total number of atoms $N$: 
$N_{0} = N - \sum_{k \neq 0} n_{k}$, 
where $n_{k} + 1/2 = \left|\phi_{k}\right|^{2}$ and $\phi_{0} = \sqrt{N_{0}+1/2}$ 
is taken to be real and positive so that $u$ and $v$ are properly defined as real quantities themselves. 
Note that we could also have chosen to fix the number of condensed atoms $N_{0}$ rather than the total; we have checked that this makes little difference in the simulations of interest to us. 

Having prepared the initial state 
assuming the validity of the BdG formalism, 
the nonlinearities of the system are taken into account by letting each configuration evolve under Eq.~(\ref{eq:GPE-1D}) for a period of 
time during which the background is stationary (i.e., the width $\sigma$ is constant), allowing 
the system to 
settle in a nearly stationary state before 
the sudden increase of the trapping frequency. 
After some trial and error, 
and taking heed of 
the infrared and ultraviolet constraints on the spatial discretization encountered in one-dimensional quasi-condensates (see Ref.~\cite{Mora-Castin}), 
we settled on the following ``benchmark'' 
values for the parameters: 
\begin{itemize}
\item the initial temperature is fixed at a modest value of $T_{\rm in} = m c_{\rm in}^{2} / 2 = g_{1,{\rm in}} n_{1} / 2$ 
(where $c_{\rm in}$ is the initial value of the speed of low energy phonons); 
\item the mean 1D atomic density is given by $n_{1} a_{s} = 0.6$, which is relatively large (3D effects come into play for $n_{1}a_{s} \gtrsim 1$) but which is close to that used in the first experiment reported in Ref.~\cite{Jaskula-et-al}~\footnote{To avoid any confusion, 
in this first experiment, pair production of longitudinal phonons was triggered by a sudden increase of the radial trapping frequency.  As can be understood from Eq.~(\ref{eq:Eeff}), this sudden increase induced oscillations of the radial density with an angular frequency equal to $2 \omega_{\perp}$. Their amplitude was significantly larger than that of the second experiment of~\cite{Jaskula-et-al}, which was induced by the controlled modulation of the trapping frequency.};
\item the radial oscillations commence after a sudden contraction of the trapping potential, which is fixed by $\omega_{\perp}/\omega_{\perp, {\rm in}} = \sqrt{2}$, exactly as in~\cite{Jaskula-et-al} (note that $\omega_{\perp}$ and $a_{\perp}$, with no clarifying subscript, shall always refer to their values {\it after} the sudden change of the trapping potential);
\item the length of the torus is given by $L/a_{\perp} = 128$, which is considerably larger than in experiments but which gives reasonable resolution in $k$ and typically gives two discrete values of $k$ within the resonant window~\cite{Busch-Parentani-Robertson}; 
\item the harmonic oscillator length is such that, typically, $a_{s}/a_{\perp} = 1.7 \times 10^{-3}$; combining this with the values of $n_{1} a_{s}$ and $L/a_{\perp}$ above leads to a typical total atom number of $N = 4.5 \times 10^{4}$, roughly an order of magnitude larger than in~\cite{Jaskula-et-al};
\item the grid spacing is fixed at $\Delta x / a_{\perp} = 1/2$, which when combined with the value of $L/a_{\perp}$ above gives a total number of grid points / phonon modes of 256; this ensures that $\Delta x$ is smaller than the healing length ($\xi/a_{\perp} \approx 1.15$) but considerably larger than the scattering length, as required for the validity of the TWA~\cite{Sinatra-Lobo-Castin}; 
\item the time spacing is fixed at $\omega_{\perp} \, \Delta t = 10^{-2}$, which is such that $1/\Delta t \gg \omega_{k_{\rm max}}$, the largest frequency of the phonon modes, thus ensuring that there are no spurious resonance effects due to the discretization of time. 
\end{itemize}
Note that the last two of the listed benchmark parameters are not physical but are required by the numerics.  There is thus some freedom in the choice of these parameters, which should not lead to any significant changes in the physical predictions of the simulations.  We have checked that this is indeed the case.  
We also checked that the coherence length $l_\phi(T)$~\cite{Bouchoule} 
is of the order of $10\, L$ which means that, 
for the benchmark values, one deals with a quasi-condensate before the sudden increase of the trapping potential. This point shall be further discussed in Sec.~\ref{first-order-coh}. 

In forthcoming simulations, to display the behavior of nonlinearities neglected in the BdG approximation, we shall consider three values of $a_{s}/a_{\perp}$, namely, $1.7 \times 10^{-4}$, $1.7 \times 10^{-3}$ (which is the above benchmark value), and $1.7 \times 10^{-2}$.  Instead, $T_{\rm in}/m c_{\rm in}^{2}$, $n_{1} a_{s}$, $\omega_{\perp}/\omega_{\perp,{\rm in}}$ and $L/a_{\perp}$ will remain fixed so that the three cases share the same BdG description.  As we shall clearly see, increasing the value of $a_{s}/a_{\perp}$ increases the deviations with respect to BdG predictions.

\subsection{Following the evolution of the state of longitudinal excitations
\label{observables}} 

Let us now explain which observables we shall use to follow the state after the sudden change in $\omega_{\perp}$. 
When considering {\it in situ} measurements, such as in~\cite{Schley-et-al,Steinhauer-2014,Steinhauer-2016}, one typically has access to the 1D atomic density $\hat{\rho}(t,z) = \hat{\phi}^{\dagger}(t,z) \hat{\phi}(t,z)$ in each realization. 
To have access to the population and entanglement of the phonon state, it is useful to consider the normalized equal-time two-point correlation function in $k$ space~\cite{Robertson-Michel-Parentani-1}: 
\begin{equation}
G^{(2)}(t,k; t,k^{\prime}) = \frac{1}{N} \, \left\langle \hat{\rho}_{k}(t) \hat{\rho}^{\dagger}_{k^{\prime}}(t) \right\rangle \,, 
\label{eq:G2_defn}
\end{equation}
where we have defined
\begin{equation}
\hat{\rho}_{k}(t) = \int_{0}^{L} \mathrm{d}z \, e^{-i k z} \, \hat{\rho}(t,z) \,.
\end{equation}
The usefulness of $G^{(2)}$ stems from its close relationship to the phonon state: in a statistically homogeneous state, $G^{(2)}$ is only non-zero when $k=k^{\prime}$. 
Moreover, 
when the background is stationary, it 
always has the form
\begin{equation}
G_{k}^{(2)}(t) = \left(u_{k}+v_{k}\right)^{2} \left( 1 + n_{k}^{\rm ph} + n_{-k}^{\rm ph} + 2 \, \mathrm{Re} \left[ c_{k}^{\rm ph} e^{-2 i \omega_{k} t} \right] \right) \,, 
\label{eq:G2k}
\end{equation}
where $n_{\pm k}^{\rm ph} = \left\langle \hat{\varphi}_{\pm k}^{\dagger} \hat{\varphi}_{\pm k} \right\rangle$ is the (constant) number of phonons at wave vector $\pm k$, while the complex number $c_{k}^{\rm ph} = \left\langle \hat{\varphi}_{k} \hat{\varphi}_{-k} \right\rangle$ gives the phase and the strength of the correlation between $k$ and $-k$ phonons. 
It should be noticed that $G_{k}^{(2)}$ is governed by symmetrized expectation values of operators (as $\hat{\rho}_{k}(t)$ and $\hat{\rho}^{\dagger}_{k^{\prime}}(t)$ commute with each other). 
It is thus appropriate to use the TWA to evaluate it as this method, by construction, delivers the expectation values 
of symmetrized operators.

From \eq{eq:G2k} we see that 
the time averaged value of $G^{(2)}_{k}(t)$ gives the total phonon number while the amplitude of the oscillations about the mean gives the strength of the correlations. Interestingly, the dipping of $G^{(2)}_{k}$ below its vacuum expectation value $\left(u_{k}+v_{k}\right)^{2}$ is sufficient to conclude that the two-mode state $(k,-k)$ is nonseparable~\cite{Robertson-Michel-Parentani-1,Robertson-Michel-Parentani-2}~\footnote{The sudden change produces phonons in pairs $(k,-k)$, which are entangled if they are seeded principally by vacuum fluctuations (rather than, say, an initial thermal distribution).  For mixed states, the concept of entanglement is ambiguous, and various definitions have been proposed. We refer to our former work~\cite{Robertson-Michel-Parentani-2} for a recent comparison between nonseparability and steerability. In what follows we shall use the notion of nonseparability.  Simply put, given a natural division of a system into two subsystems (here the wave vectors $k$ and $-k$), the state is nonseparable if the correlations between the two subsystems are so strong that they cannot be represented by a classically correlated state.}. 
In fact, a sufficient criterion for nonseparability is 
\begin{equation}
n_{k}^{\rm ph} n_{-k}^{\rm ph} - \left|c_{k}^{\rm ph}\right|^{2} < 0 \,. 
\label{eq:nonsep}
\end{equation}
As we shall see in the next figures, this threshold translates rather simply when following 
the evolution of $G^{(2)}_{k}$ in time. 

Note that, although the behavior of $G^{(2)}_{k}(t)$ is closely related to the phonon state, it is obtained by measuring the density of atoms.  There is thus no need to explicitly transform into the phonon basis when calculating $G^{(2)}_{k}(t)$. Moreover, $G^{(2)}_{k}(t)$ is still well-defined 
when $n_k^{\rm ph}$ becomes so large that the Bogoliubov approximation becomes invalid. In other words, it is only the reading of the $G^{(2)}_{k}(t)$ in terms of linear phonon modes which becomes invalid at late times. 

In parallel to the study of the $G^{(2)}$, it is also useful to follow the evolution of the atom 
content of the state,
especially at late time where the notion of phonons becomes inappropriate due to the progressive loss of coherence of the condensate. 
In this case, we simply use $n_{k}^{\rm at}$ and $c_{k}^{\rm at}$, which give respectively the mean occupation number of atoms of momentum $k$ and their correlation amplitude, 
in strict analogy with the above $c_{k}^{\rm ph}$ for phonons.

\section{Early-time behavior
\label{sec:early}}

In this section, we consider the early-time evolution of the system after the sudden change in the trapping frequency $\omega_\perp$, as was done in the first type of experiment reported in~\cite{Jaskula-et-al}. 
The ``early-time'' period roughly corresponds to the time during which there is an exponential growth of the phonon number near resonance, as predicted by the BdG formalism~\cite{Busch-Parentani-Robertson,Robertson-Michel-Parentani-1}.
This exponential growth is described in subsection~\ref{sub:ExponentialGrowth}, while in subsection~\ref{sub:NonsepDissipation} we consider the first observable differences from the BdG predictions due to phonon interactions, namely the loss of nonseparability of $(k,-k)$ pairs, as well as a small reduction of the exponential growth of resonant modes. In particular, we observe 
that 
these deviations from BdG 
are governed by
two dissipative rates which scale with 
different powers of the number of resonant phonons, 
$n_{\rm res}(t)$.
In other words, 
the time dependence can be eliminated by using as parameter the number $n_{\rm res}$ itself. 
Moreover, to a good approximation, the dissipative rates only depend on the combination $n_{\rm res} \times a_s/a_\perp$  
when using the benchmark value ($a_s/a_\perp=1.7 \times 10^{-3}$) and ten times smaller, 
i.e., in the weak coupling limit. 

Before presenting the results, it should be recalled that during this early period, the radial energy $E_{\rm rad}(t)$ hardly varies. Hence the results we present in this section could have been obtained by considering only Eq.~\eqref{eq:GPE-1D} with a periodically modulated $g_1(t)$, 
i.e., by ignoring the backreaction 
governed by the last term in Eq.~\eqref{eq:Veff_backreaction}.

\subsection{Parametric amplification of resonant phonon modes 
\label{sub:ExponentialGrowth}}

As seen in previous works~\cite{Busch-Parentani-Robertson,Robertson-Michel-Parentani-1}, the BdG treatment, which neglects interactions between phonons, predicts exponential growth of the number of phonons at and around wave vectors $\pm k_{\rm res}$, where for a background modulated at frequency $\omega_{p}$, $k_{\rm res}$ is determined by the relation $2\omega_{k_{\rm res}} = \omega_{p}$.  If the trapping frequency is $\omega_{\perp}$, it is straightforward to show~\cite{Robertson-Michel-Parentani-1} that $\omega_{p} = 2\omega_{\perp}$, so the resonance condition becomes $\omega_{k_{\rm res}} = \omega_{\perp}$.
It should be mentioned here that there is a finite resonant window (in $k$-space) wherein the occupation number grows exponentially in time, see Appendix~A of~\cite{Busch-Parentani-Robertson} for an analytical description of this aspect. 

\begin{figure}
\includegraphics[width=0.45\columnwidth]{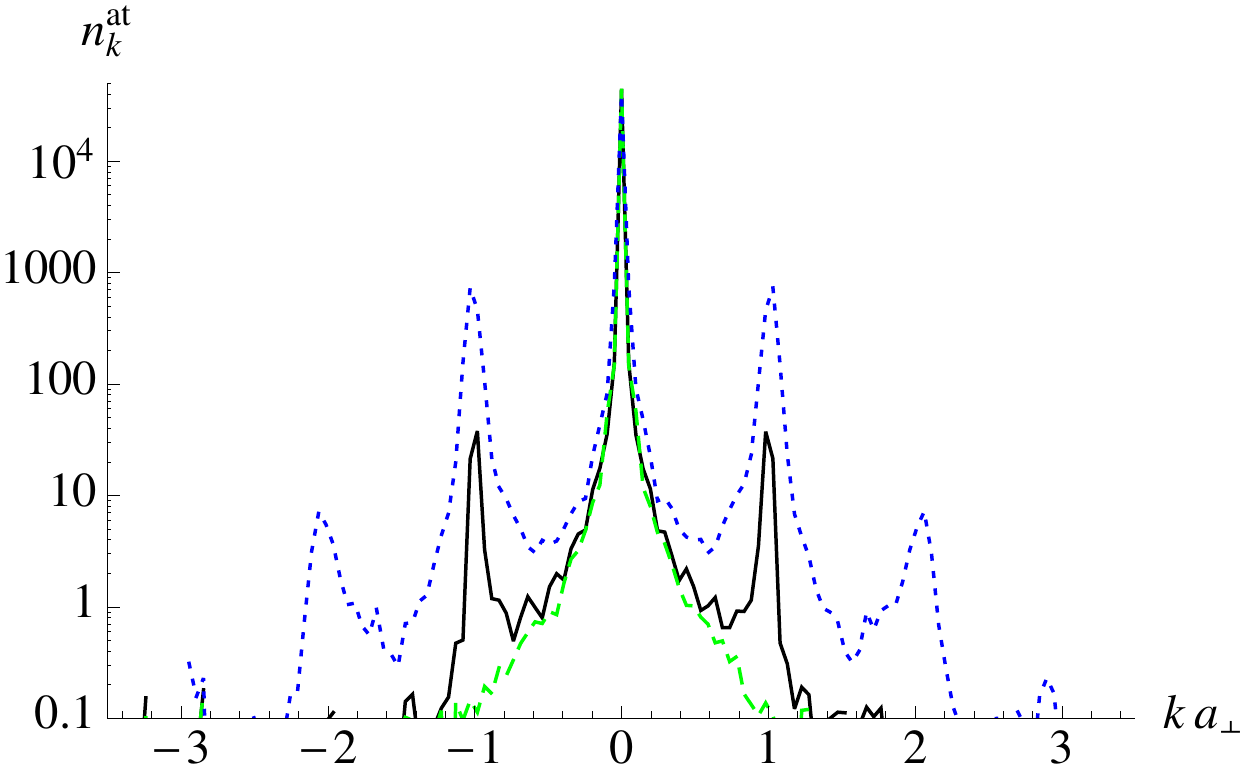} \, \includegraphics[width=0.45\columnwidth]{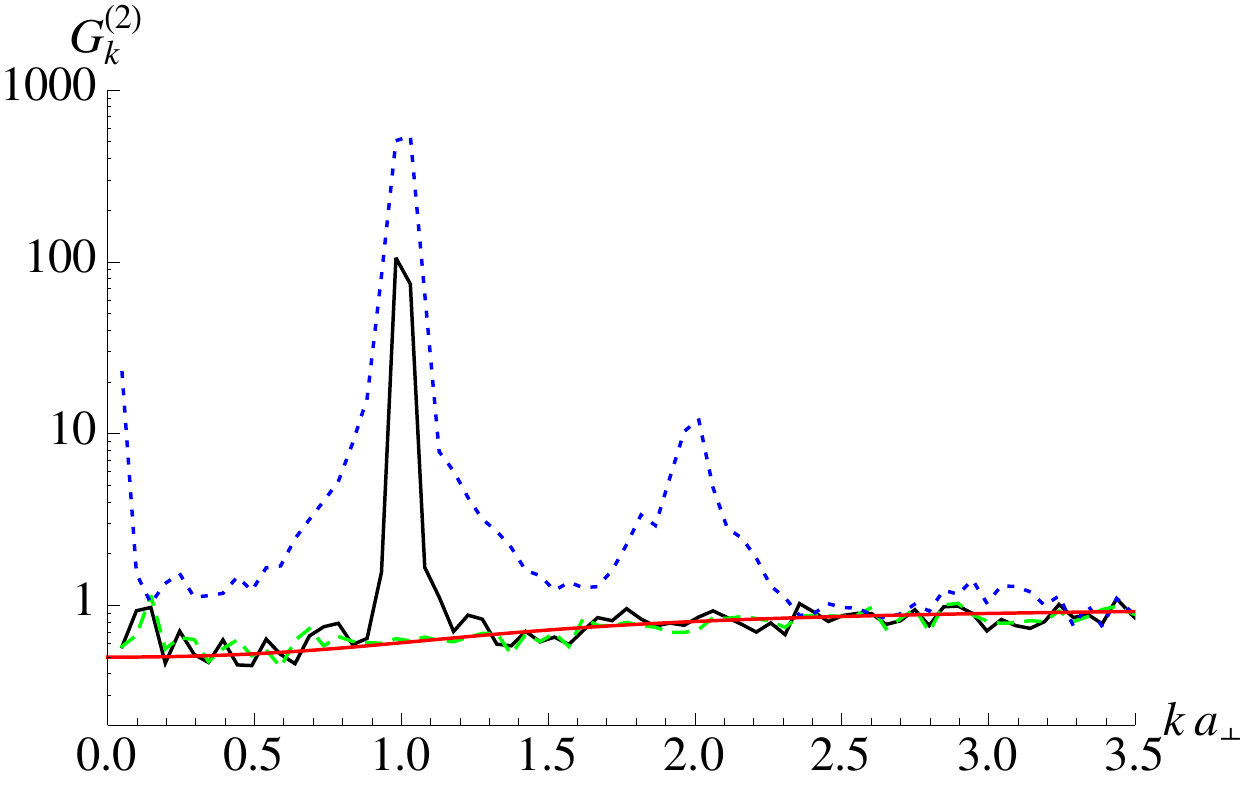}
\caption{Evolution of the whole system at early times.  Shown are the number of atoms (left panel) and the density-density correlation function (right panel) as functions of $k$ at three different times: $N_{\rm osc} = 0$ (green dashed curve), $14.3$ (black solid curves) and $28.6$ (blue dotted curves), 
where the two last values of $N$ have been chosen so as to clearly illustrate the initial growth of the resonant peak followed by the broadening and the growth of the second harmonic. 
Note that the peaks do not grow steadily, but show significant oscillations, see Figure~\ref{fig:G2} below. 
The 
parameters have their benchmark values 
given at the end of Sec.~\ref{sec:state_prep}, averaged over $100$ realizations.  The red curve on the right panel shows the form of $G^{(2)}_{k}$ at an initial temperature $T_{\rm in} = m c_{\rm in}^{2}/2$, according to the BdG theory.  The black solid curves show the growth of the resonant peak where $\omega_{k} = \omega_{\perp}$, as predicted by BdG; the blue dotted curves show deviations from BdG through the appearance of harmonics and the broadening of the peaks. 
\label{fig:earlytime}}
\end{figure}

\begin{figure}
\includegraphics[width=0.45\columnwidth]{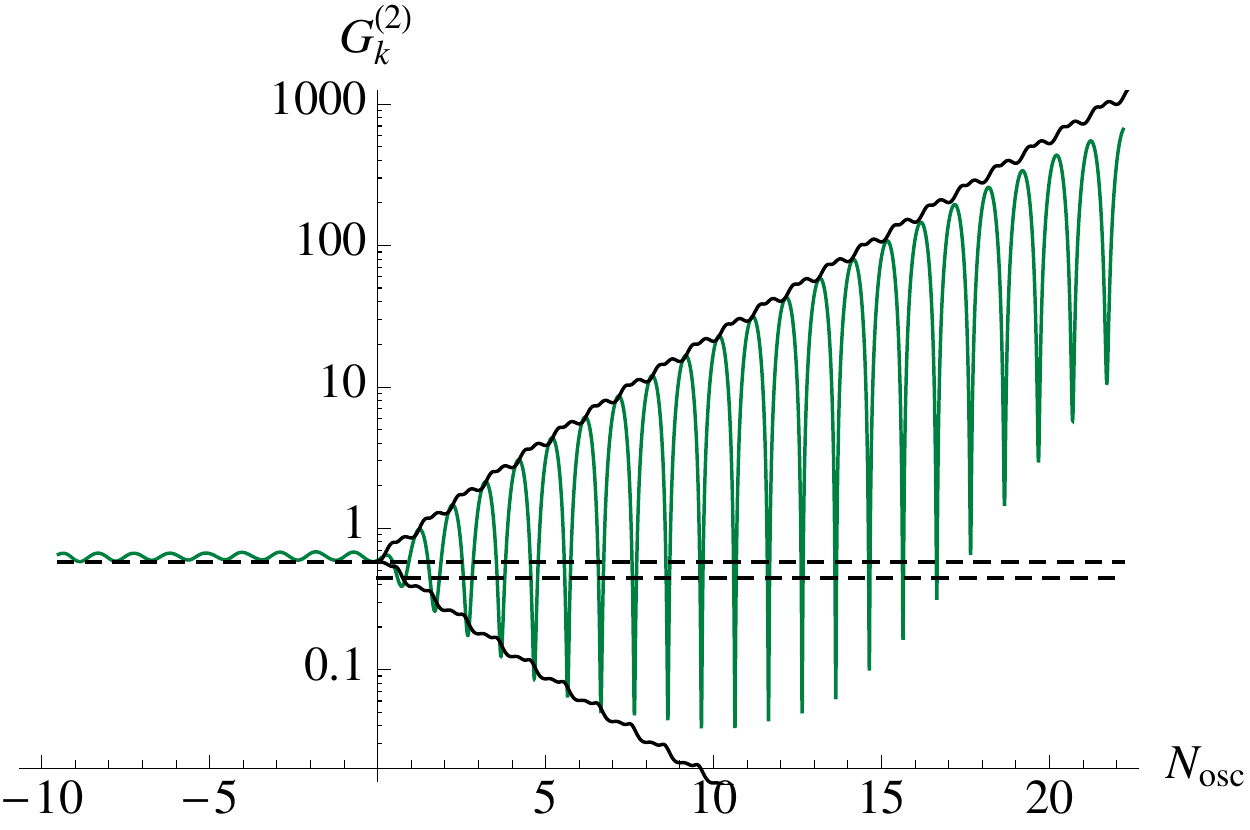} 
\caption{Evolution of $G^{(2)}_{k_{\rm res}}$ at early times, for $k$ within the resonant window. The same benchmark parameters as in Fig.~\ref{fig:earlytime} are used here.  $N_{\rm osc} < 0$ corresponds to the initial state before the sudden change in $\omega_{\perp}$; it shows oscillations in $G^{(2)}_{k_{\rm res}}$ because, after averaging over $100$ realizations, the effective correlation amplitude $c_{k}^{\rm ph}$ is small but non-zero.  The solid black curves indicate the evolution of the maxima and minima of $G^{(2)}_{k_{\rm res}}$ as predicted by BdG.  The dashed horizontal lines show the minimum and maximum values of $(u_{k}+v_{k})^{2}$ reached during the oscillations of the condensate (note that this starts at its maximum value when $\omega_{\perp} > \omega_{\perp, {\rm in}}$, so that only the maximum is shown for $N_{\rm osc} < 0$).  When the minima of the oscillations of $G^{(2)}_{k_{\rm res}}$ lie below the dashed lines (here, for $1 \lesssim N_{\rm osc} \lesssim 17$), the two-mode phonon state $(k,-k)$ is nonseparable.  The increase of the minima after $N_{\rm osc} \approx 10$ and the subsequent loss of nonseparability are the first observed deviations from the BdG prediction.  We also observe a decrease in the maxima from their predicted values. 
\label{fig:G2}}
\end{figure}

In Figure~\ref{fig:earlytime} are shown the atomic number $n^{\rm at}_{k}$ and the density-density correlation $G^{(2)}_{k}$, as a function of $k$ for three different times during the early stage of the evolution, and with the system parameters set to the benchmark values given at the end of Sec.~\ref{sec:state_prep}.  We clearly see, at early time, the growth of the resonant peaks at $k a_{\perp} \approx \pm 1$.  
As shall be seen later (particularly in Figs.~\ref{fig:G2} and~\ref{fig:history_Nk}), the growth of the peaks is not monotonic, but shows significant oscillations.  When considering the atomic number $n_{k}^{\rm at}(t)$, the oscillations follow from the atom-atom interactions.  Instead, when considering $G^{(2)}_{k}(t)$, the oscillations reveal the strong correlations between phonons of opposite wave number, as can be seen by the $c_{k}^{\rm ph} \neq 0$ term in Eq.~(\ref{eq:G2k}).  (The particular times have been chosen so as to avoid the narrow dips displayed by $G^{(2)}_{k}(t)$ and clearly visible in Fig.~\ref{fig:G2}.)  
We also see, at later times, the growth of peaks at the harmonics $k a_{\perp} \approx \pm 2$ and $\pm 3$.  These are due to the 
large number of phonons in the resonant mode causing 
the solution of the wave equation to become 
nonlinear 
(see Appendix~\ref{app:nonlinear}). Notice also that the peaks have broadened at later time. 
This effect shall be further studied in the next Section. 
Interestingly, a similar 
sequence of events was 
recently described in the context of an unstable two-component BEC system, see Ref.~\cite{Zache-Kasper-Berges} and footnote~\ref{fn:ZKB}.

We now focus on values of $k$ within the resonant window~\cite{Busch-Parentani-Robertson}. In Figure~\ref{fig:G2} is shown, for the same simulation as above, the evolution of $G^{(2)}_{k_{\rm res}}(t)$ as a function of time (parameterized by 
$N_{\rm osc}$, the number of oscillations of the condensate since the sudden change of $\omega_{\perp}$).  
Here the oscillatory nature of the peak growth is manifest. 
We have also included the values of $G^{(2)}_{k_{\rm res}}(t)$ 
before the sudden change, where the steady oscillations show that the phonon number and correlation are essentially constant.  (The non-vanishing of the correlation seems to be mostly due to a lack of statistics.)  After the change, we see a steady exponential growth in the mean number of phonons and the correlation, as predicted by the BdG formalism.  We also note that the minimum of $G^{(2)}_{k_{\rm res}}(t)$ is clearly well below its vacuum expectation value (shown in dashed) for a significant duration, showing 
(as explained in Sec.~\ref{observables}) 
that the two-mode state $(k,-k)$ is nonseparable during this time.

However, Figure~\ref{fig:G2} also shows a significant departure from the predictions of BdG in that there is a clear turning point around $N_{\rm osc} = 10$ in the minima of $G^{(2)}_{k_{\rm res}}(t)$.  Whereas BdG predicts that the minima tend asymptotically to zero, the actual results show that the minima increase again, eventually returning above the vacuum expectation value at $N_{\rm osc} \approx 17$.  Nonseparability of the two-mode phonon state $(k,-k)$ is thus lost around this time.  
Therefore, unlike in BdG~\cite{Busch-Parentani-Robertson}, the fully nonlinear theory does not allow the system to reach a nonseparable state for any initial temperature if one only waits for a long enough time: the evolution of the state towards nonseparability is progressively slowed down, and after a certain time decoherence effects cause the left-hand side of Eq.~(\ref{eq:nonsep}) to increase. We also observe in Figure~\ref{fig:G2} the first signs of saturation, in that the slope of the maxima of $G^{(2)}_{k_{\rm res}}(t)$ appears to be decreasing by the time we reach $N_{\rm osc} = 22$.

\begin{figure}
\includegraphics[width=0.45\columnwidth]{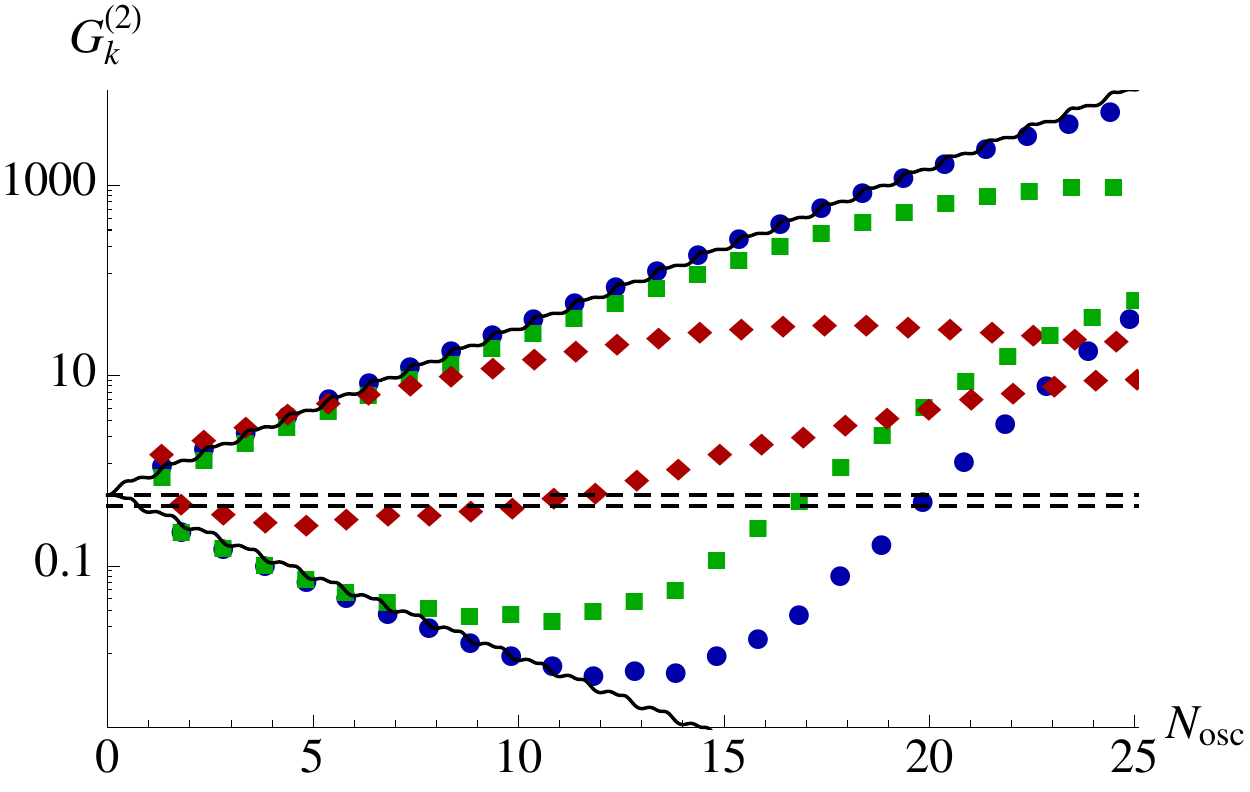}
\caption{Maxima and minima of $G^{(2)}_{k_{\rm res}}$ for $k$ within the resonant window and for three different values of $a_{s}/a_{\perp}$.  For clarity, the full profiles of $G^{(2)}_{k_{\rm res}}$ are removed here, and only the maxima and minima of its oscillations are plotted. The parameters are the same as in Figs.~\ref{fig:earlytime} and~\ref{fig:G2} for the green squares, namely $a_{s}/a_{\perp}= 1.7 \times 10^{-3}$.  Instead, red diamonds and blue circles show the extrema of $G^{(2)}_{k_{\rm res}}$ 
for $a_{s}/a_{\perp}=1.7 \times 10^{-2}$ and $1.7 \times 10^{-4}$ respectively.  As in Fig.~\ref{fig:G2}, 
the dashed horizontal lines show the maximum and minimum values reached by $(u_{k}+v_{k})^{2}$, and the solid black curves show the BdG predictions for the maxima and minima of $G^{(2)}_{k_{\rm res}}$, both of which are common to the three cases since $a_{s}/a_{\perp}$ drops out of the BdG description.  We see clearly that lower values of $a_{s}/a_{\perp}$ correspond to later deviations from the BdG prediction and thus to 
a later loss of nonseparability. 
\label{fig:G2minmax}}
\end{figure}

It is instructive to further characterize the deviations between our numerical observations with the outcome obtained using the BdG approximation. To this end, in Figure~\ref{fig:G2minmax} we dispense with the full evolution of $G^{(2)}_{k_{\rm res}}(t)$ and plot only the maxima and minima of its oscillations.  
This is done for three different simulations 
which coincide using the BdG description, as they differ only in the value of $a_{s}/a_{\perp}$; indeed, 
under our scheme of adimensionalization, 
this ratio drops out from the BdG equation~\cite{Robertson-Michel-Parentani-1}.  
The chosen values of this ratio are $a_{s}/a_{\perp} = 1.7 \times 10^{-4}$ (blue circles), $1.7 \times 10^{-3}$ (green squares) and $1.7 \times 10^{-2}$ (red diamonds); correspondingly, the total number of atoms takes the values $N = 4.5 \times 10^{5}$, $4.5 \times 10^{4}$ and $4.5 \times 10^{3}$.  
The middle case (shown in green squares) corresponds exactly to that shown in Fig.~\ref{fig:G2}.  Also shown, in solid black, is the common prediction of the BdG formalism.  
As can be seen, the departures from BdG occur earlier for larger values of $a_{s}/a_{\perp}$ (smaller values of $N$). 
This is clear for both types of departure, namely, the loss of nonseparability (indicated by the lower dots), and the reduction of the increase in $n_{k}$ and $\left|c_{k}\right|$ (indicated by the upper dots).  
We verify, therefore, that the BdG description is better when $a_s/a_\perp$ is smaller, which means that the self-interactions so far ignored are weaker and that there is a larger number of condensed atoms (relative to the total)~\footnote{The observed 
damping of phonons with respect to 
the BdG predictions 
is {\it a priori} 
rather surprising. Indeed, it is well known that the Landau-Beliaev 
damping vanishes on-shell in one-dimensional systems, as do 
some higher-order interactions~\cite{Ristivojevic-Matveev}. 
We conjecture that the significant damping seen here stems from the high occupation number of soft phonons found is quasi-condensates with a finite temperature.  We are currently investigating these effects and plan to present the results in a forthcoming paper. 
We thank Andrea Trombettoni for discussions on this issue, and for pointing out the above reference.}.

\subsection{Visibility of nonseparability and effective dissipation
\label{sub:NonsepDissipation}}

Let us now consider the results of Figure~\ref{fig:G2minmax} 
from a more phenomenological standpoint.  First, we discuss 
the ``visibility'' of nonseparability.  By this we mean that at large $n_{\pm k}^{\rm ph}$, even within the BdG description, when the two-mode state $(k,-k)$ is maximally entangled it is already very difficult to verify its nonseparability due to the necessity of taking the difference between two large numbers, namely $n_{k}^{\rm ph} n_{-k}^{\rm ph}$ and $\left|c_{k}^{\rm ph}\right|^{2}$ in Eq.~(\ref{eq:nonsep}).  One can appreciate the difficulty by examining Figure~\ref{fig:G2}: as $n_{\pm k}^{\rm ph}$ increases, the amount of time $G^{(2)}_{k_{\rm res}}(t)$ spends below $\left( u_{k}+v_{k} \right)^{2}$ decreases, and the precision required of the measuring apparatus to determine that $G^{(2)}_{k_{\rm res}}$ does indeed dip below the threshold becomes greater.  So, while it is theoretically true that nonseparability is lost when weak nonlinearities come into play, 
it may no longer be relevant by that time, so that (as far as nonseparability is concerned) very little has been lost in practice.  We also note here that, when considering the atom content of the state using 
time-of-flight (TOF) 
measurements, similar problems with visibility of nonseparability are encountered (see the upper right plot of Fig.~\ref{fig:g2} in Appendix~\ref{app:g2} and Fig.~17 of Ref.~\cite{Robertson-Michel-Parentani-1} for the BdG description of the same observable where there is no actual loss of nonseparability). 

To make this notion more concrete, we define the following ``visibility parameter'':
\begin{equation}
\tilde{\eta}_{k} = \frac{\left| c_{k} \right|^{2}}{\left(\bar{n}_{k} + 1/2\right)^{2}} \,, \qquad \mathrm{where} \qquad \bar{n}_{k} = \frac{1}{2}\left( n_{k} + n_{-k} \right) \,.
\label{eq:eta_defn}
\end{equation}
This is an appropriate definition when using the 
TWA because the extracted value of $\left(\bar{n}_{k}+1/2\right)^{2}$ is 
necessarily positive, so $\tilde{\eta}_k$ is always well-defined, and 
necessarily smaller than $1$. 
Indeed, since $\bar{n}_{k}^{2} \geq n_{k} n_{-k}$, Eq.~(\ref{eq:nonsep}) implies that a sufficient criterion for nonseparability is $\tilde{\eta}_{k} -  \tilde{\eta}_{k, {\rm th}} > 0$, where
\begin{equation}
\tilde{\eta}_{k, {\rm th}} = \frac{\bar{n}_{k}^{2}}{\left(\bar{n}_{k} + 1/2\right)^{2}} \,.
\label{eq:eta_th_defn}
\end{equation}
Explicitly, the sufficient condition for nonseparability becomes
\begin{equation}
\frac{\left|c_{k}\right|^{2} - \bar{n}_{k}^{2}}{\left(\bar{n}_{k}+1/2\right)^{2}} > 0 \, .
\end{equation}
Being a relative difference, this quantity is more experimentally relevant, and it is for this reason we refer to it as the ``visibility''.  We say that nonseparability is ``more visible'' 
when there is a larger difference between $\tilde{\eta}_{k}$ and $\tilde{\eta}_{k,{\rm th}}$.

\begin{figure}
\includegraphics[width=0.45\columnwidth]{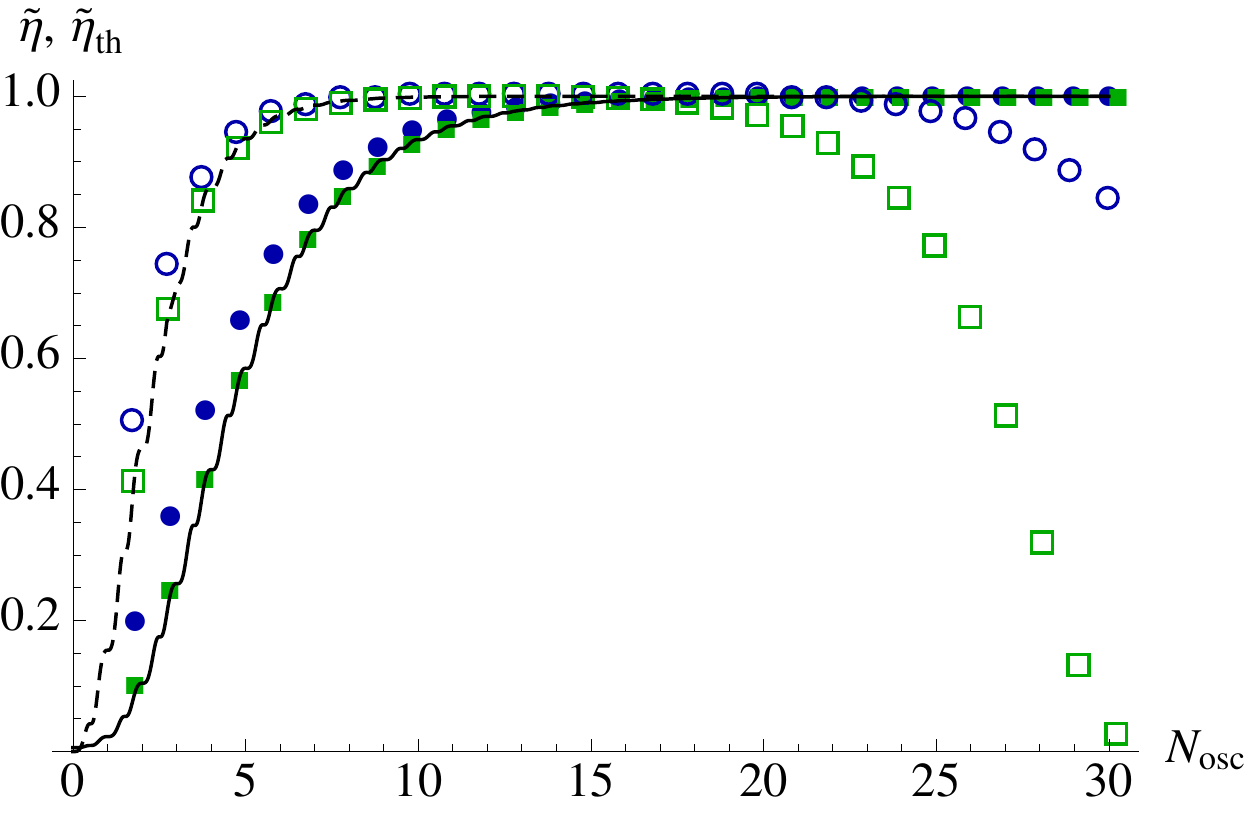}
\caption{Visibility of nonseparability.  Shown is the parameter $\tilde{\eta}_{k}$ of Eq.~(\ref{eq:eta_defn}) for $k$ at resonance 
(dashed curve and open markers), and its nonseparability threshold $\tilde{\eta}_{k,{\rm th}}$ of Eq.~(\ref{eq:eta_th_defn}) 
(solid curve and filled markers); 
the bipartite state $(k,-k)$ is nonseparable whenever $\tilde{\eta}_{k} > \tilde{\eta}_{k,{\rm th}}$. We use here only two of the simulations shown in Fig.~\ref{fig:G2minmax}: those corresponding to $a_{s}/a_{\perp} = 1.7 \times 10^{-4}$ (blue circles) and $1.7 \times 10^{-3}$ (green squares). 
The data points are extracted from the maxima and minima of $G^{(2)}_{k_{\rm res}}$ shown in Fig.~\ref{fig:G2minmax}, while the black 
curves plot the BdG prediction for the given initial temperature $T_{\rm in} = m c_{\rm in}^{2}/2$.  This presentation of the data is valuable as the separation between $\tilde{\eta}_{k}$ and $\tilde{\eta}_{k,{\rm th}}$ can be taken as a measure of the visibility of 
nonseparability.
\label{fig:eta}}
\end{figure}

In Figure~\ref{fig:eta}, the visibility parameter for $k$ at resonance 
is plotted for two of the simulations represented in Fig.~\ref{fig:G2minmax}: the blue circles correspond to $a_{s}/a_{\perp} = 1.7 \times 10^{-4}$, while the green squares correspond to $a_{s}/a_{\perp} = 1.7 \times 10^{-3}$.  The open points represent $\tilde{\eta}_{k}$ itself, while the filled points represent the nonseparability threshold $\tilde{\eta}_{k,{\rm th}}$.  We observe that, to begin with, the growth of $\tilde{\eta}_{k}$ and $\tilde{\eta}_{k,{\rm th}}$ agrees quite well with the predictions of BdG (shown in black), the discrepancies shown by the blue circles being likely due to a lack of statistics.  
Note that even the BdG prediction shows decreasing visibility, as $\tilde{\eta}_{k}$ and $\tilde{\eta}_{k,{\rm th}}$ become arbitrarily close, and are almost indistinguishable for $N_{\rm osc} \gtrsim 15$. 
When including nonlinear effects, for both 
cases shown, the actual loss of nonseparability occurs {\it after} the loss of its visibility, and is thus no great loss in practical terms.  Even the idealized BdG prediction shows maximum visibility for $N_{\rm osc} \sim 4$, and both 
simulations corroborate this result. 

Let us now turn to the actual loss of nonseparability, 
which is clearly seen in Fig.~\ref{fig:G2minmax} at 
$N_{\rm osc} \sim 17$ and $\sim 20$ (for the two cases considered here).  This loss can be considered a manifestation of an effective dissipative mechanism 
(see Ref.~\cite{Busch-Parentani-Robertson}) due to phonon-phonon interactions.  To further study this loss, we define the following effective dissipation rates describing (purely phenomenologically) the damping of $n_k$, $\left|c_k\right|$ and $\tilde{\eta}_{k}$ with respect to their BdG predictions:
\begin{eqnarray}
\left(n_{k}+\frac{1}{2}\right)^{2} & = & \left(n_{k}+\frac{1}{2}\right)^{2}_{\rm BdG} \, \mathrm{exp}\left( - \int \Gamma_{n} \, \mathrm{d}t \right) \,, \nonumber \\
\left|c_{k}\right|^{2} & = & \left|c_{k}\right|^{2}_{\rm BdG} \, \mathrm{exp}\left( - \int \Gamma_{c} \, \mathrm{d}t \right) \,, \nonumber \\
\tilde{\eta}_{k} & = & \tilde{\eta}_{k,{\rm BdG}} \, \mathrm{exp}\left( - \int \Gamma_{\tilde{\eta}} \, \mathrm{d}t \right) \,.
\label{eq:Gamma-defn}
\end{eqnarray}
In these expressions, $k$ is again 
understood to be within the resonant window.  Note that, from the definition of $\tilde{\eta}_{k}$ (see Eq.~(\ref{eq:eta_defn})), we have $\Gamma_{\tilde{\eta}} = \Gamma_{c} - \Gamma_{n}$.

\begin{figure}
\includegraphics[width=0.5\columnwidth]{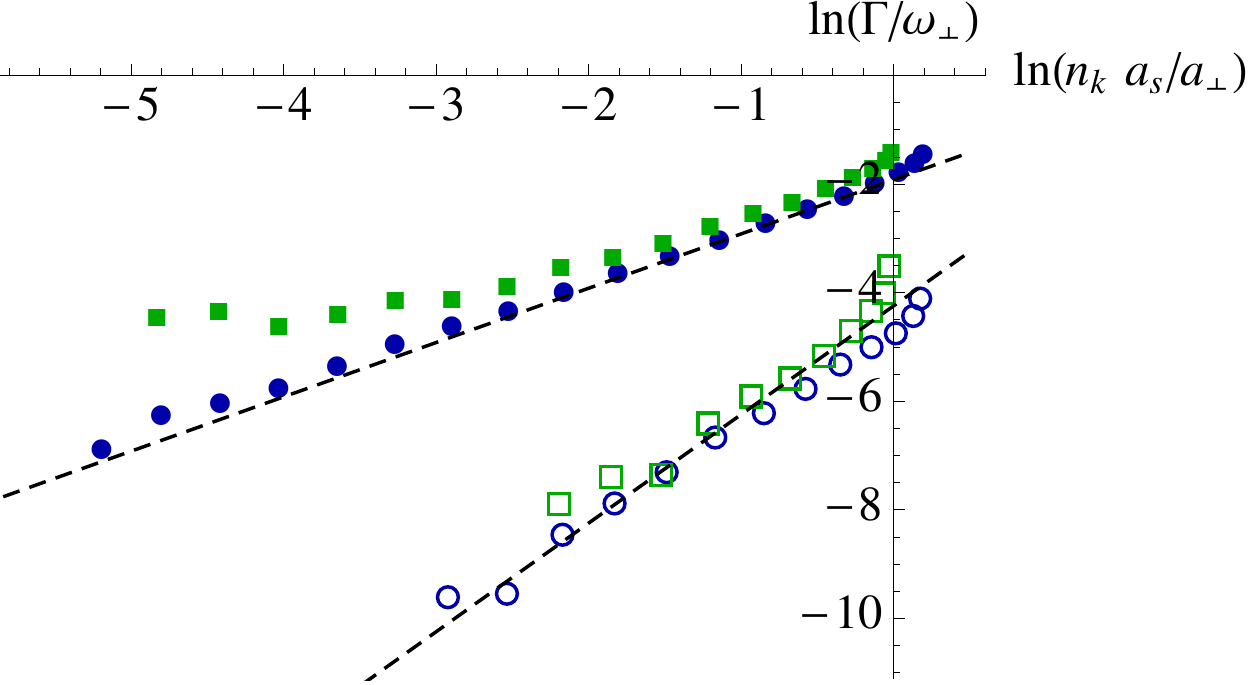} 
\caption{Effective dissipation.  Plotted here are the effective dissipative rates describing the deviations of the data of Fig.~\ref{fig:G2minmax} from the BdG prediction, for the same two simulations shown in Fig.~\ref{fig:eta}: $a_{s}/a_{\perp} = 1.7 \times 10^{-4}$ (blue circles) and $1.7 \times 10^{-3}$ (green squares).  The filled and open markers, respectively, show $\Gamma_{n}$ and $\Gamma_{\widetilde{\eta}}$ of Eqs.~(\ref{eq:Gamma-defn}), both adimensionalized by $\omega_{\perp}$, as functions of $n_{k}\,a_{s}/a_{\perp}$.  Using this combination, as noticed in the main text, the evolution of $\Gamma_{n}$ and $\Gamma_{\tilde{\eta}}$ hardly depends on the value of $a_{s}/a_{\perp}$.  The dashed lines have fitted intercepts but fixed slopes of $1$ and $2$ on the log-log plane, and show that, for a significant fraction of the evolution, both of Eqs.~(\ref{eq:Gamma-n}) are satisfied. 
\label{fig:Gamma}}
\end{figure}

The extracted values of $\Gamma_{n}$ and $\Gamma_{\tilde{\eta}}$, obtained for the same two simulations represented in Fig.~\ref{fig:eta}, are shown in Figure~\ref{fig:Gamma}, for the period during which the resonant peak grows exponentially.  ($\Gamma_{c}$ turns out to be very close to $\Gamma_{n}$, and has thus not been shown.)  Instead of plotting them as functions of time, the dissipation rates are plotted as functions of $n_{k} \, a_{s}/a_{\perp}$. 
Interestingly, on this plane, the results of the two simulations lie very close to each other, which indicates that, for a significant fraction of the evolution, the effective dissipation rates are simply functions of $n_{k} \, a_{s}/a_{\perp}$.  Moreover, the slopes of the two curves on the log-log plane are close to $1$ and $2$, and so 
to a good approximation~\footnote{
Figure~\ref{fig:Gamma} 
shows indeed some deviations from the behavior of \eq{eq:Gamma-n} both at early and late time.  We conjecture that the approximately constant value of $\Gamma_{n}$ observed for the benchmark case (green filled squares) at low values of $n_{k}$ is indicative of a standard dissipative rate per phonon of wavenumber $k$, while the linearity of $\Gamma_{n}$ in $n_{k}$ found when $n_{k}\,a_{s}/a_{\perp}$ becomes larger than $e^{-3}$ shows that the dissipation there is predominantly ``induced'' by the macroscopic value of $n_{k} \gtrsim 30$.
} we have: 
\begin{eqnarray}
\Gamma_{n}/\omega_{\perp} & \propto & n_{k} \, a_{s}/a_{\perp} \,, \nonumber \\
\Gamma_{\tilde{\eta}}/\omega_{\perp} & \propto & \left(n_{k} \, a_{s}/a_{\perp} \right)^{2} \,.
\label{eq:Gamma-n}
\end{eqnarray}
These numerical observations call for a physical explanation based on a quantum mechanical treatment (the Keldysh formalism~\cite{Kamenev}) of Eq.~(\ref{eq:GPE-1D}) following the analysis of Ref.~\cite{Ristivojevic-Matveev}. We are currently studying these effects. 

The deviations with respect to the BdG treatment shown in
Figs.~\ref{fig:G2}-\ref{fig:Gamma} (and Fig.~\ref{fig:g2}), 
obtained numerically by applying the TWA to \eq{eq:GPE-1D} when $g_1(t)$ is periodically modulated, constitute the main results of this paper. To our knowledge they have not yet been reported in the literature. 

\section{Late-time behavior
\label{sec:late}}

In this section, we turn 
to the behavior of the system at the end of and after 
the exponential growth, when nonlinear effects are strong 
and the BdG treatment loses all validity. 
For this very reason,  
we now 
abandon the phononic for the atomic point of view.  
Unlike in the previous Section, 
the second kind of nonlinearity (governed by the last term of Eq.~\eqref{eq:Veff_backreaction})
here 
plays an important role as the radial energy $E_{\rad}(t)$ now 
significantly decreases. Yet this decrease is adiabatic in the sense that $d({\rm ln} \bar{E}_{\rm rad}(t))/dt \ll \omega_{\perp}$, where $\bar{E}_{\rad}(t)$ 
is the mean 
of $E_{\rad}(t)$ over one oscillation period $\pi/\omega_\perp$. 
(We here use this time average in order 
to extract the secular effect since, 
as can be understood from \eq{eq:dotElong},
the instantaneous value of $E_{\rad}(t)$ displays rapid oscillations directly linked to those of $\sigma(t)$.) 

The interested reader will find in Appendix~\ref{app:g2} a description of the full evolution of the system (both early- and late-time behavior) in terms of $g_{2}(k)$ (see Eq.~(\ref{eq:g2_defn})), which is the observable commonly used after 
TOF 
experiments~\cite{Carusotto-BH,Jaskula-et-al,Boiron-et-al}. 

\subsection{Spectrum and density-density correlator}

\begin{figure}
\includegraphics[width=0.2708\columnwidth]{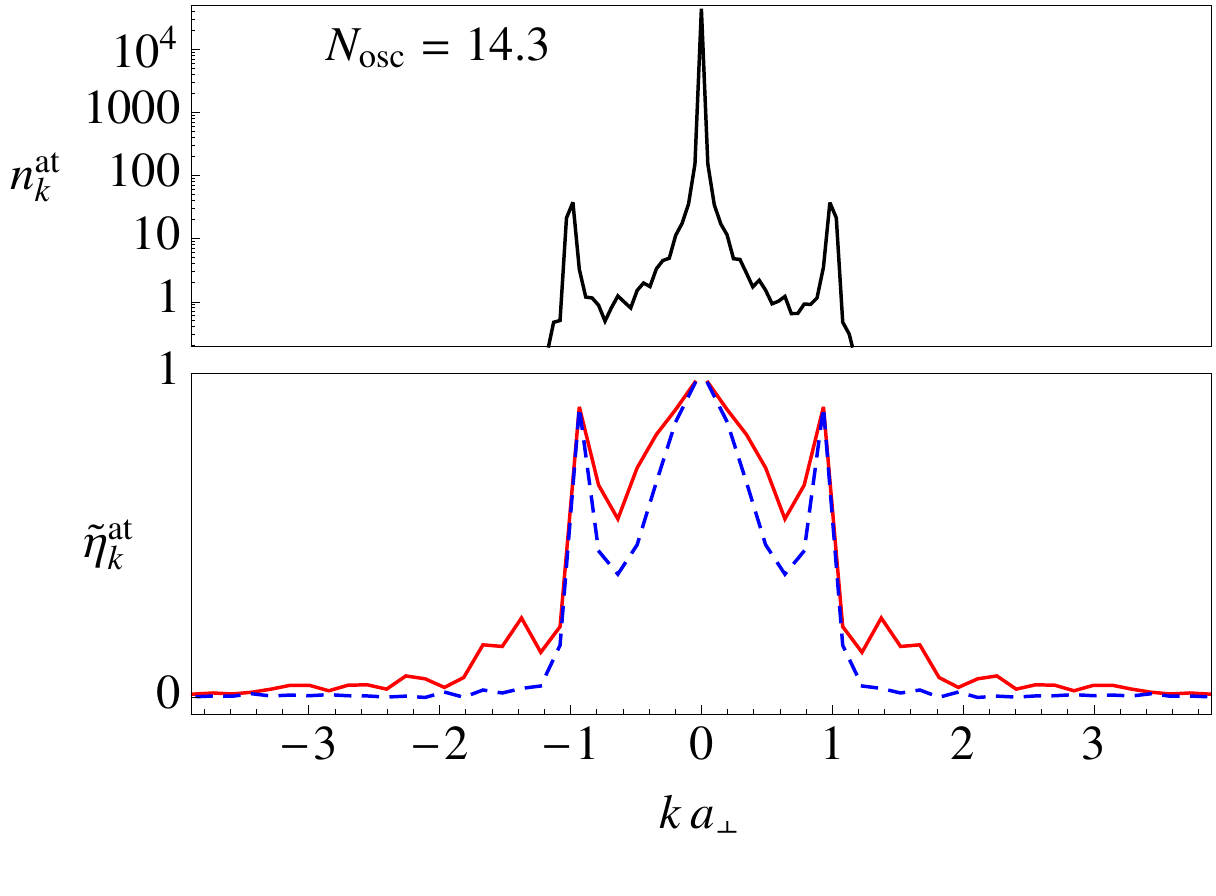} \includegraphics[width=0.229\columnwidth]{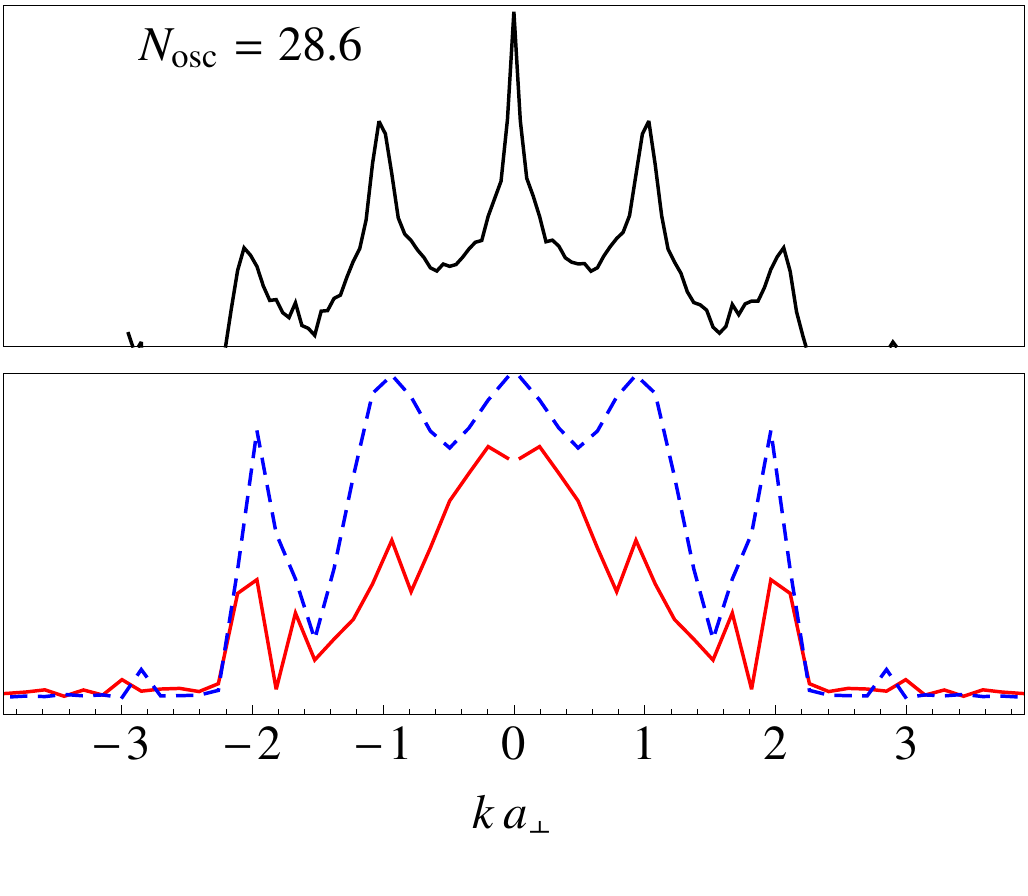} \includegraphics[width=0.229\columnwidth]{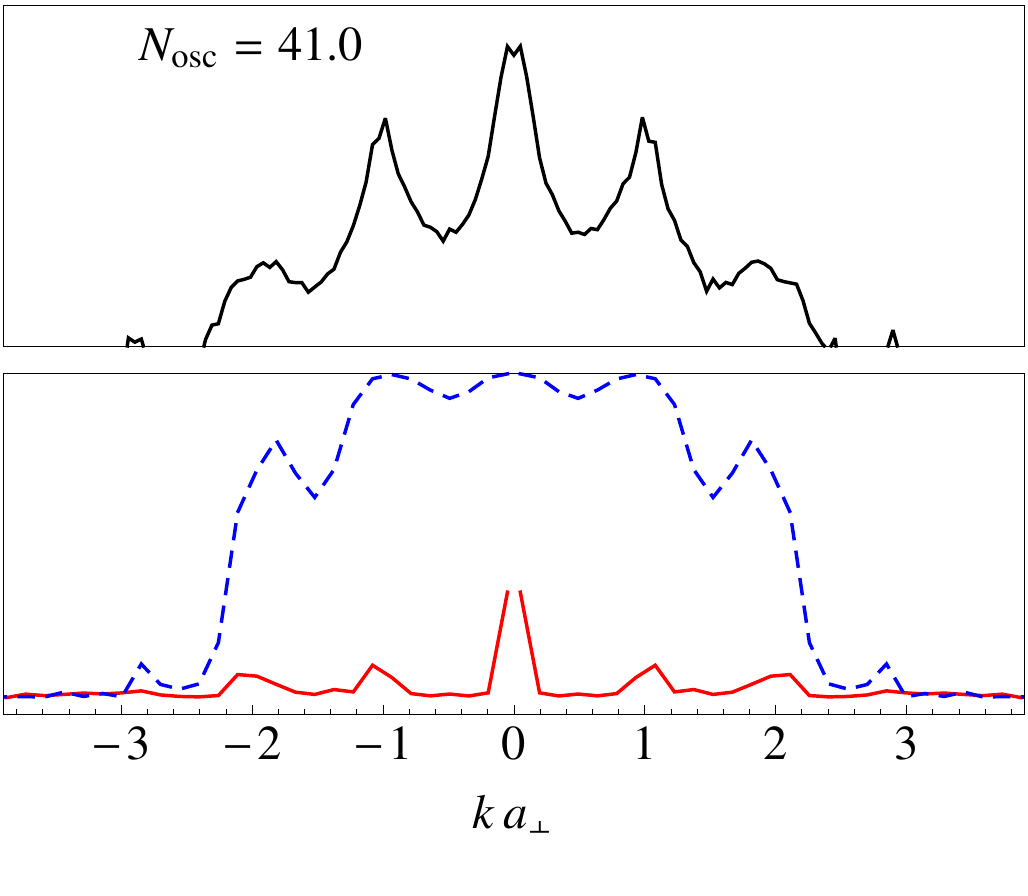} \includegraphics[width=0.229\columnwidth]{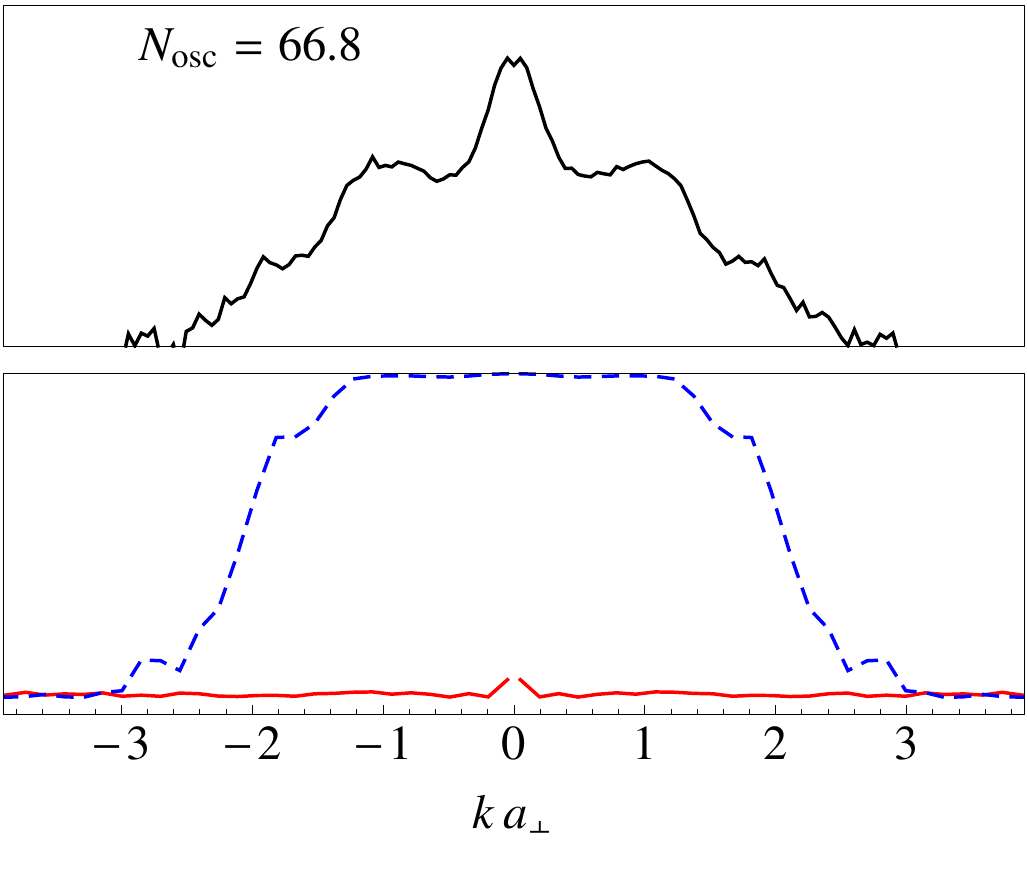}
\caption{Evolution of the system at late times.  In the top row is shown the number of atoms as a function of $k$ at four different times, the first two of which ($N_{\rm osc} = 14.3$ and $28.6$) are the latest times shown in Fig.~\ref{fig:earlytime}; the third and fourth plots thus show the continuation of the simulation in Fig.~\ref{fig:earlytime}.  One sees clearly the continued broadening of the peaks until they have essentially been smoothed out.  In the bottom row is shown, in solid red, the (atomic) visibility parameter of Eq.~(\ref{eq:eta_defn}) for atomic expectation values $n_{\pm k}^{\rm at}$ and $c_{k}^{\rm at}$, and in dashed blue, the corresponding nonseparability threshold of Eq.~(\ref{eq:eta_th_defn}).  
The curves on the lower row have been smoothed by binning the data into groups of 3. 
Roughly speaking, $\tilde{\eta}_{k}^{\rm at}$ gives the fraction of atoms at each $k$ which occur as members of $(k,-k)$ pairs.  It is clearly seen that, as the peaks broaden, the correlations between $k$ and $-k$ gradually disappear.
\label{fig:history_nk_eta}}
\end{figure}

\begin{figure}
\includegraphics[width=0.3\columnwidth]{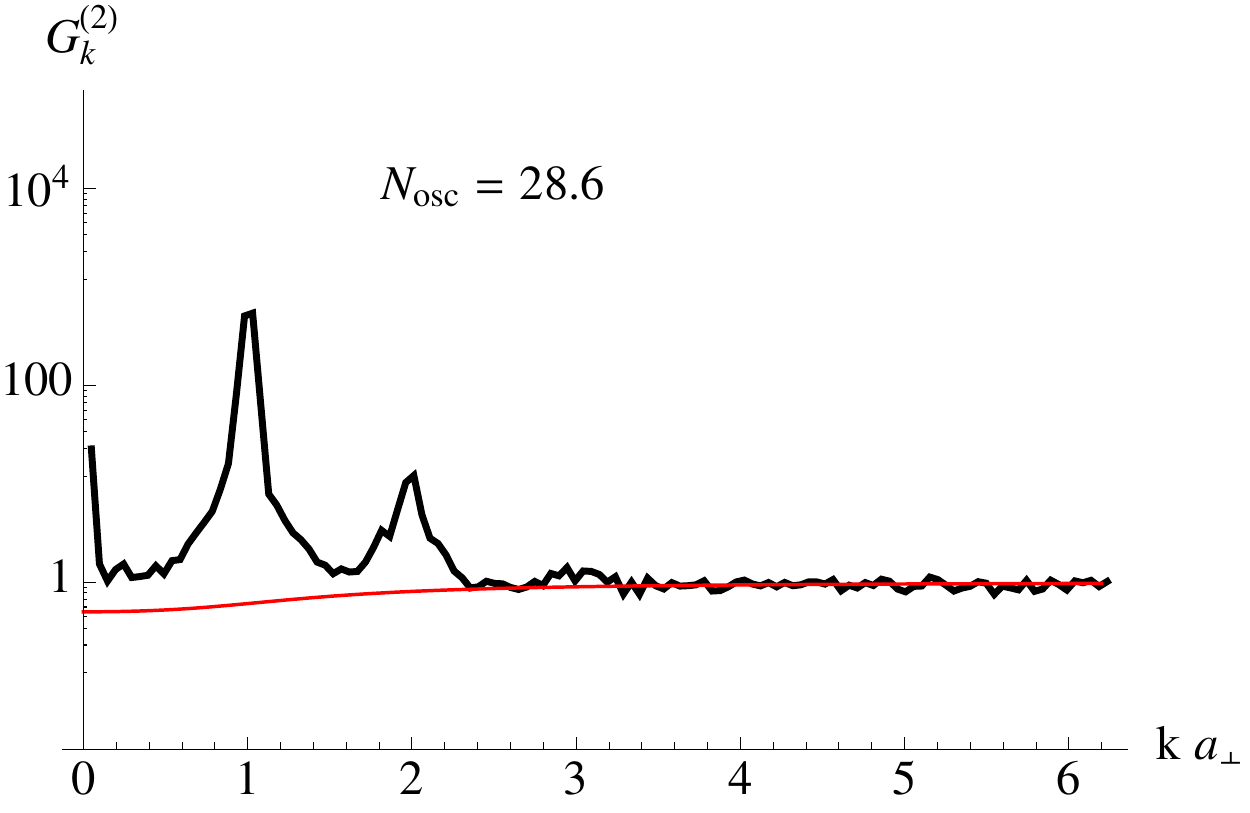} \, \includegraphics[width=0.3\columnwidth]{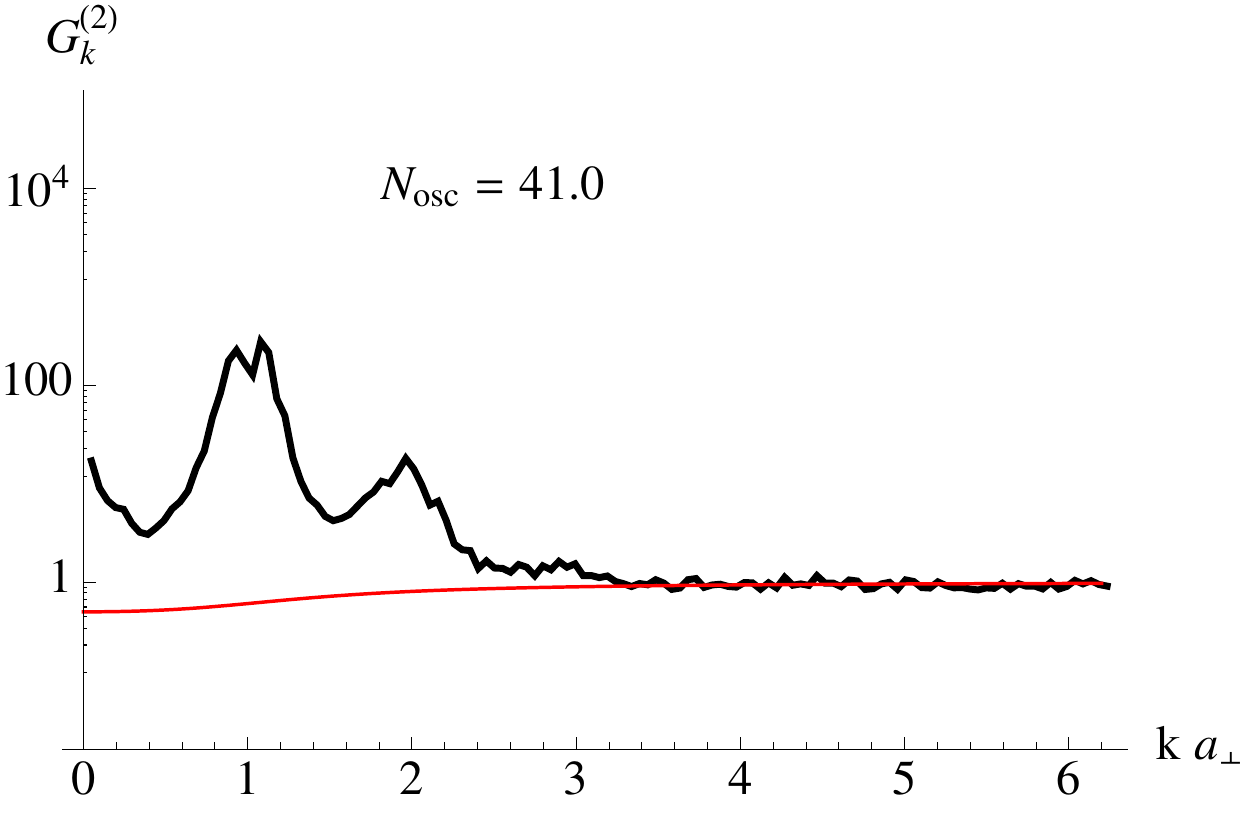} \, \includegraphics[width=0.3\columnwidth]{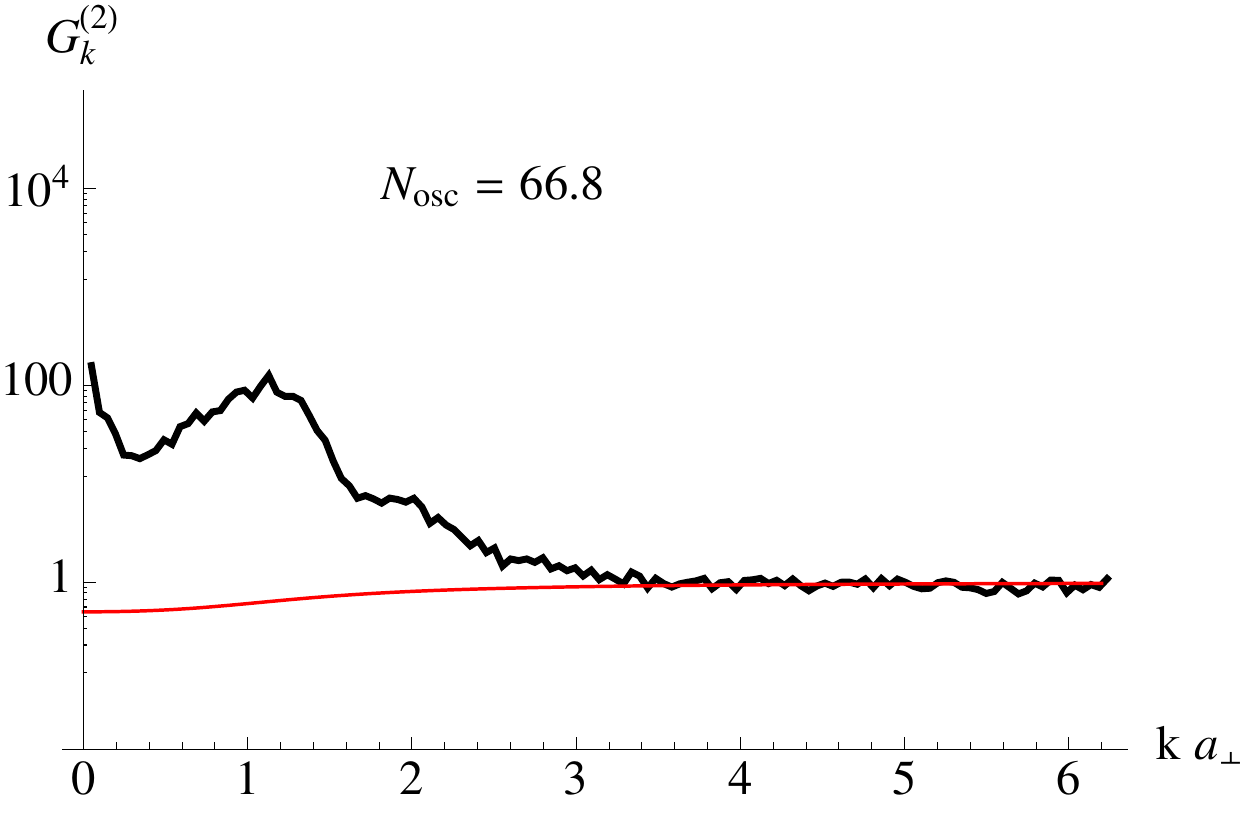}
\caption{Evolution of $G^{(2)}_{k}$ at late times.  Shown here 
are three snapshots of $G^{(2)}_{k}(N)$ corresponding to the same simulation, and the same three late times, as 
in Fig.~\ref{fig:history_nk_eta}.  The red curve is the same in each plot, and corresponds to the initial form of $G^{(2)}_{k}$ at the initial temperature $T_{\rm in} = m c_{\rm in}^{2}/2$, according to the BdG theory (exactly as in the left panel of Fig.~\ref{fig:earlytime}).  We clearly observe a 
broadening of the peaks 
similar to that 
visible in the atom number spectra of Fig.~\ref{fig:history_nk_eta}. 
\label{fig:history_G2}}
\end{figure}

Figures~\ref{fig:history_nk_eta} and~\ref{fig:history_G2} show the continuation of the two plots of Fig.~\ref{fig:earlytime}: Figure~\ref{fig:history_nk_eta} shows 
the atom number spectrum (as well as $\tilde{\eta}_{k}$ of Eq.~(\ref{eq:eta_defn})), while Figure~\ref{fig:history_G2} shows the corresponding density-density correlation function of Eq.~(\ref{eq:G2_defn}). 
We had already seen in Fig.~\ref{fig:earlytime} that the later stages of the exponential growth were marked by a broadening of the peaks.  In Fig.~\ref{fig:history_nk_eta} we see that the peaks continue to broaden, to the extent that, at very late time, they are almost completely washed out, having merged into a single, very broad, and nearly structureless peak centered at $k=0$.  This broadening is accompanied by a decrease in $\tilde{\eta}_{k}$, which is essentially zero at very late time.  This means that the $(k,-k)$ correlations are very small with respect to the corresponding expectation numbers; roughly speaking, $\tilde{\eta}_{k}$ is the fraction of the atoms at wave vector $k$ which occur in $(k,-k)$ pairs.  We have already seen the loss of $\tilde{\eta}_{k}$ for $k$ near $k_{\rm res}$ (see Fig.~\ref{fig:eta}), but now we see that this occurs for all $k$.

In Figure~\ref{fig:history_G2}, the same information is represented in terms of the {\it in situ} observable $G^{(2)}_{k}(N)$, which also clearly shows the broadening of the peaks and their gradual merging into a single wide peak. 
As in Fig.~\ref{fig:earlytime}, the values of $N$ have been chosen to clearly illustrate the different stages of the evolution, since there are still sudden dips similar to those displayed in Fig.~\ref{fig:G2} that should be avoided. 
In fact, the gradual 
disappearance of the correlations between $(k,-k)$ pairs manifests itself 
through the reduction of the dips' amplitude 
when increasing 
$N$ at fixed 
$k$. 

Note also that the high-$k$ sector ($k a_{\perp} \gtrsim 3$) remains very close to its initial (vacuum) state even after around 70 oscillations. This agreement for large $k$ after 70 oscillations
provides an {\it a posteriori} justification of our use of the TWA for describing the first kind of nonlinearity 
encoded in \eq{eq:GPE-1D} while using the corrected potential of \eq{eq:Veff_backreaction} to account for the damping of radial oscillations.~\footnote{\label{TWA-justific}It is known that the TWA is unable to properly account for the thermalization~\cite{Sinatra-Lobo-Castin,Mora-Castin}, as well as being unreliable to describe some spontaneous processes~\cite{Iacopo-last-paper-on-TWA}. 
It thus behooves us 
to argue for its reliability 
in the present context. The justification is different depending on 
whether one considers 
early-time phenomena (presented in Sec.~\ref{sec:early}) 
or those now considered. 
At early time, 
the deviations with respect to the BdG predictions are small and can be treated to leading 
order. In this regime, since we start with a Gaussian ensemble, there is no reason to doubt that the TWA is able to capture these effects. At late 
time instead, the physics is dominated by the exponentially large number of resonant phonons, and 
the TWA is still reliable because it is known to work well in the large occupation number regime, see e.g.~\cite{Zache-Kasper-Berges}. Yet, 
after 80 oscillations or so, some of our simulations gave signs 
that the TWA can no longer be trusted (see in particular the end of Sec.~\ref{late-time}), for reasons probably related to those mentioned in~\cite{Sinatra-Lobo-Castin,Mora-Castin}. 
We therefore stop the numerical integration and make no claim about the state at later time.} 

\begin{figure}
\includegraphics[width=0.3\columnwidth]{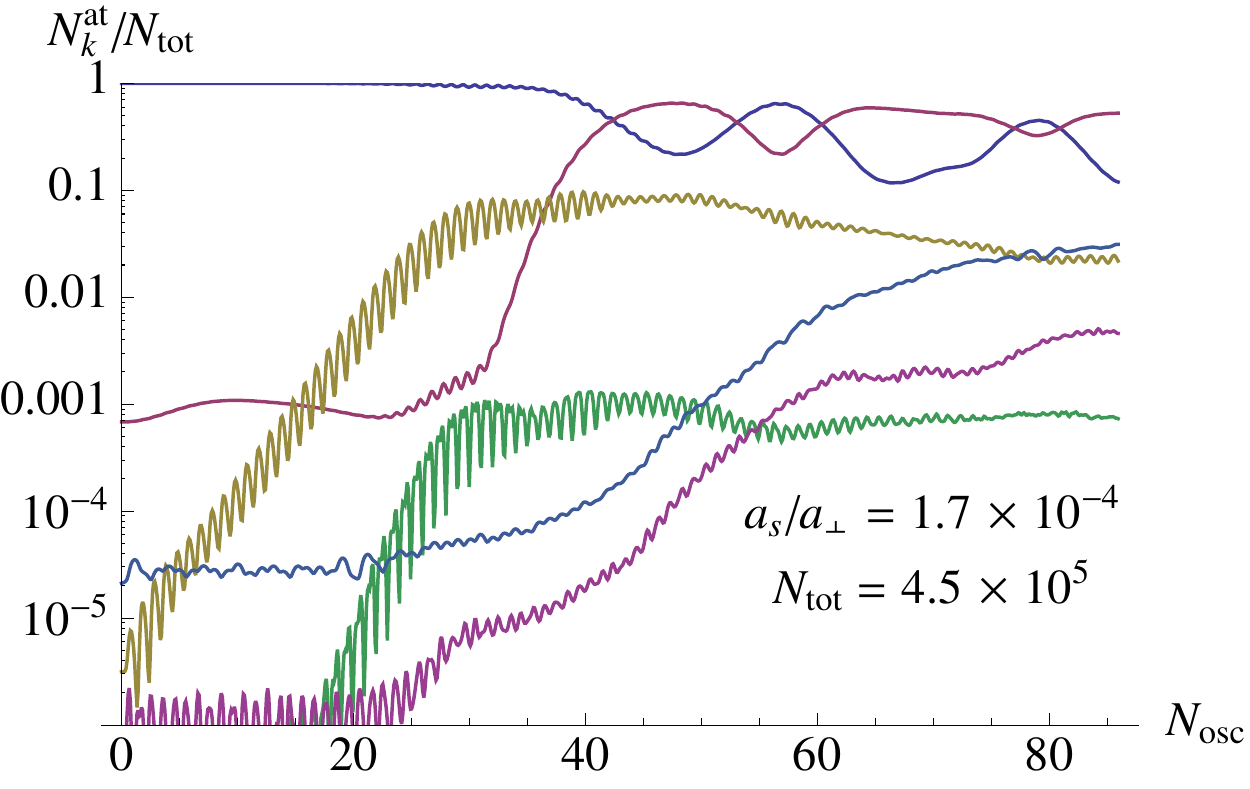} \, \includegraphics[width=0.3\columnwidth]{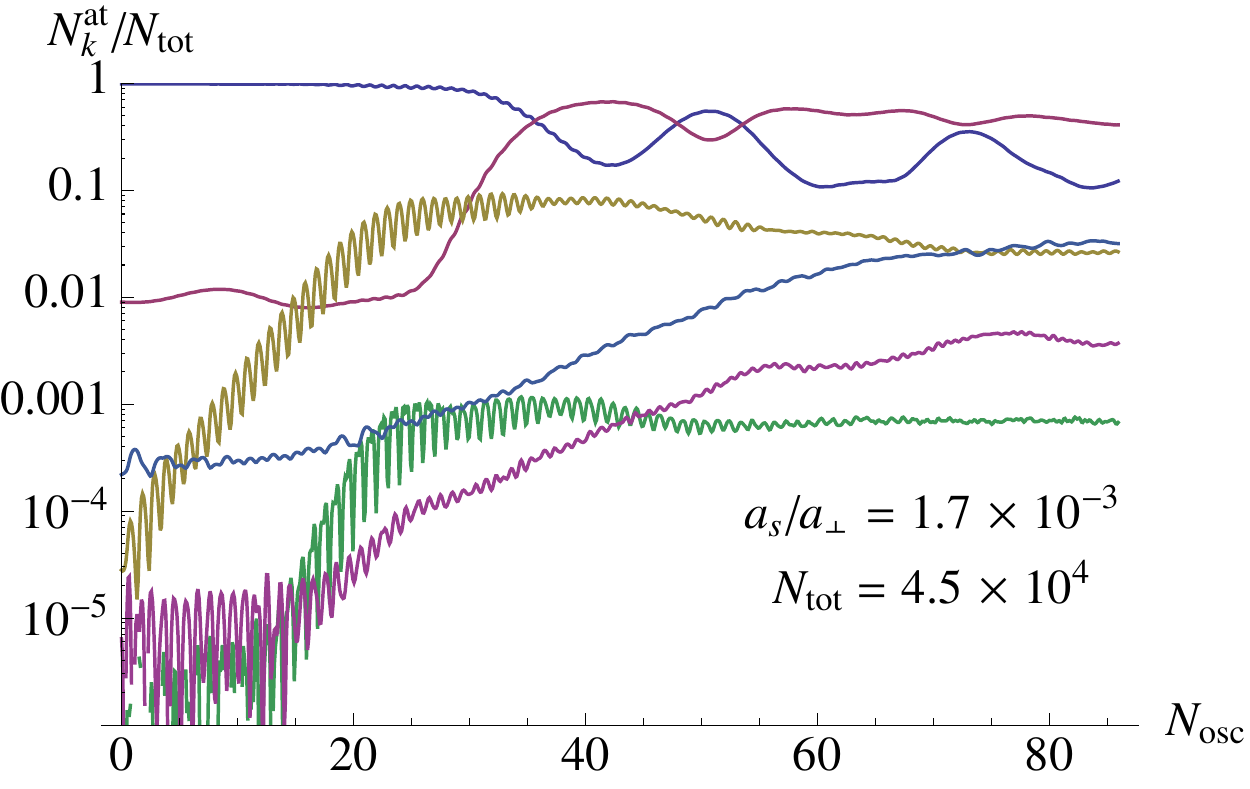} \, \includegraphics[width=0.3\columnwidth]{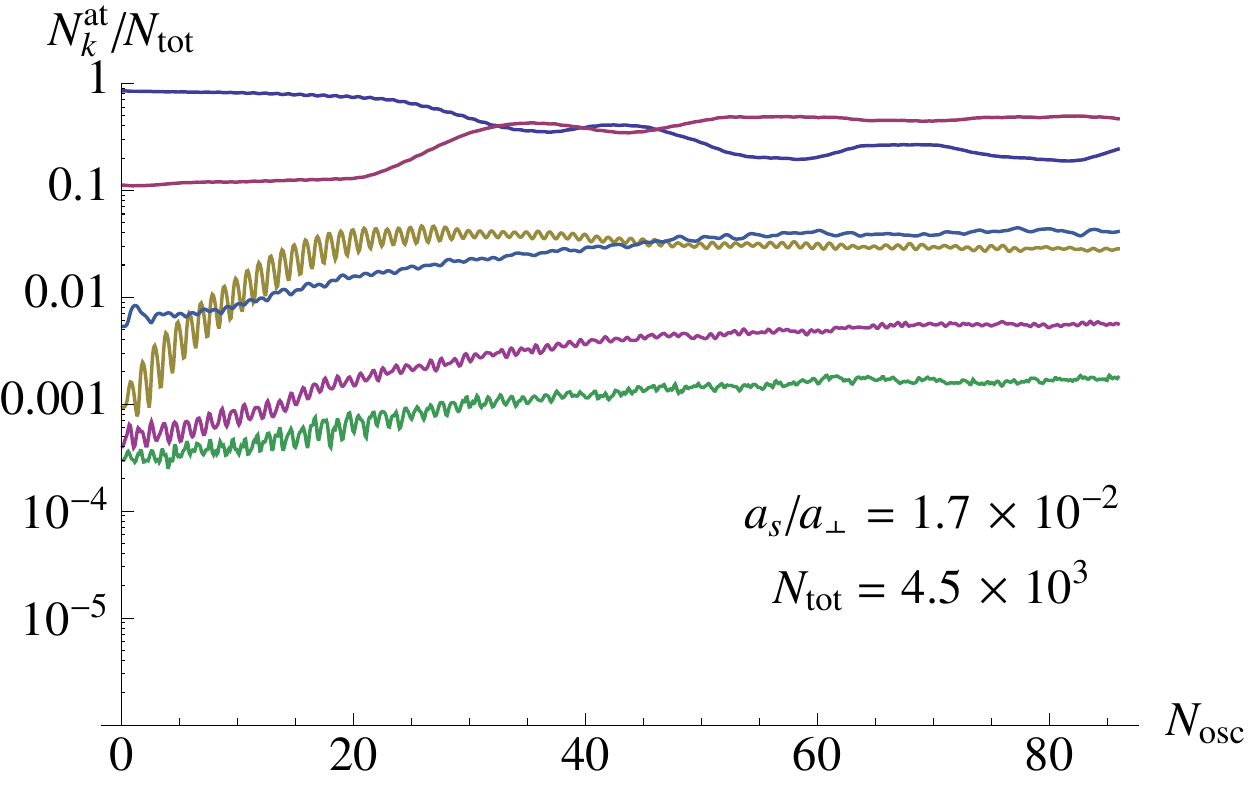}
\caption{Evolution of the fraction of the number of atoms within various ranges of $k$.  These correspond to the same three simulations shown in Fig.~\ref{fig:G2minmax}, with the three different values of $a_{s}/a_{\perp}$ (along with the corresponding values of the total number of atoms, $N_{\rm tot}$) written explicitly.  On each plot, the uppermost blue curve shows the content of the $k=0$ mode only, while the red curve just below it contains the two $k \neq 0$ modes on either side 
(i.e. $k = \pm 2\pi/L$ and $k = \pm 4\pi/L$). 
The other curves 
correspond to $k_{\rm res}/2$ (light blue), $k_{\rm res}$ (yellow), $3 k_{\rm res}/2$ (purple) and $2 k_{\rm res}$ (green), where for each 
we have included the central mode, two modes on either side, and their counterparts $k \to -k$ (i.e. 10 modes in total).  Note that, while the evolution varies quite drastically with $a_{s}/a_{\perp}$, the final fraction of atoms in each $k$-range is essentially independent of $a_{s}/a_{\perp}$.
\label{fig:history_Nk}}
\end{figure}

The loss of the peak structure is clearly demonstrated in Figure~\ref{fig:history_Nk}, which shows the evolution over all time of the logarithm of the fraction of atoms $n_k^{\rm at}$ within a set of chosen wave vector intervals.  At early time, there is a very clear preference for the peak at $k_{\rm res}$ to increase exponentially (as clearly indicated by the yellow curves), while the others remain largely stationary. 
After a certain time, the peak at $2 k_{\rm res}$ also grows exponentially, as was already seen in Fig.~\ref{fig:earlytime}.
These exponential growths saturate and just after the saturation time we see a marked growth in the occupation number of the non-resonant modes. 
Interestingly, the fraction of atoms in the peak at $k_{\rm res}$ is found to be around $10\%$ at saturation for the three values of $a_{s}/a_{\perp}$ we used, in agreement with the rough estimate discussed above Eq.~(49) and used in Figure~8 of~\cite{Robertson-Michel-Parentani-1}. 
Finally, all occupation numbers become (roughly) stationary, and are larger at smaller 
wave vectors as would be the case 
in a thermal bath.

\subsection{Energy and entropy
\label{energy-entropy}} 

The late-time behavior shows 
significant variation in the macroscopic properties of the system.
We have already seen that conservation of energy implies a backreaction effect, in which the radial oscillations are necessarily damped by the production of longitudinal phonons that they induce.
Moreover, the loss of peak structure observed in Figs.~\ref{fig:history_nk_eta}-\ref{fig:history_Nk} suggests an increase in the entropy of the system.

\begin{figure}
\includegraphics[width=0.45\columnwidth]{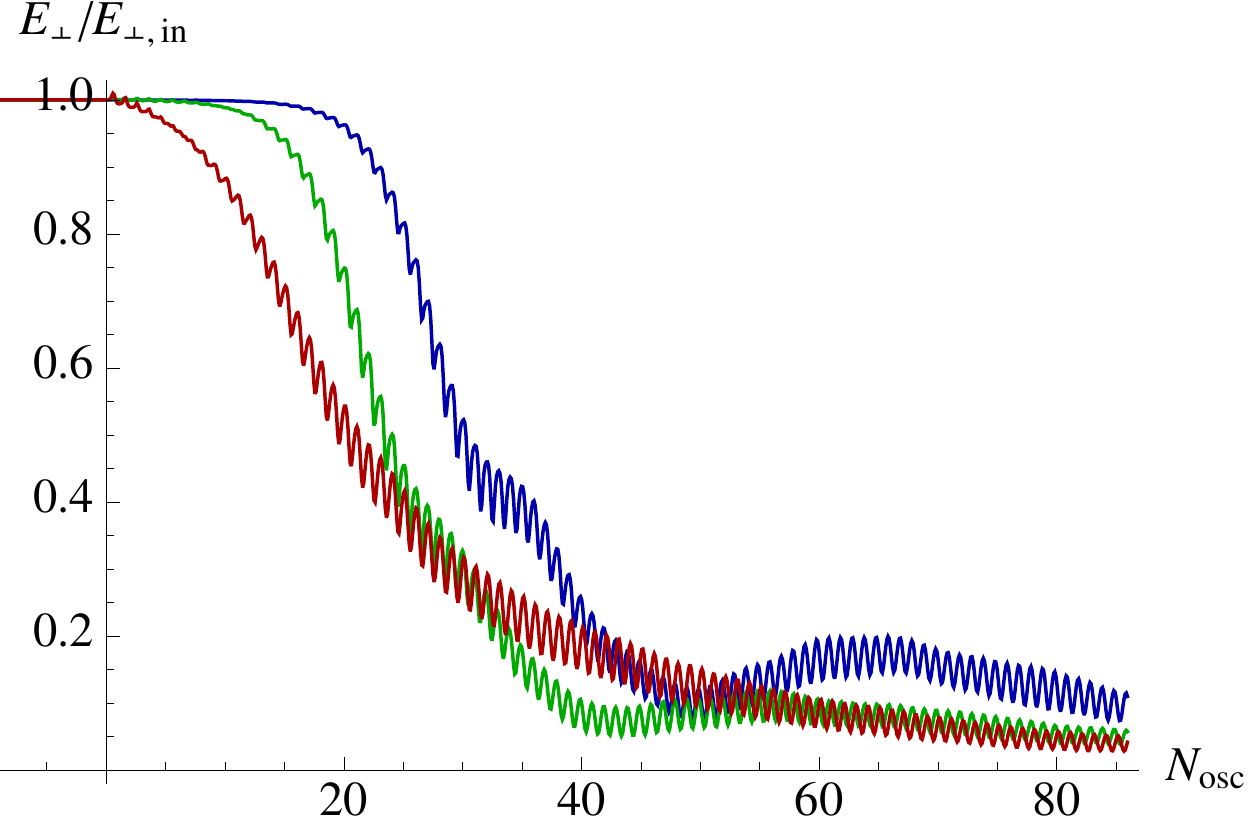} \, \includegraphics[width=0.45\columnwidth]{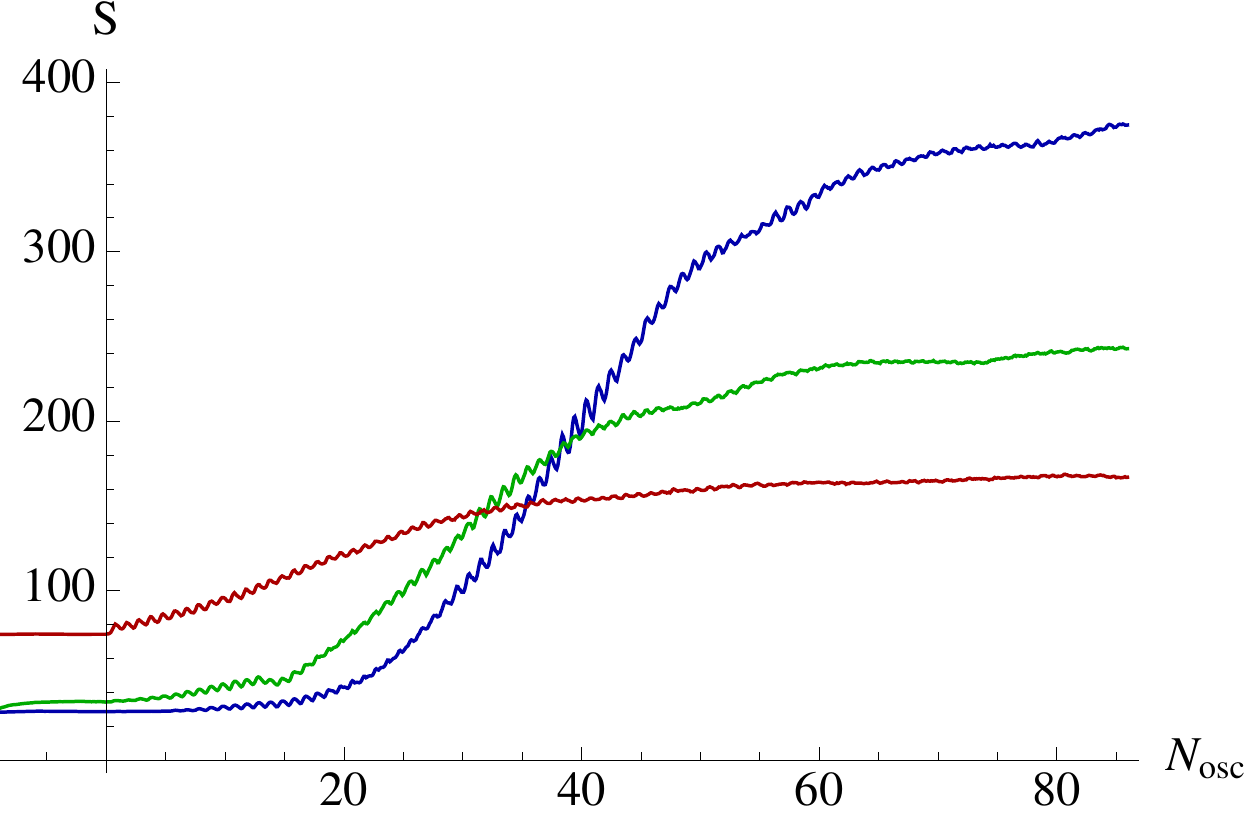}
\caption{Evolution of energy and entropy.  On the left is shown the energy of the radial oscillations of the condensate as a function of time, while on the right is shown the entropy as a function of time, for the same parameters as in all previous figures (see the caption of Fig.~\ref{fig:earlytime}) and the same values of $a_{s}/a_{\perp}$ as in Fig.~\ref{fig:history_Nk}: $1.7 \times 10^{-4}$ (blue), $1.7 \times 10^{-3}$ (green) and $1.7 \times 10^{-2}$ (red).  Note that the energy plots (particularly the blue and green curves) show three distinct phases of the evolution, which roughly correspond to an initial phase of constant entropy (where BdG is valid), an intermediate phase where the entropy is increasing, and a final phase where the entropy is roughly constant and the energy in the oscillations decreases very slowly.
\label{fig:Eperp_S}}
\end{figure}

The evolution of the energy and the entropy, for the same three simulations represented in Fig.~\ref{fig:history_Nk}, is shown in Figure~\ref{fig:Eperp_S}.
On the left is plotted, 
as a function of time and 
as a fraction of its initial value, 
$E_{\perp} \equiv E_{\rm rad} - E_{{\rm rad}, 0}$, where $E_{{\rm rad}, 0}$ is the radial energy of Eq.~(\ref{eq:GPE-radial-energy}) in the special case when $\sigma = a_{\perp}$ is stationary.  $E_{\perp}$ can thus be thought of as the energy of the radial oscillations which is available for conversion into longitudinal phonons (at least initially, for we do not include here the variation of the potential that describes the backreaction).
We could also have shown the reduction of the amplitude of the $\sigma(t)$ oscillations, 
as done in Fig.~(2) of~\cite{Figueroa2017JCAP} in the cosmological preheating scenario, see also footnote~\ref{BECvsCOSMO}. Comparing with Fig.~\ref{fig:history_Nk}, we see that the saturation of the exponential growth occurs when roughly $50\%$ of the initial oscillation energy has been exhausted.  

The entropy~\cite{Campo-Parentani-2005} we consider is formed only from the elements of the covariance matrix, i.e., the $c$-numbers $n_{k}$ and $c_{k}$ we have used thus far~\footnote{For simplicity, we here assume that the two-mode state is isotropic, i.e. that $n_{k} = n_{-k}$.  This is true for the homogeneous states we consider, but it will only be approximately true when averaging over a finite number of realizations.}. 
Amongst its virtues, its value is independent of whether we use the phononic or the atomic expectation values to compute it; explicitly, it is given by
\begin{equation}
S_{\rm cov} = \sum_{k} S_{{\rm cov}, k} = \sum_{k} \left[ \left(n^{\rm eff}_{k} + 1\right) \, \mathrm{ln}\left( n^{\rm eff}_{k}+1 \right) - n^{\rm eff}_{k} \, \mathrm{ln}\left( n^{\rm eff}_{k} \right) \right] \,,
\label{eq:entropy}
\end{equation}
where $n^{\rm eff}_{k}$ is defined such that
\begin{equation}
\left( n^{\rm eff}_{k} + \frac{1}{2} \right)^{2} = \left( n_{k} + \frac{1}{2} \right)^{2} - \left| c_{k} \right|^{2} \,,
\end{equation}
and where $n_{k}$ and $c_{k}$ can refer either to phonons or atoms. 
$n^{\rm eff}_{k}$ thus vanishes when the two-mode state $(k,-k)$ is pure, i.e. when $\left| c_{k} \right|^{2} = n_{k} \left(n_{k}+1\right)$, and the contribution of $(k,-k)$ to the entropy is then zero.

The growth of this entropy with time is shown in the right panel of Figure~\ref{fig:Eperp_S}, for the same three simulations represented in Fig.~\ref{fig:history_Nk}.  The monotonic nature of $S_{\rm cov}$ (up to small oscillations) is apparent, although at late time the growth rate is reduced, this reduction being more pronounced for a larger number of atoms (i.e. a smaller value of $a_{s}/a_{\perp}$). Of particular note is the observation that the rate of depletion of the oscillation energy tends to decrease when the system enters the broadening phase of increasing total entropy.  We interpret this behavior as indicating that the entropy increase is mainly due to energy redistribution among the phonons which is governed by the first kind of nonlinearity, and not to the second kind which concerns the damping of the coherent condensate oscillations due to production of longitudinal phonons (as studied e.g. in~\cite{Fedichev-Shlyapnikov-Walraven}).  The disconnection of these two kinds of nonlinear process is very clear in the case studied in Appendix~\ref{app:amplitude}: see the red curve on the left panel of Fig.~\ref{fig:Eperp_S_amp}, where the smallness of the damping of the coherent oscillations manifests itself by the constancy of the radial oscillation energy during the phase where the peaks are broadening and the entropy growth rate is maximal.  Finally, by comparing Figs.~\ref{fig:history_Nk} and~\ref{fig:Eperp_S}, one notices that the significant increase in entropy occurs at around the same time that the peaks in $n_{k}$ and $G^{(2)}_{k}$ broaden, when the many-peak structure starts to degenerate into a single broad peak.

\subsection{First-order coherence}\label{first-order-coh}

Another relevant quantity that sheds light on the late-time evolution of the system is the equal-time first-order coherence function
\begin{equation}
g_{1}(t,x; t,x^{\prime}) = \left\langle \psi^{\dagger}(t,x) \, \psi(t,x^{\prime}) \right\rangle \,.
\end{equation}
Since the system is spatially homogeneous, this should be a function of the distance $\Delta x = x - x^{\prime}$ rather than of $x$ and $x^{\prime}$ separately.
We exploit this fact by fixing $\Delta x$ and averaging over $x$, as well as over all realizations. 

\begin{figure}
\includegraphics[width=0.45\columnwidth]{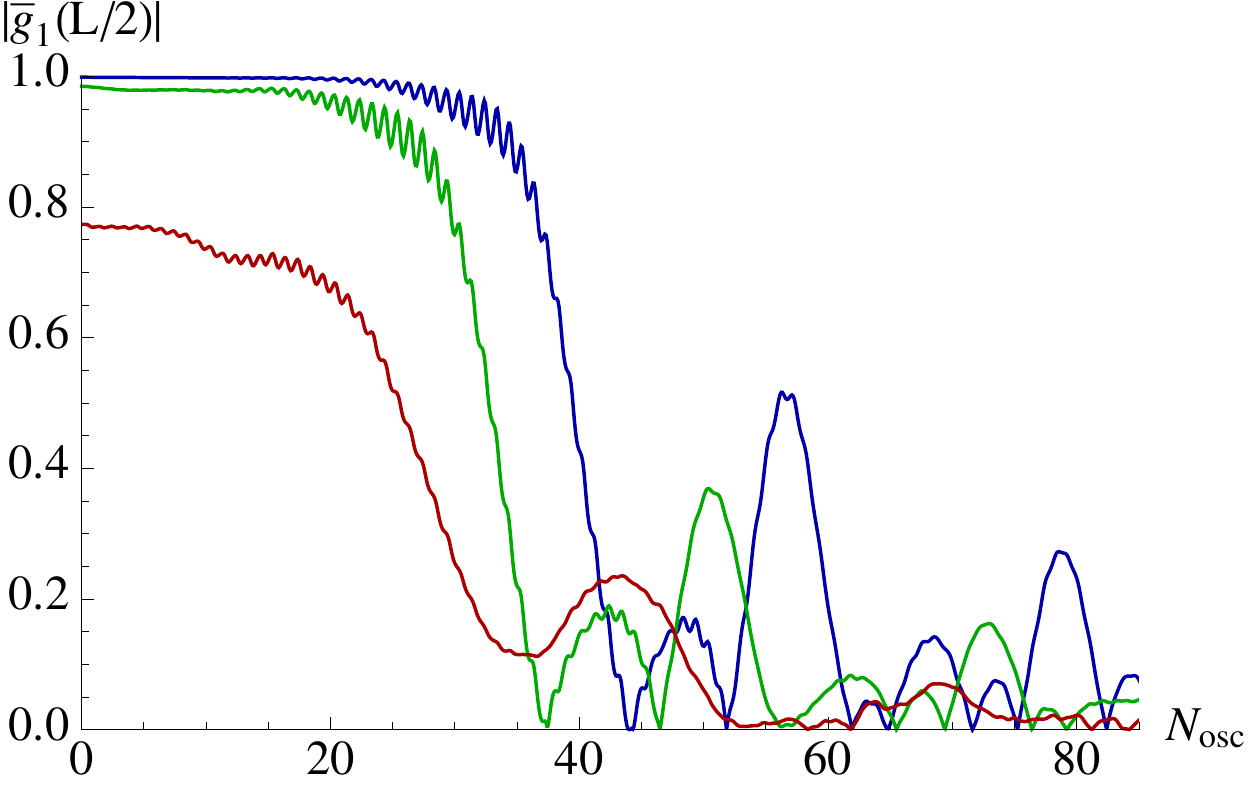} \, \includegraphics[width=0.45\columnwidth]{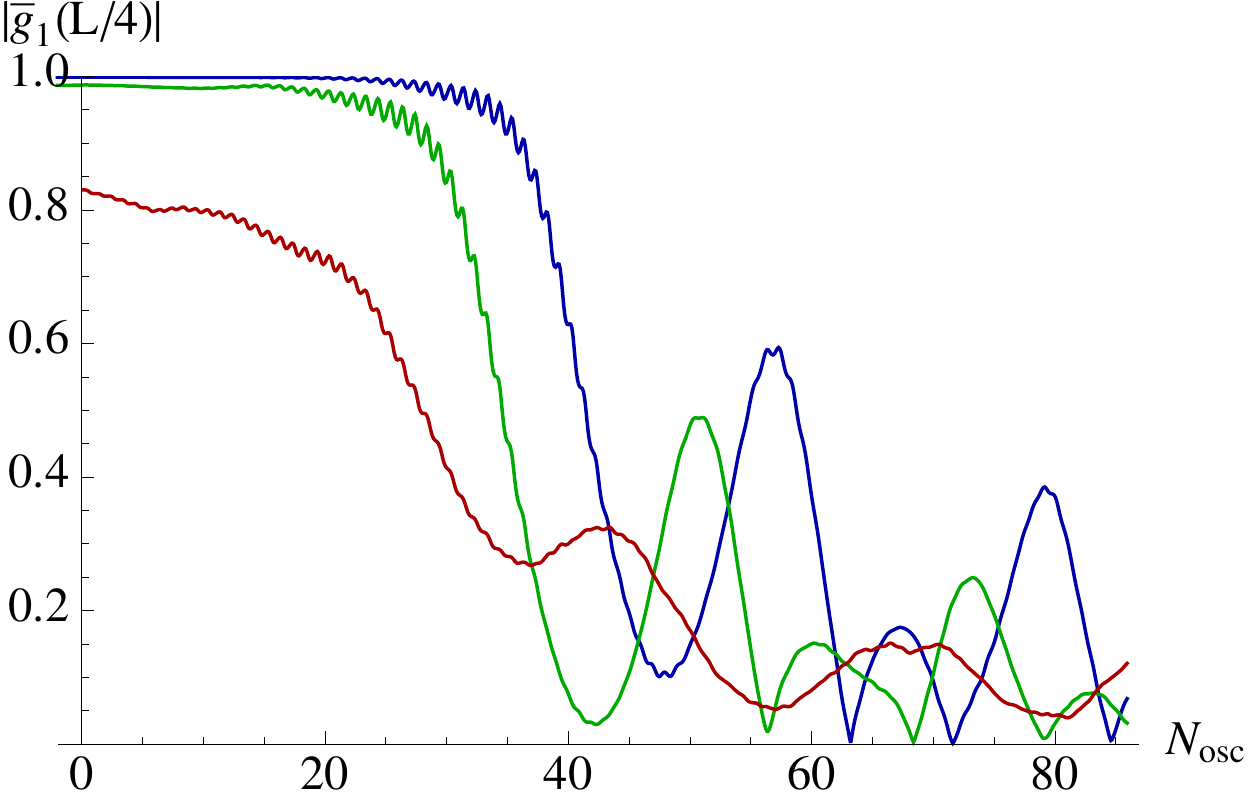}
\caption{The (absolute value of the) first-order coherence function $g_{1}(\Delta x, t) = \left\langle \hat{\psi}^{\dagger}(x,t) \hat{\psi}(x+\Delta x, t) \right\rangle$, as a function of time, for the same parameters as in previous plots and the same three values of $a_{s}/a_{\perp}$: $1.7 \times 10^{-4}$ (blue), $1.7 \times 10^{-3}$ (green) and $1.7 \times 10^{-2}$ (red).  On the left $\Delta x = L/2$, so that the correlation is between antipodal points on the torus; on the right $\Delta x$ is only half this value, yet the curves behave in a similar manner.  We have exploited homogeneity of the state by averaging over $x$ as well as over the number of realizations (here there are $100$ realizations for each of the three curves). 
Of particular note is the sharp decrease in $g_{1}$ at a well-defined time, which occurs later for smaller $a_{s}/a_{\perp}$ and which corresponds to the loss of $(k,-k)$ correlations (through the decrease of the parameter $\tilde{\eta}$ in Figs.~\ref{fig:eta} and~\ref{fig:history_nk_eta}) and the increase of the entropy on the right of Fig.~\ref{fig:Eperp_S}.  Note also that, whereas the blue and green curves 
start with $\left|g_{1}\right|$ close to $1$, the red curve starts with $\left|g_{1}\right|$ significantly lower (around $0.8$), in 
agreement with the fact that the correlation length $l_\phi(T)/L \sim 1$ for this case. 
\label{fig:g1}}
\end{figure}

The evolution of $g_{1}(t,\Delta x)$ for the same three simulations represented in Figs.~\ref{fig:history_Nk} and~\ref{fig:Eperp_S} is shown in Figure~\ref{fig:g1}.
Its most salient feature is the rather sudden drop that coincides with the broadening of the peaks (Fig.~\ref{fig:history_nk_eta}) and the increase of 
the entropy (right panel of Fig.~\ref{fig:Eperp_S}). 
This occurs both for $\Delta x = L/2$ and $\Delta x = L/4$, showing that the effective coherence length reduces markedly and quite suddenly from the entire length of the condensate to a value somewhat less than $L/4$.

To relate our observations to known theoretical 
results concerning the coherence length in one-dimensional quasi-condensates, we give here the three values of $l_\phi(T)$, the correlation length at temperature $T$ defined by $g_{1}(t,l_{\phi}(T)) = 1/e$, 
see Eq.~(14) in~\cite{Bouchoule}. 
Namely, for $a_s/a_\perp = 1.7\times 10^{-4}$, $1.7\times 10^{-3}$ and $1.7\times 10^{-2}$, 
we correspondingly have $l_\phi(T)/L = 100$, $10$, and $1$. 
The last value indicates that for the largest 
value of $a_s/a_\perp$ we consider, on distances comparable to $L$ 
the spatial correlation will be imperfect even before the sudden change, in agreement with the early value of $g_{1}$ shown by the red curve in Figure~\ref{fig:g1}. 

It is also 
worth pointing out that, after 
the sudden drop 
of $g_{1}(t, \Delta x)$, 
the observed coherence length 
can be used to define 
an effective temperature via 
the following 
expression 
given in~\cite{Bouchoule}: 
\begin{equation}
l_{\phi}(T) = \frac{\hbar^{2} n_{1}}{m T} = n_{1} \xi^{2} \frac{m c^{2}}{T} \,,
\label{eq:lphi} 
\end{equation}
where we have set $k_{B} = 1$.
We find $T_{\rm eff}/mc_{\rm fin}^2 \approx 400$, $50$ and $3$ for $a_{s}/a_{\perp} = 1.7\times 10^{-4}$, $1.7\times 10^{-3}$ and $1.7\times 10^{-2}$, respectively. 
We shall see that these values are in qualitative agreement with 
those one can extract 
from the low $k$ behaviors of 
final plots of $G^{(2)}_{k}$ shown in Figure~\ref{fig:final_G2_S} below.

\subsection{Late-time evolution
\label{late-time}} 

\begin{figure}
\includegraphics[width=0.45\columnwidth]{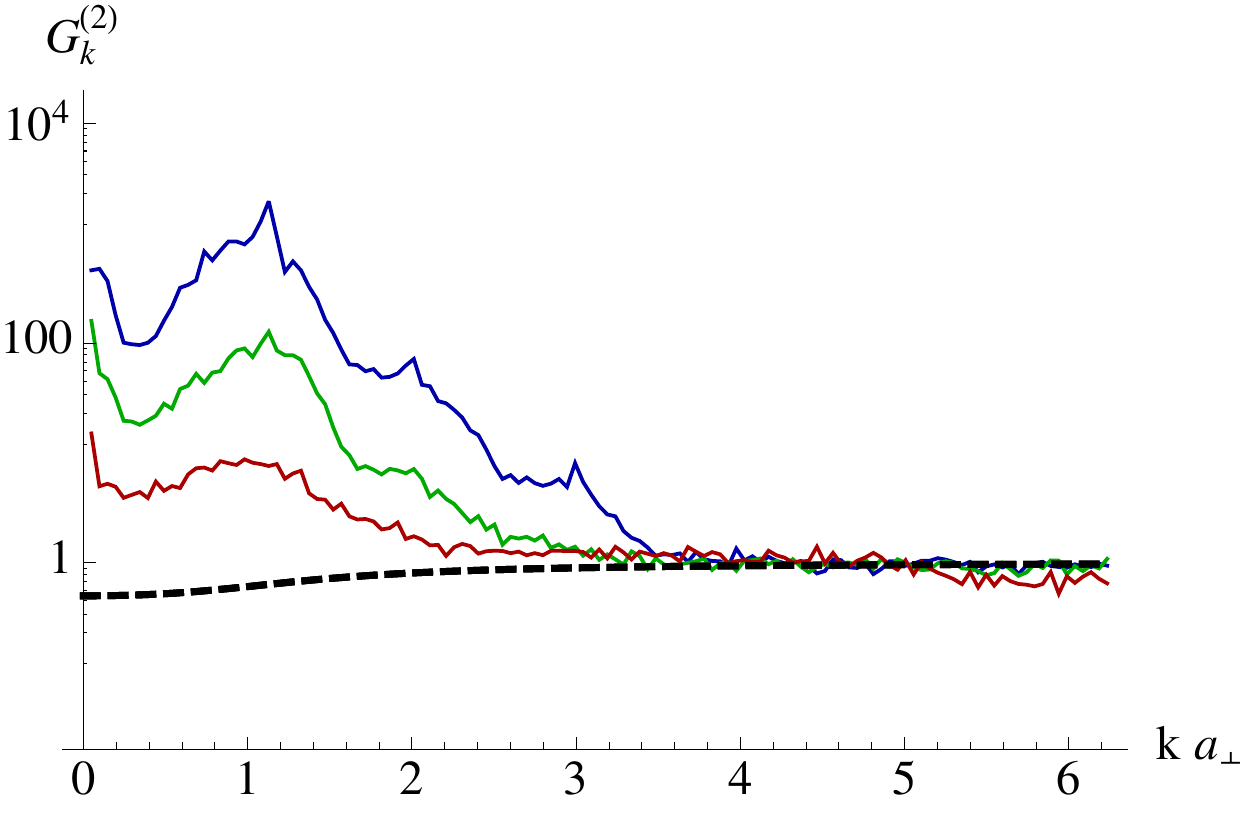} \, \includegraphics[width=0.45\columnwidth]{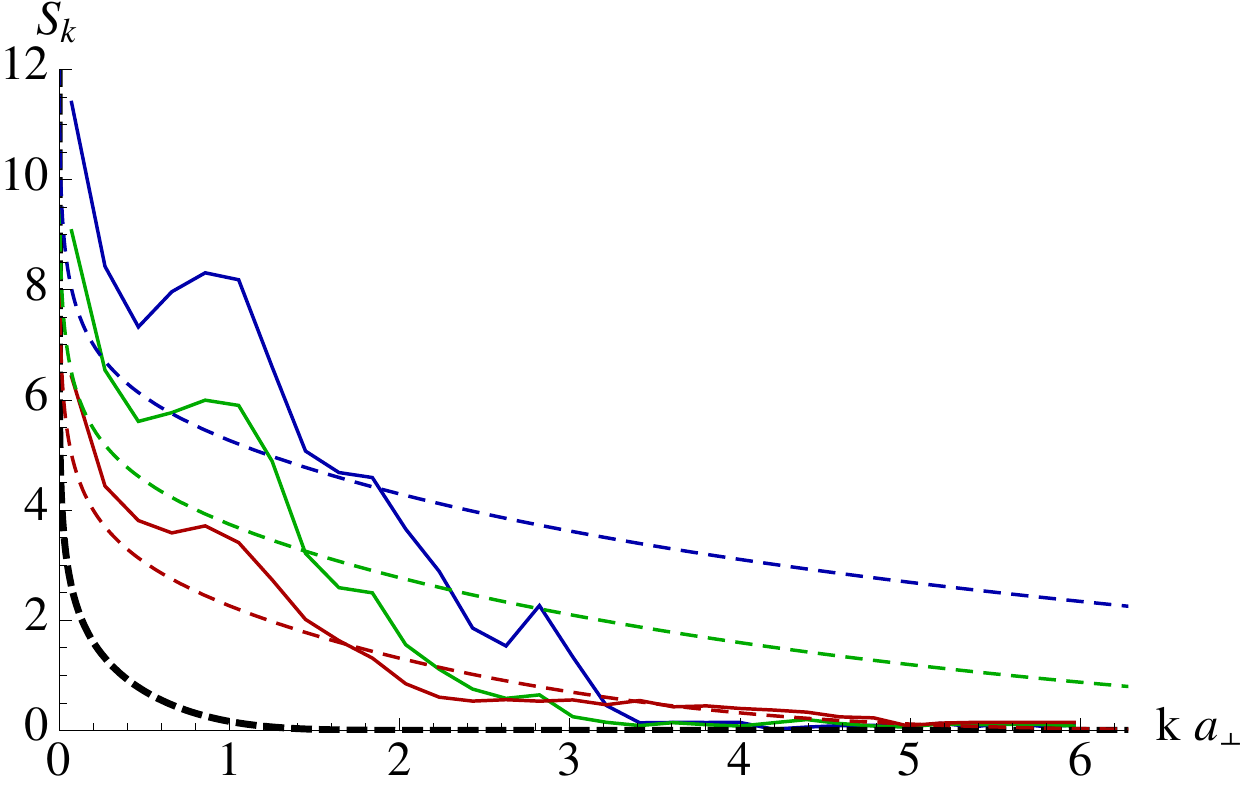} 
\caption{The final (i.e. at $N_{\rm osc} = 66.8$, the last time shown in Figs.~\ref{fig:history_nk_eta} and~\ref{fig:history_G2}) profiles of the density-density correlation function $G^{(2)}_{k}$ (left panel) and the entropy $S_{k}$ (right panel), for the same parameters as in previous plots and the same three values of $a_{s}/a_{\perp}$: $1.7 \times 10^{-4}$ (blue), $1.7 \times 10^{-3}$ (green) and $1.7 \times 10^{-2}$ (red).  The dashed black curves show the theoretical profiles for the initial state at temperature $T_{\rm in} = m c_{\rm in}^{2}/2$. On the right plot, the dashed colored curves show the theoretical predictions for the entropy in a thermal state, with a final temperature $T_{\rm fin}$ determined by the total available energy in the oscillations of the condensate after the sudden change of $\omega_\perp$. 
The corresponding values of $T_{\rm fin}/m c_{\rm fin}^{2}$ are 
$94$ (blue), $21$ (green) and $4.7$ (red). 
Note that $S_{k}$ has been binned into groups of $4$ modes each with respect to the raw numerical data, while the $G^{(2)}_{k}$ has not been binned at all. 
\label{fig:final_G2_S}}
\end{figure}

After the increase in entropy that coincides with the broadening of the peaks
and the drop in the first-order coherence, 
the system continues to evolve, but at a much slower rate.  In Figure~\ref{fig:final_G2_S} are plotted the final profiles of $G^{(2)}_{k}$ and $S_{k}$, along with the thermal predictions at the intial temperature. 
In the plot of the entropy, we also show the thermal prediction at the ``final'' temperature that would correspond to the total available energy in the system. 
Notice that remnants of the resonant peaks are still visible and 
continue to broaden; 
we expect that, if we allow the simulation to run for a long enough duration, they 
will eventually disappear.

There is a clear trend for the 
late time profiles of $G^{(2)}_{k}$ and $S_{k}$ to increase with decreasing $a_{s}/a_{\perp}$ or, equivalently, with increasing total number of atoms $N$.  Indeed, 
$G^{(2)}_{k}$ appears to be directly proportional to $N$ and, since $G^{(2)}_{k}$ approaches $T_{\rm eff}/mc^{2}$ 
in the limit $k \to 0$~\cite{Robertson-Michel-Parentani-1},
it can be concluded that the  
late time effective temperatures 
should be 
proportional to $N$. This trend is 
corroborated by the effective 
temperatures given by the drop of the coherence lengths discussed 
at the end of Sec.~\ref{first-order-coh}. 

This trend is also
predicted by a straightforward mapping of the available energy in the oscillations of the condensate to 
what should be 
the final temperature $T_{\rm fin}$ of the system. 
The steady increase with $N$ of the corresponding thermal entropy profiles
are shown in dashed in the right panel of Fig.~\ref{fig:final_G2_S}. 
It should be noticed that the values of $T_{\rm fin}$ are significantly lower than those of the effective temperatures
$T_{\rm eff}$, in accord with the fact that the energy is mainly distributed in the low frequency modes when we stopped our numerical integration. 
This out-of-equilibrium repartition is most clearly seen in the plot of the entropy, where we see an 
excess at low $k$ and a deficit at high $k$.  
It indicates 
that energy redistribution from the low-$k$ to the high-$k$ regime is still taking place.  Indeed, when the number of condensed atoms is sufficiently large (blue and green curves), the high-$k$ modes are still essentially in their ground state, in qualitative agreement with the findings of Refs.~\cite{Micha-Tkachev-1,Micha-Tkachev-2} where it was noticed that high-$k$ modes thermalize slowly.

Instead, when the number of condensed atoms is small (red curve), $S_{k}$ appears to be close to the equilibrium curve after 67 oscillations.  We also notice that, for the same system (red curve), $G^{(2)}_{k}$ dips below the thermal curve 
of BdG at high $k$.  This behavior seems to be generic when a significant fraction of the atoms are not condensed, as it appeared in the large set of simulations (based on the TWA) we have performed but not shown.  
This phenomenon is probably 
related to the well-known fact that the TWA is unable to properly describe the thermalization of an atomic cloud~\cite{Sinatra-Lobo-Castin,Mora-Castin}, see footnote~\ref{TWA-justific}.

\section{Summary and conclusions 
\label{sec:conclusion}}

We studied the sequence of dynamical processes taking place in an elongated 
effectively 
one-dimensional  condensed atomic cloud when the trapping frequency $\omega_\perp$ governing the two narrow perpendicular directions is suddenly increased.  This causes the radial 
atomic density to oscillate with a high frequency equal to twice 
the final value of $\omega_\perp$. These coherent oscillations induce a modulation of the frequency of longitudinal excitations which in turn leads to an exponential amplification of the phonon modes in a frequency band centered around $\omega_\perp$.  In our numerical simulations, the initial temperature of the homogeneous phonon bath is taken to be relatively low (equal to half the initial value of $m c^{2}$) so that, in effect, the resonant modes are initially in their ground state. 

We used the large scale separation between the longitudinal length $L$ and the perpendicular width $a_\perp$ of the cloud, namely $L/a_{\perp} = 128$, to identify two kinds of non-linearity 
that are treated 
in a self-consistent manner.
The first 
describes the mutual interactions 
of longitudinal excitations, which propagate 
in a time-dependent homogeneous background governed by the scale factor $\sigma(t)$ describing 
the radial oscillations. 
We restrict ourselves to statistically homogeneous states, and we use the truncated Wigner approximation to numerically solve the corresponding nonlinear 
field equation. The second kind of nonlinearity 
concerns the adiabatic reduction 
of the radial oscillations caused by the increase in 
the mean energy of the resonant modes. This second type is 
governed by the ``semi-classical'' 
equation of motion 
for $\sigma(t)$, in the same spirit as what is done in semi-classical gravity~\cite{RWaldbook}.
To obtain this equation, which here reduces to an ODE, 
we have taken the ensemble average over the various realizations of longitudinal excitations, and because 
our ensemble is statistically homogeneous, this 
is expressed via 
the spatial integral 
of their energy density. 
Then, by construction, the total energy of the system is conserved. The two equations of motion are numerically integrated in time in a single code. In practice, the identification of the two kinds of nonlinearity 
is implemented by postulating that the three-dimensional wave function factorizes, see Eq.~(\ref{eq:factorization}). It should be stressed that this 
identification 
and the 
subsequent numerical integration closely follow 
the procedure which is used in numerical studies of the preheating 
scenario in cosmology~\cite{Figueroa2017JCAP}. 

Having adopted this description, we first paid attention to the initial phonon state. We set the random initial conditions of phonon fluctuations well before the sudden jump so as to let the 
quasi-condensate (governed by a one-dimensional Gross-Pitaevskii equation) 
settle into 
a nearly stationary state. After the sudden jump of $\omega_\perp$ we observed, as expected, an exponential growth of resonant phonon modes and the entanglement (nonseparability) of the two-mode phonon states comprising opposite wave vectors. Both of these observations are in good agreement with the predictions obtained using the BdG equation. Yet we rapidly 
observed two 
deviations with respect to this linear treatment. 
The main one concerns the loss of nonseparability of $(k,-k)$ phonon pairs while the number of these phonons is still exponentially growing, see Fig.~\ref{fig:G2}. The other 
deviation is the 
progressive reduction of the (exponential) growth rate of the phonon occupation number.

Moreover, we numerically verified that the strength 
of these two deviations (at any 
given time) is reduced 
when decreasing the ratio $a_{s}/a_{\perp}$. 
This can be explained by noting that, while in our scheme of adimensionalization (in which, in particular, the number $n_{1}a_{s}$ is held fixed) 
$a_s/a_\perp$ does not enter in the BdG description, 
it {\it does} govern 
the fraction of depleted atoms with respect to the total atomic number. 
We numerically observed that, for a significant period during the exponential growth of the resonant peak, the reduction of the growth rate of the phonon occupation number scales linearly with $n_k \, a_s/a_\perp$ (where $n_k$ is the number of resonant phonons), while the parameter governing the loss of nonseparability increases like 
$\left(n_k \, a_s/a_\perp\right)^{2}$, 
see Fig.~\ref{fig:Gamma}. 

In parallel to the study of the loss of nonseparability, we addressed the important issue of the {\it visibility} of nonseparability, i.e., the ability to distinguish separable from nonseparable states given some observables. As previously noticed~\cite{Robertson-Michel-Parentani-1}, we recovered that both {\it in situ} measurements of the two-point correlation function and statistical properties of the atomic numbers after TOF (see Appendix~\ref{app:g2}) are unable to distinguish between these two classes of states when the mean occupation number (of phonons or atoms) becomes larger than $\sim 10$. With our ``benchmark'' values for the system parameters, 
the visibility is maximal after only $\sim 4$ oscillations of the atomic cloud. 

In the second part of the paper, we studied the late-time behavior, where nonlinear effects are essential in 
the evolution of the system. Since the BdG description is no longer valid, 
it is then appropriate to use atomic (rather than phononic) occupation numbers. 
As clearly seen in Fig.~\ref{fig:history_nk_eta}, the expected saturation of the exponential growth of resonant atoms is accompanied by a series of interesting effects. Firstly, we observed new peaks at 
harmonics of the wave number $k_{\rm res}$ (and not of the frequency $\omega_{k_{\rm res}}$), 
whose appearance and amplitudes are 
explained in Appendix~\ref{app:nonlinear}. Secondly, we observed a 
rapid broadening of all peaks, including the central one at $k=0$. 
Thirdly, the coherence of resonant phonon pairs of opposite wave vector is essentially washed out, as can be seen from the lower panels in Fig.~\ref{fig:history_nk_eta}.  These observations are corroborated by the temporal behavior of the two-point function shown in Fig.~\ref{fig:history_G2}, and 
by the time-dependence 
of the atomic occupation 
numbers in various wave number bands, see Fig.~\ref{fig:history_Nk}. 
We conjecture that these decoherence effects are due to 
frequent exchanges with the large bath of soft phonons which is known to be present 
in one-dimensional quasi-condensates at a finite temperature, and which is 
the origin of their 
finite correlation length. 

The analysis of late-time 
effects is completed by a study of the energetic and entropistic aspects.  One clearly sees that there is an almost complete energy transfer from the cloud oscillations to the various longitudinal excitations. 
However, some of our observations suggest an interplay between the growth of energy and entropy, in that when one varies rapidly, the other less so, and {\it vice versa}; see in particular the blue curves in Fig.~\ref{fig:Eperp_S} and the red curves in Fig.~\ref{fig:Eperp_S_amp} (in Appendix~\ref{app:amplitude}). 
This seems to indicate 
that the damping of the coherent radial oscillations is 
governed by nonlinear processes which are distinct from those responsible for the broadening of the peaks. 
In addition, 
by computing the equal-time first-order coherence function evaluated at some large distance
comparable to the length of the torus, 
we observed that the cloud is well-described by a quasi-condensate up to a certain moment which roughly coincides with the moment at which the peaks broaden
and the entropy increases. 
After this moment, the spatial coherence is suddenly lost.  As could have been expected, we observed that this time occurs later when there is a larger fraction of condensed atoms initially, 
i.e., when $a_{s}/a_{\perp}$ is smaller. 

Altogether these results suggest 
that the system is on its way to thermalization.  They also indicate that the last stages of the energy redistribution, both within the bath of longitudinal phonons and 
from the coherent radial oscillations to the phonon bath, 
are rather slow.  These observations are clear when examining the occupation of high-frequency modes.  From the right panel of Fig.~\ref{fig:final_G2_S}, it is seen that after many oscillations of the condensate, the high-frequency modes are still far away from their thermal values, in agreement with the slow increase of the entropy observed at late time in the right panel of Fig.~\ref{fig:Eperp_S}. 

These slow late-time processes, here observed in a one-dimensional system, are very reminiscent of the outcome of 
studies of the three-dimensional processes (involving turbulence and vortices) of the (p)reheating 
scenario of primordial cosmology~\cite{Kofman-Linde-Starobinsky,Micha-Tkachev-1}, where it was also seen that high-frequency modes thermalize long after the broadening of the spectrum involving low-frequency modes.  We hope that the dynamical origins of these similarities will be clarified in the near future.

\section*{Acknowledgments}

We thank Chris Westbrook, Denis Boiron, Iacopo Carusotto, Andrea Trombettoni, Sergey Sibiryakov and Daniel Figueroa for interesting discussions. 
We also thank the authors of Ref.~\cite{Zache-Kasper-Berges} for bringing our attention to their recent work, and the two anonymous referees whose useful remarks helped us in clarifying the manuscript. 
This work was supported by the French National Research Agency through the Grant No. ANR-15-CE30-0017-04 associated with the project HARALAB.
RP wishes to warmly thank the SISSA for hosting him within its ``Excellence Visiting Programme'', during which the revised version of this work was completed.

\begin{appendices}
\numberwithin{equation}{section}


\section{Alternative description of whole history in terms of $g_{2}(k)$
\label{app:g2}}

\begin{figure}
\includegraphics[width=0.45\columnwidth]{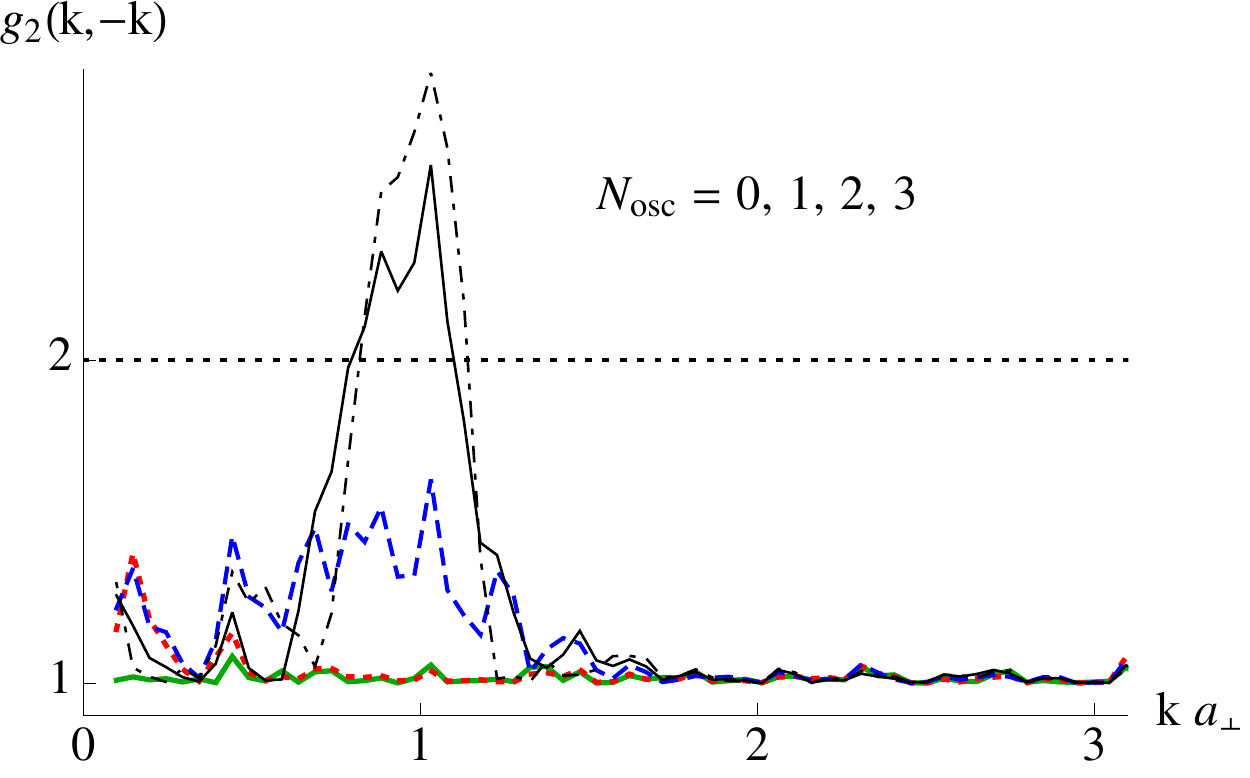} \, \includegraphics[width=0.45\columnwidth]{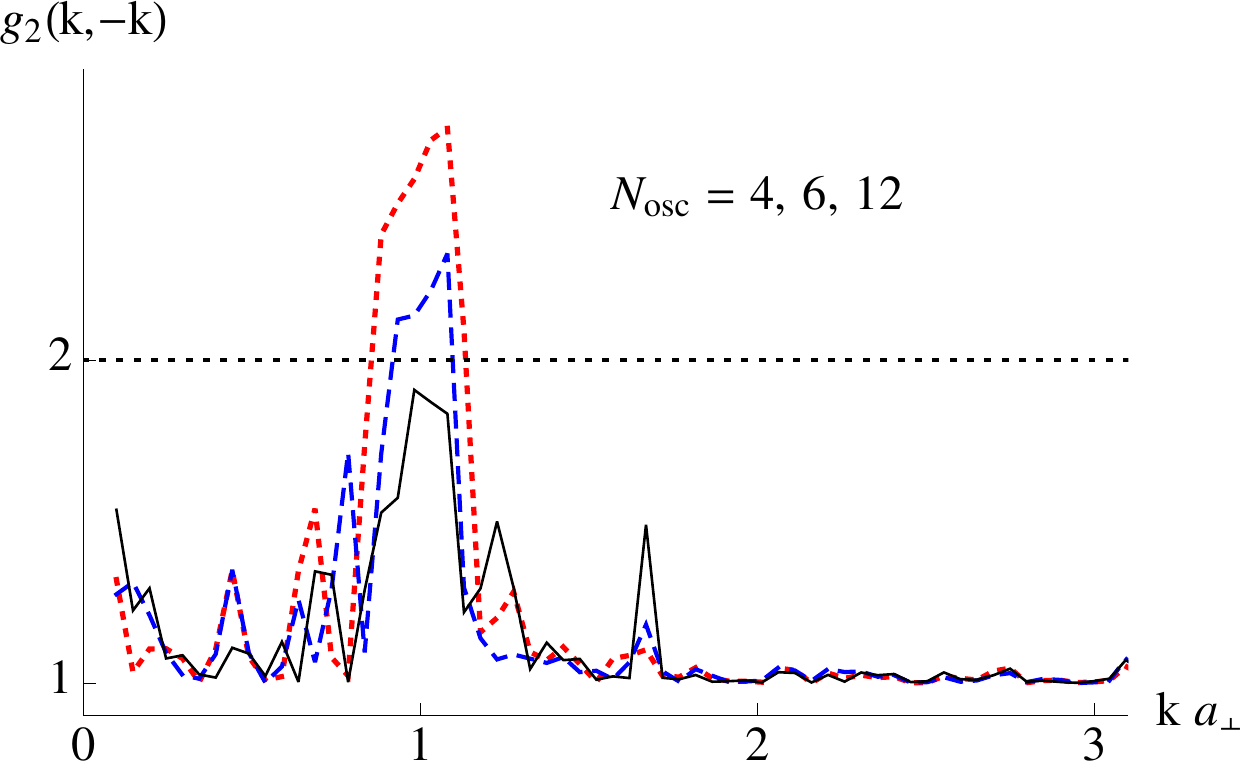} \\
\includegraphics[width=0.45\columnwidth]{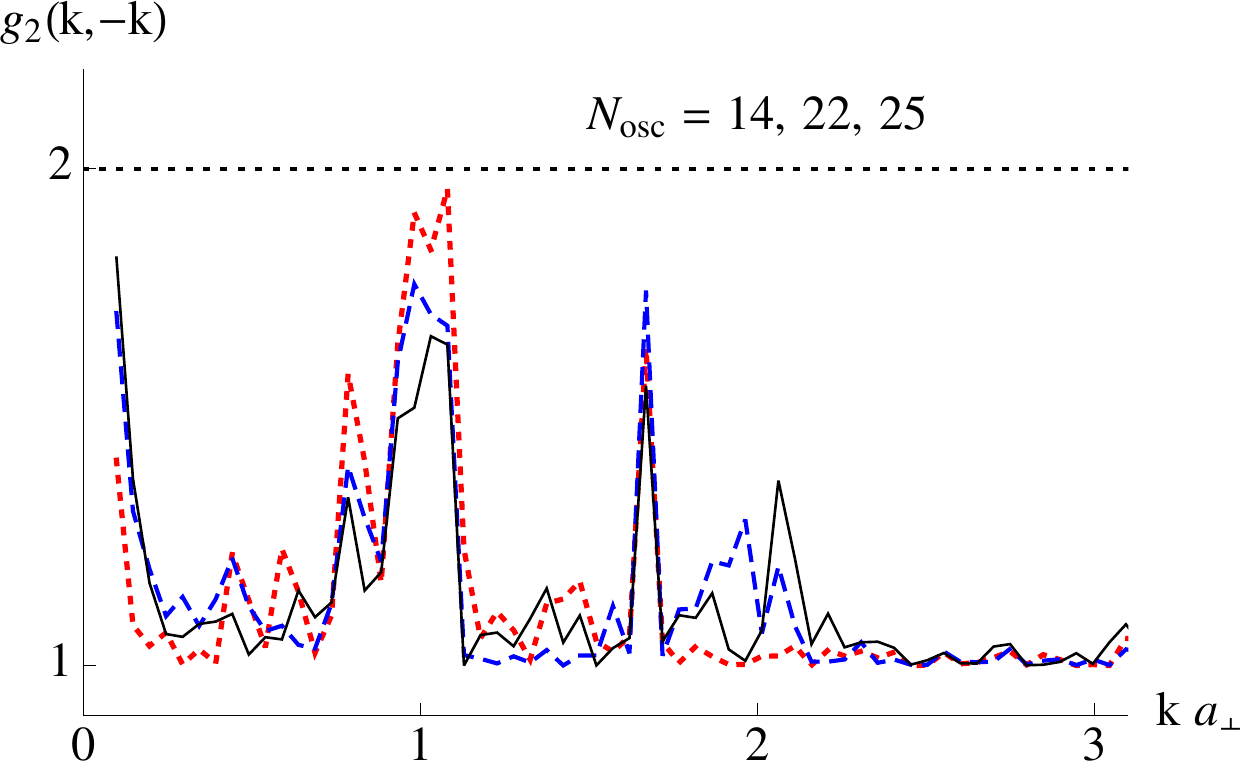} \, \includegraphics[width=0.45\columnwidth]{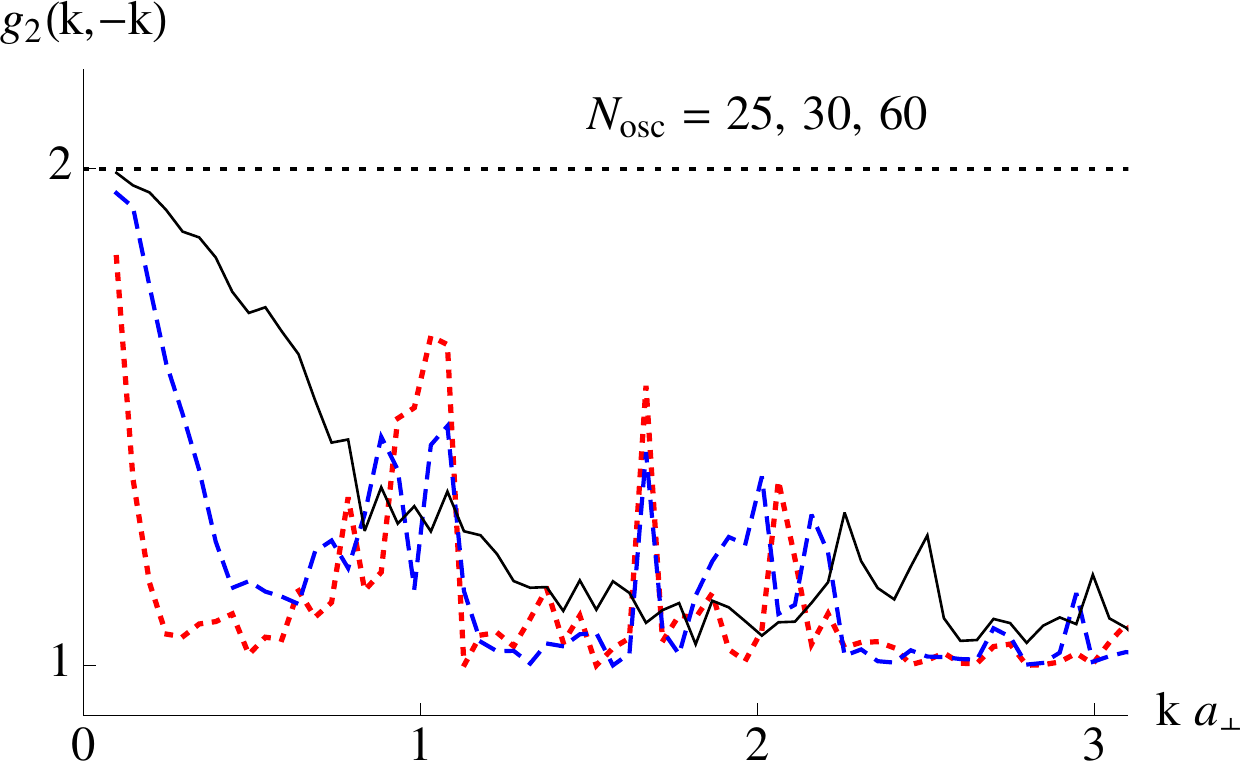}
\caption{Evolution in time of $g_{2}(k; t)$, with the upper and lower plots respectively showing the early- and late-time behavior described in the main text.  In each plot, the curves are time-ordered as follows: (solid green), dotted red, dashed blue, solid black, (dot-dashed black).  Note that the solid green and dot-dashed black curves appear only in the first (upper left) plot.  The green curve corresponds to the input state, assumed to be a thermal state of phonons at temperature $T = mc^{2}/2$, at $N_{\rm osc} = -10$, i.e. before the sudden change in the trapping frequency.  The red curve in the first plot shows the form of $g_{2}$ at $N_{\rm osc} = 0$, after having evolved the system according to the full quartic Hamiltonian.
\label{fig:g2}}
\end{figure}

We remind the reader that after TOF, the observables are the statistical properties of the atomic occupation 
numbers with wave number $k$, see e.g.~\cite{Kheruntsyan-et-al}.  For simplicity, we here assume that the expansion of the cloud is such that the atom occupation numbers after 
the opening of the trap are equal to the phonon occupation numbers beforehand, which amounts to assuming that the trap is opened adiabatically with respect to the relevant atom or phonon frequencies.  
Since the natural expansion rate of the cloud on the switching-off of the harmonic potential is on the order of $\omega_{\perp}$, 
this assumption is valid for atoms of wave numbers close to or higher than the resonant window.  (The reader interested in 
the residual effect induced by a more accurate description of the expansion of the cloud during the opening of the trap is invited to consult Ref.~\cite{Robertson-Michel-Parentani-1}.) 

We recall that the $g_2(k)$ function, which has been used in Refs.~\cite{Carusotto-BH,Jaskula-et-al,Boiron-et-al}, is given by 
\begin{equation}
g_{2}(k) = \frac{\left\langle \hat{\phi}_{k}^{\dagger} \hat{\phi}_{-k}^{\dagger} \hat{\phi}_{-k} \hat{\phi}_{k} \right\rangle}{ \left\langle \hat{\phi}_{k}^{\dagger} \hat{\phi}_{k} \right\rangle \left\langle \hat{\phi}_{-k}^{\dagger} \hat{\phi}_{-k} \right\rangle} = \frac{\left(n^{\rm at}_{k}\right)^{2} + \left| c^{\rm at}_{k} \right|^{2}}{\left(n^{\rm at}_{k}\right)^{2}} \,.
\label{eq:g2_defn}
\end{equation}
In the second equality, we assumed that the state is isotropic ($n_{k}^{\rm at} = n_{-k}^{\rm at}$) and Gaussian, so that the expectation value of the quartic operator can be expressed in terms of expectation values of quadratic operators via Wick contraction.  This is much the same philosophy adopted when using $S_{\rm cov}$ as a measure of the entropy in Eq.~(\ref{eq:entropy}), and it here means that the connected part of the four-point function has not been taken into account.  The study of non-Gaussianities of the quantum state is 
beyond the scope of the present paper. 

In Figure~\ref{fig:g2}, we have adopted four plots in order to distinguish four stages illustrating the successive processes at play.  Before describing 
them, we state the following preliminary facts.   
Firstly, we have added a regulator on the average squared number of phonons (equal to $1/10$) so as to avoid large fluctuations of $g_{2}$ for low occupation numbers which occur at high $k$.  These fluctuations are 
due to the finite number of realizations, here (as throughout this paper) 
equal to $100$. 

Secondly, as far as $k$ is concerned, we have partially smoothed out the curves by taking the following weighted average:
\begin{equation}
\bar{g}_{2}(k,t) \equiv \frac{1}{8} \left( g_{2}(k-\delta,t) + 6 \, g_{2}(k,t) + g_{2}(k+\delta,t) \right) \,,
\label{eq:ksmooth}
\end{equation}
where $\delta$ is equal to $2\pi/L$ and 
$L = 128 \, a_{\perp}$.  We have adopted this smoothing-out 
because it preserves the detailed properties while erasing high resolution oscillations that are present even in 
vacuum, 
the latter observation indicating 
that these oscillations partially stem from the finite number of realizations as they are present even before the onset of the condensate oscillations at $N_{\rm osc} = 0$.

Finally, we emphasize that the curves are snapshots taken at a series of specified times. 
When comparing curves at two times 
separated by about $\pi/2\omega_{\perp}$, significant modifications on the order of $10\%$ are observed for all values of $k$ except for $k a_{\perp} \lesssim 1/4$.  Moreover, without the smoothing-out of 
Eq.~(\ref{eq:ksmooth}), these modifications can be significantly larger (more than a factor of $2$ with respect to those of the smoothed 
$\bar{g}_{2}$).

Let us now turn to the curves themselves. 
The two upper plots 
represent the early-time evolution.  In the left panel, 
we see show the growth of the maximum value of $g_{2}$ at very early times.  For the 
modes within the resonant window 
(of which there are two in the present simulation), the maximum $g_{2,{\rm max}} \sim 3$ occurs at $N_{\rm osc} = 3$, and hardly varies from $N_{\rm osc} = 2$ to $N_{\rm osc} = 4$.  Note that the maximum value is well above the nonseparability threshold 
$g_{2} = 2$.  Also notice 
that the black curve at $N_{\rm osc} = -10$ and the red curve at $N_{\rm osc} = 0$ 
describe the nonlinear evolution of the phonon vacuum under Eq.~(\ref{eq:GPE-1D}).  (Had we used the BdG equation these two curves would coincide and describe the thermal phonon state at temperature $m c^{2}/2$.)  The discrepancy between these 
curves decreases when increasing the total number of atoms at fixed $n_{1} a_{s}$.

In the upper right 
plot is shown the loss of visibility due to the exponential growth of the occupation number, and has nothing to do with the loss of coherence, as can be understood from the fact that this evolution is accurately described by the linear BdG equation.  The sideband oscillations are also predicted by the BdG equation, see Fig.~17 of~\cite{Robertson-Michel-Parentani-1}.  Furthermore, we notice the appearance of a 
peak at 
$k a_{\perp} \sim 1.7$ associated with the second harmonic in frequency.  Here, its presence is due to the anharmonicity of the oscillation of the condensate, and is thus also in agreement with BdG. 

Instead, the two lower plots show consequences of nonlinear effects absent from 
the BdG description.  The left one displays both the loss of the visibility of nonseparability and the loss of nonseparability itself occurring near $N_{\rm osc} \sim 17$.  These two observations are in full agreement with what is displayed in Fig.~\ref{fig:G2}.  We also notice the growth of the peak at $k a_{\perp} \sim 2$, which is present due to the nonlinear nature of the resonant mode at large amplitudes as discussed in Appendix~\ref{app:nonlinear}.  The last plot shows the gradual disappearance of all peak structure for $k a_{\perp} \gtrsim 1$, and therefore the approach to a thermal-like state, in agreement with the growth of $g_{2}$ for $k a_{\perp} \lesssim 1$.
However, as noted in the main text, the TWA becomes unreliable at very late time, and moreover, the use of the phonon mode basis becomes ambiguous when nonlinearities are strong; we must therefore be cautious when interpreting the last plot of Fig.~\ref{fig:g2}.  


\section{Varying the amplitude of the oscillations
\label{app:amplitude}}

Variation of the oscillation amplitude can be achieved by varying the ratio of the initial to the final trapping frequency.  Through its effect on the amplitude of the oscillations of the scale factor $\sigma$, this controls the amplitude of the oscillations of $\omega_{k}^{2}/\left\langle \omega_{k}^{2} \right\rangle$, the relevant quantity from the point of view of the phonon response.  As shown in~\cite{Busch-Parentani-Robertson}, this affects the system in two ways: it increases the growth rate of the resonant modes, and it increases the size of the resonant window in $k$-space so that more of the discrete modes on the torus are parametrically amplified.

\begin{figure}
\includegraphics[width=0.45\columnwidth]{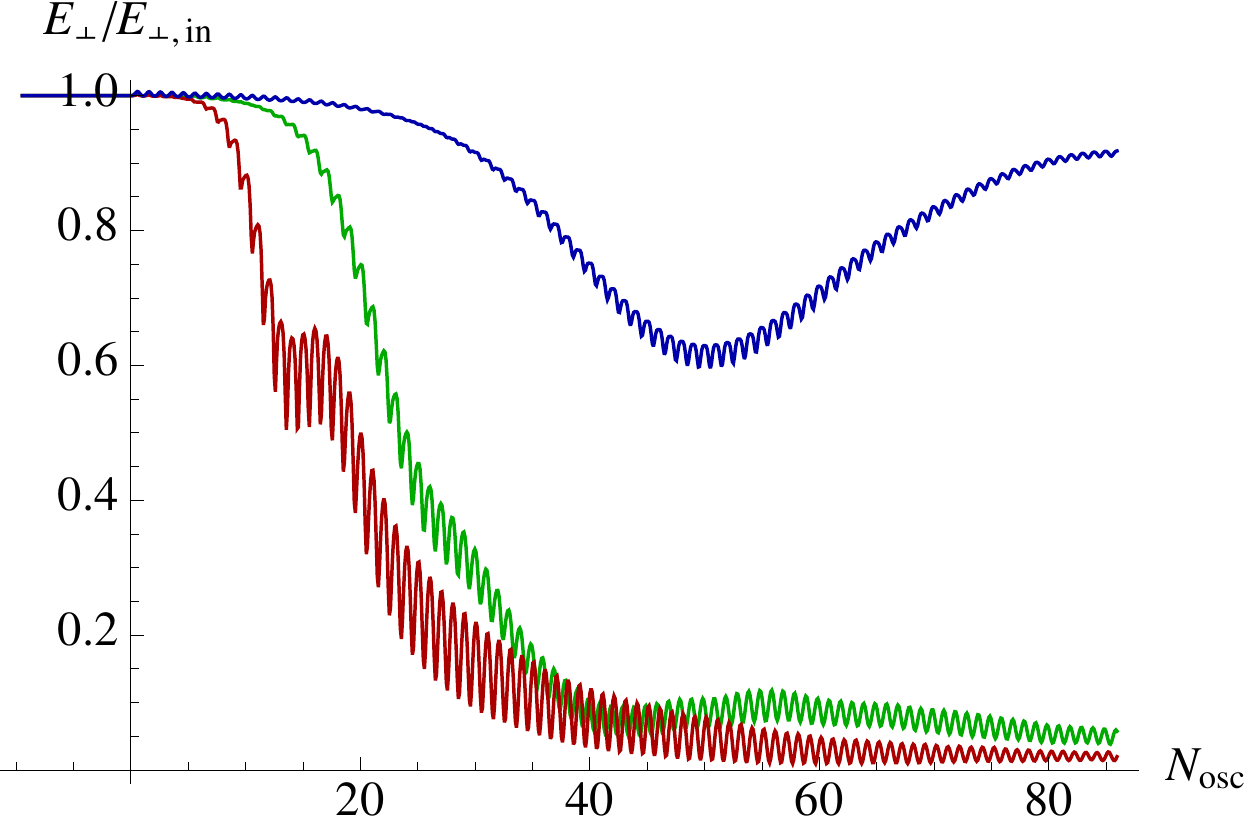} \, \includegraphics[width=0.45\columnwidth]{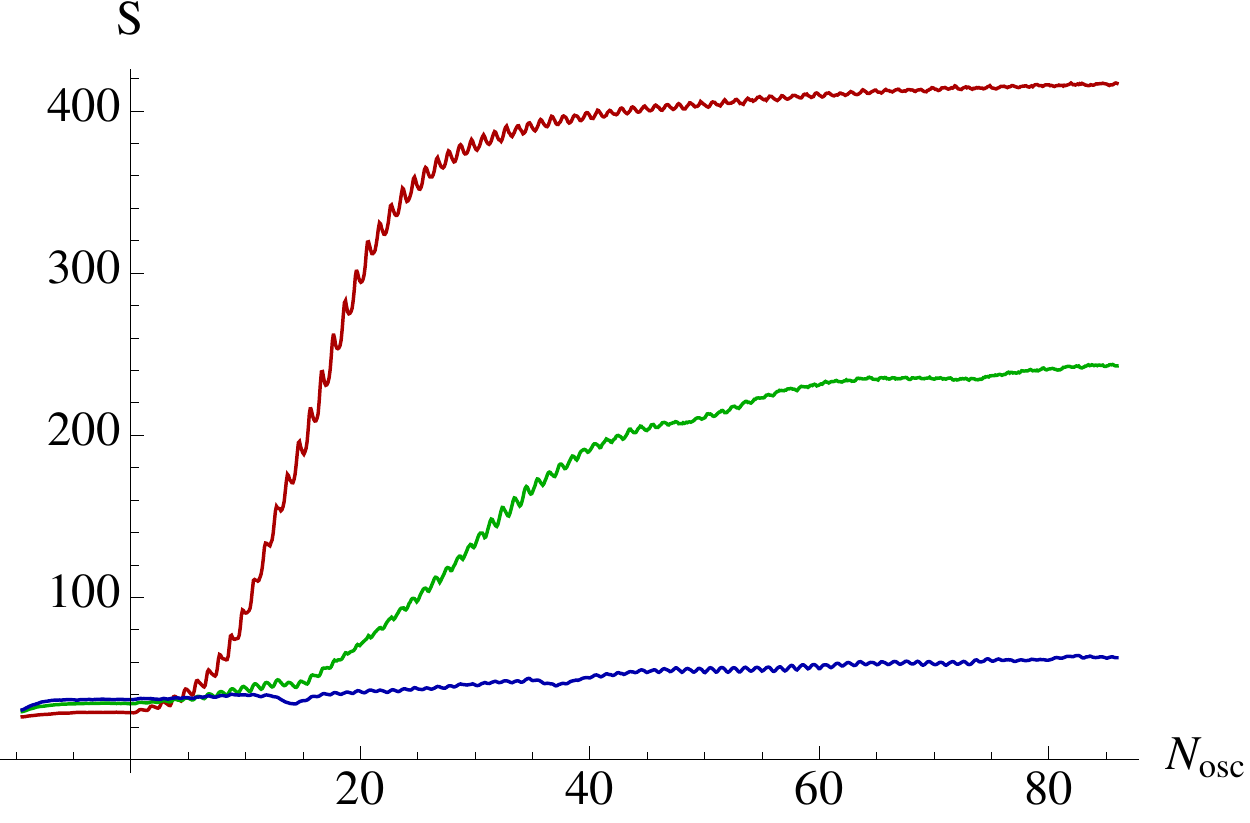}
\caption{Evolution of energy and entropy, with varying $\omega_{\perp}/\omega_{\perp, {\rm in}}$.  On the left is shown the energy of the radial oscillations of the condensate as a function of time, while on the right is shown the entropy as a function of time, for the same parameters as in all previous figures except that $a_{s}/a_{\perp}$ is fixed at $1.7 \times 10^{-3}$ while $\omega_{\perp}/\omega_{\perp, {\rm in}}$ takes the values $1.2$ (blue), $\sqrt{2}$ (green) and $2$ (red).  The number of 
discrete 
resonant modes is $1$ for the blue curve, $2$ for the green curve and $4$ for the red curve.  
\label{fig:Eperp_S_amp}}
\end{figure}

In Figure~\ref{fig:Eperp_S_amp} are shown the time evolution of the energy in the radial oscillations (see Sec.~\ref{energy-entropy} for the definition of $E_{\perp}$) 
and the entropy $S_{\rm cov}$ of Eq.~(\ref{eq:entropy}).  This is just as in Fig.~\ref{fig:Eperp_S}, except that $a_{s}/a_{\perp}$ is fixed at the benchmark value of $1.7 \times 10^{-3}$ while $\omega_{\perp}/\omega_{\perp,{\rm in}}$ varies, taking the values $1.2$ (blue), $\sqrt{2}$ (green) and $2$ (red).  These values lead to different numbers of discrete modes occurring within the resonant window: $1$ (blue), $2$ (green) and $4$ (red).  As expected, the increased rate of phonon production at larger amplitudes leads to the radial energy being used up more quickly.  There are, however, some interesting features: the red curve shows a plateau in the energy lasting over about $4$ oscillations at the transition between early-time and late-time behavior (i.e. where the peaks broaden and the entropy grows most rapidly); while the blue curve shows long-time oscillations in the radial energy, meaning 
that energy is recuperated by the radial oscillations at the expense of longitudinal phonons.  The latter is most likely a pathological feature due to there being only one discrete mode at resonance, while the former is a particularly clear indicator that the entropy increase at broadening is primarily due to energy redistribution among the longitudinal phonon modes themselves, having little to do with backreaction on the oscillating cloud, i.e. on the damping of the coherent radial oscillations. We tested this interpretation by performing an extra simulation with the last term removed from Eq.~(\ref{eq:Veff_backreaction}) so that no backreaction occurs. 
This new simulation displayed essentially the same plateau (both in height and duration) in the longitudinal energy acquired by the phonons, thereby confirming that 
this particular 
process is unrelated to the reduction of the radial oscillation energy. 

Despite the interesting features in the evolution of the energy, the entropy (shown in the right panel of Fig.~\ref{fig:Eperp_S_amp}) behaves essentially as expected, increasing sooner and most rapidly for larger oscillation amplitudes.  It is quite clear that the blue curve evolves so slowly that it is very far from its final state even after $\sim 85$ oscillations, and the small dips in entropy it shows near $N_{\rm osc} \sim 14$ and $\sim 37$ are again likely due to there being only one discrete resonant mode.


\section{Harmonics in nonlinear solutions 
\label{app:nonlinear}}

In this appendix, we estimate the order of the amplitude of the harmonics for small nonlinear perturbations.  To this end, we look for solutions of the 1D Gross-Pitaevskii equation (GPE) 
of the form $\phi(t,z) = \phi_0 (t,z) \s \left( 1 + u(t,z) \right)$, where $\phi_0$ is a solution corresponding to a homogeneous flow and $u$ is a small but finite perturbation.  Here we work in units where $\hbar = 1$, and in the presence of a nonvanishing background velocity $v_{0}$. 
The 1D GPE becomes:
\begin{equation}\label{eq:NLBdG}
i \s (\partial_t + v_0 \s \partial_z) u = - \frac{\partial_z^2 u}{2 \s m} + g_1 \s \rho_0 \s (u + u^* + 2 \s u \s u^* + u^2 + u^2 \s u^*),
\end{equation}
where $v_0 = \mathrm{Im}((\partial_z \phi_0)/\phi_0)$ and $\rho_0 = \left| \phi_0 \right|^2$.  Notice that this differs from the BdG equation by the inclusion of the nonlinear terms.  Let us look for solutions of the form:
\begin{equation}
u(t,z) = \sum_{n \in \mathbb{Z}} u_n \s \exp \left( i \s n \s (k \s z - \omega \s t) \right),
\end{equation}
where $k$ and $\omega$ are two real numbers.  Plugging this into Eq.~\eqref{eq:NLBdG} gives, for all $n \in \mathbb{Z}$:
\begin{equation}\label{eq:NLBdG_rec}
\left( n \s (\om - v_0 \s k) - \frac{n^2 \s k^2}{2 \s m} \right) u_n - g_1 \s \rho_0 \s (u_n + u_{-n}^*) = 
	g_1 \s \rho_0 \sum_{l \in \mathbb{Z}} (2 \s u_{n+l} \s u_l^* + u_{n+l} \s u_{-l})
	+ g_1 \s \rho_0 \sum_{(l,p) \in  \mathbb{Z}^2} u_{n+l} \s u_{p-l} \s u_p^*. 
\end{equation}

Let $\ep$ be a small parameter. We consider solutions where $u_0$ and $u_{\pm 1}$ are of order $\ep$ or smaller. 
When working to linear order, we can set $u_{\pm n} = 0$ for all $n \geq 2$ and we recover the solutions of the BdG equation.  
When computing the nonlinear corrections, the right-hand side of Eq.~\eqref{eq:NLBdG_rec} will be in $O(\ep^2)$ for $n = \pm 2$, in $O(\ep^3)$ for $n = \pm 3$, and so on. 
In fact, one can easily see (using that $\abs{n+l} + \abs{l} \geq \abs{n}$ for any $(n,l) \in \mathbb{Z}^2$) that the system is consistent, in the sense that the linear terms in $u_n$ are of the same order as the leading nonlinear term, if $u_l = O(\ep^{\abs{l}})$ for any $l \in \mathbb{Z}$. 
Assuming there is no fortuitous cancellation, the $n^\text{th}$ peak will thus have an amplitude in $O(\ep^{\abs{n}})$, where $\ep$ is the order of magnitude of the amplitude of the first peak. 

\begin{figure}
\includegraphics[width=0.45\columnwidth]{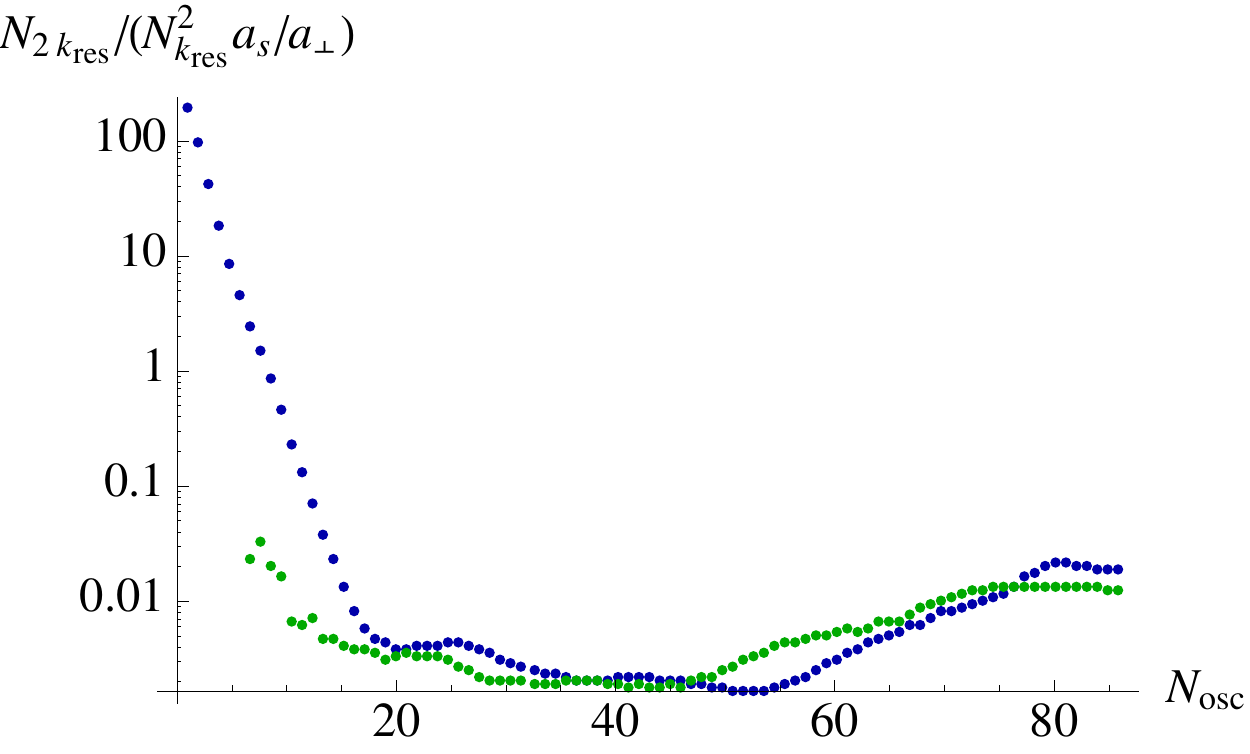} \, \includegraphics[width=0.45\columnwidth]{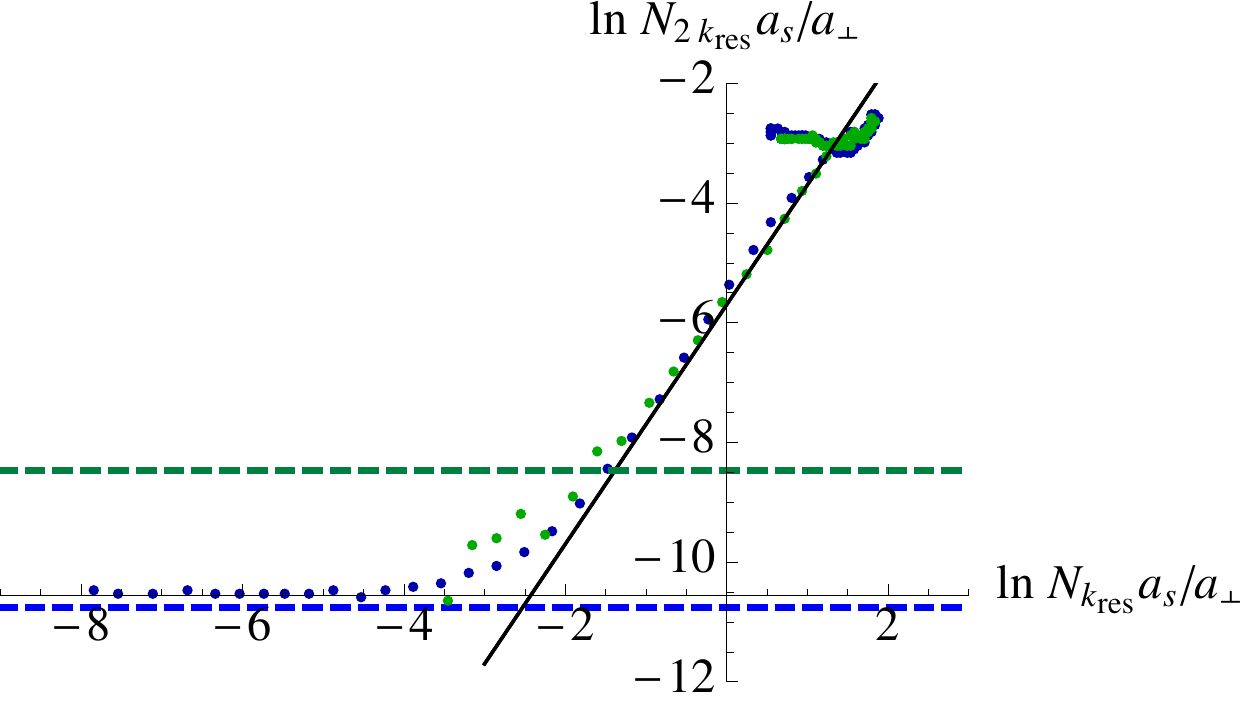} 
\caption{Emergence of nonlinear solutions.  On the left is plotted $N_{2 k_{\rm res}}/\left(N_{k_{\rm res}}^{2}\,a_{s}/a_{\perp}\right)$ as a function of time, 
where $N_{k}$ is the number of atoms summed over $k$ and $-k$, including the two nearest neighbors on either side (just as was done in Fig.~\ref{fig:history_Nk}), and having averaged over the oscillation period to make the curves smooth.  On the right is plotted $\mathrm{ln}\left( N_{2 k_{\rm res}} a_{s}/a_{\perp} \right)$ 
as a function of $\mathrm{ln}\left( N_{k_{\rm res}} a_{s}/a_{\perp} \right)$.  
The blue and green points correspond, respectively, to numerical results for $a_{s}/a_{\perp} = 1.7 \times 10^{-4}$ and $1.7 \times 10^{-3}$,
and we note that these two cases agree quite closely at intermediate and late times. 
In the right plot, the black line has 
slope $2$, and its intercept has been 
fitted to the intermediate-time behavior of the numerical data. 
The horizontal dashed lines show 
the quantum depletion times $a_{s}/a_{\perp}$. 
The deviations seen particularly for the green dots are due to a lack of statistics. 
\label{fig:SecondHarmonic}}
\end{figure}

This prediction is validated by the numerical results presented in Figure~\ref{fig:SecondHarmonic}, which shows the amplitude of the second harmonic at $k = 2 k_{\rm res}$ relative to that of the fundamental at $k = k_{\rm res}$,
for two of the simulations considered in the main text (the ``benchmark'' case with $a_{s}/a_{\perp} = 1.7\times 10^{-3}$ and that with $a_{s}/a_{\perp}$ ten times smaller).  
The key result is that there is a period during which the peak at $k=2 k_{\rm res}$ grows exponentially with an amplitude proportional to the square of the amplitude of the peak at $k=k_{\rm res}$, in agreement with the nonlinear theory described above.  

We also note that the evolution can be divided into three regimes.  During the first regime, $N_{k_{\rm res}}$ grows exponentially while $N_{2k_{\rm res}}$ remains essentially constant because of the smallness of nonlinearities.  In the mean, the initial value of $N_{2k_{\rm res}}$ is simply 
the quantum depletion, i.e., the 
vacuum expectation value $v_{2k_{\rm res}}^{2}$, where $v_{k}$ is the antidiagonal element of the Bogoliubov $SU(1,1)$ matrix entering Eq.~(\ref{eq:phi_varphi}).  This expectation value is represented by the dashed 
horizontal lines 
in the right plot of Fig.~\ref{fig:SecondHarmonic}~\footnote{The discrepancies seen 
at early time are 
due to a lack of statistics (here 100 simulations).  
The lack of early-time data for the green dots is due to the fact that the ensemble average of $N_{2k_{\rm res}}$ is a small negative number for this particular run, and therefore cannot be shown when representing $\mathrm{ln} N_{2k_{\rm res}}$.}.  The second regime is that during which both grow exponentially with the square relationship mentioned above.  Moreover, we note that, when rescaling each $N_{k}$ by $a_{s}/a_{\perp}$, the plots in this regime are practically the same for the two values of $a_{s}/a_{\perp}$, as was already noticed in Fig.~\ref{fig:Gamma}.  The third and final regime, which is also common to both values of $a_{s}/a_{\perp}$, is where the peaks become saturated, no longer growing in time and migrating slightly from the black line 
in the right panel.


\section{Exact solutions for a cylindrically symmetric condensate
\label{app:cylind_solns}}

We here provide a fairly detailed account of the $z$-independent solutions of the Gross-Pitaevskii equation 
in a cylindrically harmonic potential, following the treatment of Ref.~\cite{Kagan-Surkov-Shlyapnikov}.  The usefulness is in the derivation of approximate expressions for $A\left(n_{1}a_{s}\right)$ and $G\left(n_{1}a_{s}\right)$ appearing in Eq.~(\ref{eq:Veff_backreaction}).

\subsection{Adimensionalization}

Under the same assumptions adopted in Sec.~\ref{sub:factorization}, we arrive at Eq.~(\ref{eq:GPE-radial}) as our starting point.  However, it is convenient to work with adimensionalized quantities.  To this end, we choose an arbitrary fixed reference length $a_{\perp,0}$, and define its associated frequency $\omega_{\perp,0}=\hbar/m a_{\perp,0}^{2}$.  This allows us to define the following adimensionalized quantities:
\begin{equation}
T \equiv \omega_{\perp,0} t \,, \qquad R \equiv \frac{r}{a_{\perp,0}} \,, \qquad \Omega(T) \equiv \frac{\omega_{\perp}(t)}{\omega_{\perp,0}} = \frac{a_{\perp,0}^{2}}{a_{\perp}^{2}(t)} \equiv \frac{1}{A_{\perp}^{2}(T)} \,, \qquad \chi(R,T) \equiv a_{\perp,0} \, \psi(r,t) \,, 
\label{adim}
\end{equation}
upon which Eq.~(\ref{eq:GPE-radial}) becomes 
\begin{equation}
i \partial_{T}\chi = \left[-\frac{1}{2R}\partial_{R}R\partial_{R} + \frac{1}{2} \Omega_{\perp}^{2}(T) R^{2} + 2 n_{1} a_{s} \left|\chi\right|^{2} \right] \chi \,,
\label{GPE-adim}
\end{equation}
with $\chi$ normalized according to the adimensionalized form of the first of Eqs.~(\ref{eq:normalization}), i.e. 
\begin{equation}
\int_{0}^{\infty} R \left|\chi\left(R,T\right)\right|^{2} \dd R = 1 \,.
\label{R-norm}
\end{equation}

\subsection{Time-independent case}

Whenever $\Omega_{\perp}$ is constant, a stationary state of the time-independent GPE can be written as $\chi(R,T) = e^{-i M T} \widetilde{\chi}(R)$, where 
\begin{equation}
M \widetilde{\chi} = \left[ -\frac{1}{2R}\partial_{R}R\partial_{R} + \frac{1}{2} \Omega_{\perp}^{2} R^{2} + 2 n_{1} a_{s} \left|\widetilde{\chi}\right|^{2} \right] \widetilde{\chi} \,.
\label{GPE-stationary}
\end{equation}
Here, $M$ is the adimensionalized chemical potential, i.e. $M \equiv \mu/\hbar\omega_{\perp,0}$.  There are many solutions of Eq.~(\ref{GPE-stationary}), but we always assume the ground state, in which (for the given parameters) $M$ takes its lowest value.  For a given value of $n_{1} a_{s}$, we take the solution at $\Omega_{\perp}\equiv 1$ as a reference solution, i.e. we define $M_{0}$ and $\widetilde{\chi}_{0}$ such that 
\begin{equation}
M_{0} \widetilde{\chi}_{0} = \left[ -\frac{1}{2R}\partial_{R}R\partial_{R} + \frac{1}{2} R^{2} + 2 n_{1} a_{s} \left|\widetilde{\chi}_{0}\right|^{2} \right] \widetilde{\chi}_{0} \,.
\label{GPE-stationary-ref}
\end{equation}
The dependence of $M_{0}$ and $\widetilde{\chi}_{0}$ on $n_{1} a_{s}$ is implicitly determined by this equation.  Straightforward algebra shows that, for a general $\Omega_{\perp} = 1/A_{\perp}^{2}$, Eq.~(\ref{GPE-stationary}) is satisfied if $M$ and $\widetilde{\chi}$ are set equal to
\begin{equation}
M = \frac{M_{0}}{A_{\perp}^{2}} \,, \qquad \widetilde{\chi}\left(R\right) = \frac{1}{A_{\perp}} \widetilde{\chi}_{0}\left(\frac{R}{A_{\perp}}\right) \,.
\end{equation}
Note that the scaling of $\widetilde{\chi}\left(R\right)$ with $A_{\perp}$ ensures that the normalization condition~(\ref{R-norm}) is respected for $\widetilde{\chi}$ if it is respected for $\widetilde{\chi}_{0}$.

\subsection{Time-dependent case}

So far we have dealt with the stationary ground state solutions described by Eq.~(\ref{GPE-stationary}).  Let us now return to the time-dependent solutions described by Eq.~(\ref{GPE-adim}).  It turns out that, for a fixed value of $n_{1} a_{s}$ and assuming cylindrical symmetry, the time-dependent solutions can also be described by a straightforward rescaling of the reference solution $\widetilde{\chi}_{0}$, where the scale factor is now time-dependent.  Let us introduce the dimensionless scale factor $\Sigma(T)$, and make the following ansatz:
\begin{equation}
\chi\left(R,T\right) = \frac{1}{\Sigma(T)} \widetilde{\chi}_{0}\left(\frac{R}{\Sigma(T)}\right) \, \mathrm{exp}\left( i \theta_{0}(T) + i \frac{\Sigma^{\prime}(T)}{\Sigma(T)} \frac{R^{2}}{2} \right) \,.
\label{t-dep-ansatz}
\end{equation}
Plugging this into Eq.~(\ref{GPE-adim}) and using Eq.~(\ref{GPE-stationary-ref}) for the reference solution, we find that it reduces to the following:
\begin{equation}
\left[ \frac{\theta_{0}^{\prime}(T)}{\Sigma(T)} + \Sigma^{\prime\prime}(T) \frac{R^{2}}{2\Sigma^{2}(T)} + \frac{M_{0}}{\Sigma^{3}(T)} + \Omega_{\perp}^{2}(T) \Sigma(T) \frac{R^{2}}{2\Sigma^{2}(T)} - \frac{1}{\Sigma^{3}(T)} \frac{R^{2}}{2\Sigma^{2}(T)} \right] \widetilde{\chi}_{0}\left(\frac{R}{\Sigma(T)}\right) = 0 \,.
\end{equation}
Since this equation must hold for all $R$, the coefficients of $R^{0}$ and $R^{2}$ must vanish separately, yielding the following:
\begin{equation}
\theta_{0}^{\prime}(T) = - \frac{M_{0}}{\Sigma^{2}(T)} \,, \qquad \Sigma^{\prime\prime}(T) = -\Omega_{\perp}^{2}(T) \Sigma(T) + \frac{1}{\Sigma^{3}(T)} \,.
\label{theta-sigma-Gora}
\end{equation}
The second of these equations fully determines the time-evolution of the scale factor $\Sigma$ once initial conditions have been specified.  Note that it can be written in the form
\begin{equation}
\Sigma^{\prime\prime}(T) = -\frac{\partial}{\partial\Sigma} \widetilde{V}_{\rm eff}\left(\Sigma(T),T\right) \qquad \, {\rm where} \qquad \widetilde{V}_{\rm eff}\left(\Sigma,T\right) = \frac{1}{2} \Omega_{\perp}^{2}(T) \Sigma^{2} + \frac{1}{2 \Sigma^{2}} \,.
\end{equation}
Whenever $\Omega_{\perp}$ is time-independent, $\widetilde{V}_{\rm eff}$ has no explicit time dependence and the total (adimensional) effective energy
\begin{equation}
\widetilde{E}_{\rm eff} = \frac{1}{2} \dot{\Sigma}^{2} + \widetilde{V}_{\rm eff}\left(\Sigma\right) = \frac{1}{2} \dot{\Sigma}^{2} + \frac{1}{2} \Omega_{\perp}^{2} \Sigma^{2} + \frac{1}{2\Sigma^{2}}
\label{Eeff}
\end{equation}
is conserved.  (Note that, since $\Sigma = \sigma/a_{\perp,0}$, it is straightforward to show that $\widetilde{E}_{\rm eff} = E_{\rm eff} / \hbar \omega_{\perp,0}$, where $E_{\rm eff}$ is the effective energy of Eq.~(\ref{eq:Eeff}).)  Given this fact, and assuming $\Omega_{\perp}$ is $T$-independent, it can be shown that
\begin{equation}
\partial_{T}^{2} \left( \Sigma^{2} - \frac{\widetilde{E}_{\rm eff}}{\Omega_{\perp}^{2}} \right) = -\left(2\Omega_{\perp}\right)^{2} \left( \Sigma^{2} - \frac{\widetilde{E}_{\rm eff}}{\Omega_{\perp}^{2}} \right) \,,
\end{equation}
and hence that $\Sigma^{2}$ varies sinusoidally in time with frequency $2 \Omega_{\perp}$.  We can thus write the following general form for $\Sigma^{2}(T)$:
\begin{equation}
\Sigma^{2}(T) = \frac{\widetilde{E}_{\rm eff}}{\Omega_{\perp}^{2}} + A^{2} \, \mathrm{cos}\left(2 \Omega_{\perp} T + \phi\right) \,.
\label{Sigma2soln}
\end{equation}
The effective energy and $A^{2}$ are algebraically related.  Straightforward algebra yields the following relation:
\begin{eqnarray}
2 \widetilde{E}_{\rm eff} \Sigma^{2} & = & \frac{1}{4} \left(\partial_{T}\left(\Sigma^{2}\right)\right)^{2} + \Omega_{\perp}^{2} \left(\Sigma^{2}\right)^{2} + 1 \nonumber \\
& = & 1 + \Omega_{\perp}^{2} A^{4} - \frac{\widetilde{E}_{\rm eff}^{2}}{\Omega_{\perp}^{2}} + 2 \widetilde{E}_{\rm eff} \Sigma^{2} \,.
\end{eqnarray}
We thus have
\begin{equation}
\widetilde{E}_{\rm eff} = \Omega_{\perp} \sqrt{1 + \Omega_{\perp}^{2} A^{4}} \qquad \iff \qquad A^{2} = \frac{1}{\Omega_{\perp}} \sqrt{\frac{\widetilde{E}_{\rm eff}^{2}}{\Omega_{\perp}^{2}} - 1} \,.
\end{equation}
Finally, then, we can write the general form for $\Sigma^{2}(T)$, knowing the effective energy $\widetilde{E}_{\rm eff}$ which determines both the mean value and the amplitude of the oscillations:
\begin{equation}
\Sigma^{2}\left(T; \widetilde{E}_{\rm eff}\right) = \frac{1}{\Omega_{\perp}} \left[ \frac{\widetilde{E}_{\rm eff}}{\Omega_{\perp}} + \sqrt{\frac{\widetilde{E}_{\rm eff}^{2}}{\Omega_{\perp}^{2}}-1} \,\, \mathrm{cos}\left(2\Omega_{\perp}T + \phi\right) \right] \,.
\end{equation}
Note that the minimum possible value of $\widetilde{E}_{\rm eff}$ is $\Omega_{\perp}$, at which value the amplitude of the oscillations vanishes and $\Sigma = 1/\sqrt{\Omega_{\perp}}$ is constant in time. 

\subsection{Form of the energy}

Here we consider the expression for the total energy in Eq.~(\ref{eq:GPE-3D-energy}).  Using the factorization ansatz~(\ref{eq:factorization}) with $\phi \equiv \sqrt{n_{1}}$, and the relationship $g = 4 \pi \hbar^{2} a_{s} / m$, the total energy $E_{3D} = E_{\rm rad}$ where 
\begin{equation}
E_{\rm rad} = N \, \hbar\omega_{\perp} \, a_{\perp}^{2} \int_{0}^{\infty} \dd r \, r \left(\frac{1}{2} \left| \partial_{r}\psi \right|^{2} + \frac{r^{2}}{2 a_{\perp}^{2}} \left|\psi\right|^{2} + n_{1}a_{s} \left|\psi\right|^{4} \right) \,.
\label{r-energy}
\end{equation}
Using the adimensionalized quantities of Eqs.~(\ref{adim}), this can be written as
\begin{equation}
E_{\rm rad} = N \, \hbar\omega_{\perp,0} \int_{0}^{\infty} \dd R \, R \, \left( \frac{1}{2}\left|\partial_{R}\chi\right|^{2} + \frac{1}{2}\Omega_{\perp}^{2}R^{2}\left|\chi\right|^{2} + n_{1}a_{s} \left|\chi\right|^{4} \right) \,.
\end{equation}
Finally, we plug in the exact time-dependent solution~(\ref{t-dep-ansatz}), extracting any dimensionless integrals that depend only on the form of the reference solution $\widetilde{\chi}_{0}$.  The result is
\begin{equation}
E_{\rm rad} = N \, \hbar\omega_{\perp,0} \left\{ \left( \frac{1}{2} \dot{\Sigma}^{2} + \frac{1}{2}\Omega_{\perp}^{2}\Sigma^{2} \right) \, A\left(n_{1} a_{s}\right) + \frac{1}{2\Sigma^{2}} \, B\left(n_{1} a_{s}\right) \right\} \,,
\label{Ecylind-AB}
\end{equation}
where
\begin{equation}
A\left(n_{1} a_{s}\right) \equiv \int_{0}^{\infty} \dd R \, R^{3} \left| \widetilde{\chi}_{0}(R) \right|^{2} \,, \qquad B\left(n_{1} a_{s}\right) \equiv \int_{0} \dd R \, R \left|\partial_{R}\widetilde{\chi}_{0}(R)\right|^{2} + 2 n_{1} a_{s} \int_{0}^{\infty} \dd R \, R \left|\widetilde{\chi}_{0}(R)\right|^{4} \,.
\label{eq:A-B-defn}
\end{equation}
Using Eq.~(\ref{Eeff}), we can rewrite Eq.~(\ref{Ecylind-AB}) in the form
\begin{equation}
E_{\rm rad} = N \, \hbar \omega_{\perp,0} \left\{ \widetilde{E}_{\rm eff} \, A\left(n_{1}a_{s}\right) + \frac{1}{2\Sigma^{2}} \left( B\left(n_{1}a_{s}\right) - A\left(n_{1}a_{s}\right) \right) \right\} \,.
\end{equation}
Recalling that $\widetilde{E}_{\rm eff}$ is also constant in time, we conclude that conservation of $E_{\rm rad}$ implies the identity $A\left(n_{1}a_{s}\right) \equiv B\left(n_{1}a_{s}\right)$, and hence that the total energy is
\begin{equation}
E_{\rm rad} = N \, \hbar \omega_{\perp,0} \, \widetilde{E}_{\rm eff} \, A\left(n_{1}a_{s}\right) = N \, E_{\rm eff} \, A\left(n_{1}a_{s}\right) \,,
\end{equation}
which is exactly Eq.~(\ref{eq:GPE-radial-energy}).  $A\left(n_{1}a_{s}\right)$ is thus the same here as in Sec.~\ref{sec:system}, and we have an expression for it in Eqs.~(\ref{eq:A-B-defn}), albeit an implicit one since $\left|\widetilde{\chi}_{0}(R)\right|^{2}$ is not explicitly known.  We are also able to write a similar expression for $G\left(n_{1}a_{s}\right)$ using Eqs.~(\ref{eq:g1-defn}), (\ref{eq:g1-G}) and the adimensionalized quantities of Eqs.~(\ref{adim}):
\begin{equation}
G\left(n_{1}a_{s}\right) = \int_{0}^{\infty} \dd R \, R \, \left| \widetilde{\chi}_{0}(R) \right|^{4} \,. 
\label{G-chi0}
\end{equation}

\subsection{Form of the chemical potential}

To find an explicit expression for $A\left(n_{1}a_{s}\right)$ and $G\left(n_{1}a_{s}\right)$, it will prove useful to turn our attention to the chemical potential, for which (as shown in Refs.~\cite{Gerbier,Robertson-Michel-Parentani-1}) a very good analytic approximation is known. To this end, we multiply Eq.~(\ref{GPE-stationary-ref}) by $R \widetilde{\chi}_{0}^{\star}(R)$ and integrate over $R$.  The normalization condition~(\ref{R-norm}) ensures that the integral on the left-hand side is equal to $1$, leaving just $M_{0}$.  The first term on the right-hand side can be integrated by parts, and we find
\begin{eqnarray}
M_{0} & = & \frac{1}{2} \int_{0}^{\infty} \dd R \, R \, \left| \partial_{R}\widetilde{\chi}_{0}(R) \right|^{2} + \frac{1}{2} \int_{0}^{\infty} \dd R \, R^{3} \, \left|\widetilde{\chi}_{0}(R)\right|^{2} + 2 n_{1} a_{s} \int_{0}^{\infty} \dd R \, R \, \left|\widetilde{\chi}_{0}(R)\right|^{4} \nonumber \\
& = & A\left(n_{1}a_{s}\right) + n_{1}a_{s} \, G\left(n_{1}a_{s}\right) \,, 
\label{M0-AC}
\end{eqnarray}
where we have used Eqs.~(\ref{eq:A-B-defn}) and~(\ref{G-chi0}), as well as the identity $A\left(n_{1}a_{s}\right) \equiv B\left(n_{1}a_{s}\right)$.  Since $\widetilde{\chi}_{0}(R)$ is a reference stationary (ground state) solution corresponding to $\Omega_{\perp} = 1$, it has effective energy $\widetilde{E}_{\rm eff} = 1$ and hence its total energy is
\begin{equation}
E_{0} = N \, \hbar \omega_{\perp,0} \, A\left(n_{1}a_{s}\right) = N \, \hbar\omega_{\perp,0} \, A\left(N \frac{a_{s}}{L} \right) \,.
\end{equation}
The chemical potential $\mu_{0} = \partial E_{0} / \partial N$, and on adimensionalizing and differentiating, we find
\begin{equation}
M_{0} = A\left(n_{1}a_{s}\right) + n_{1}a_{s} \, A^{\prime}\left(n_{1}a_{s}\right) \,.
\label{M0-AAprime}
\end{equation}
Comparing with Eq.~(\ref{M0-AC}), we see that we must have 
\begin{equation}
G\left(n_{1}a_{s}\right) \equiv A^{\prime}\left(n_{1}a_{s}\right)
\label{G-Aprime}
\end{equation}
as an identity. 

\subsection{Approximate form of $A\left(n_{1} a_{s}\right)$}

It has been checked (see Fig.~12 of Ref.~\cite{Robertson-Michel-Parentani-1}) that, for the ground state solution and up to a maximum error of less than $2.5\,\%$, the chemical potential of the ground state is given by
\begin{equation}
\frac{\mu}{\hbar \omega_{\perp}} \approx \sqrt{1 + 4 n_{1} a_{s}} \,.
\end{equation}
Indeed, as seen in Ref.~\cite{Robertson-Michel-Parentani-1}, this is an excellent approximation both in the Gaussian limit when $n_{1}a_{s}$ is small and in the Thomas-Fermi limit when $n_{1}a_{s}$ is large.  We can thus use this to get an approximation for $A\left(n_{1}a_{s}\right)$.  It is straightforward to show that this is solved by
\begin{equation}
A\left(n_{1}a_{s}\right) \approx \frac{1}{6 n_{1}a_{s}} \left( \left(1+ 4 n_{1} a_{s}\right)^{3/2} - 1 \right) \,.
\label{A-explicit}
\end{equation}
This approaches $1$ as $n_{1}a_{s} \to 0$, as required by the Gaussian limit; and it approaches $\frac{4}{3} \sqrt{n_{1}a_{s}}$ when $n_{1}a_{s}$ is large, as can be calculated explicitly in the Thomas-Fermi limit.  It is Eq.~(\ref{A-explicit}), and its derivative with respect to $n_{1}a_{s}$, that have been used to determine $G\left(n_{1}a_{s}\right)/A\left(n_{1}a_{s}\right)$ of Eq.~(\ref{eq:Veff_backreaction}) in the numerical simulations described in the main body of this paper. 
When working with $n_{1}a_{s} = 0.6$, as in 
our numerical simulations, one finds that 
$A\left(n_{1}a_{s}\right) = 1.46$ and $G\left(n_{1}a_{s}\right) = 0.63$.


\end{appendices}

\bibliography{biblio}

\end{document}